\documentclass[10pt]{article}
\pdfoutput=1
\usepackage{jheparxiv}
\usepackage[latin1]{inputenc}
\usepackage{amsmath}
\usepackage{dsfont}
\usepackage{mathrsfs}
\usepackage{graphicx}
\usepackage{amssymb}
\usepackage{enumerate}
\usepackage[usenames,dvipsnames,svgnames,table]{xcolor}
\usepackage{cleveref}
\usepackage[colorinlistoftodos]{todonotes}
\usepackage{blkarray}
\usepackage{tikz}
\usepackage{graphicx}

\usepackage{tikzit}

\tikzstyle{new style 0}=[fill=black, draw=black, shape=circle, radius=0.1]

\numberwithin{equation}{section}

\newcommand\Tstrut{\rule{0pt}{2.6ex}}         
\newcommand\Ttstrut{\rule{0pt}{3.8ex}}         
\newcommand\Bbstrut{\rule[-2.3ex]{0pt}{0pt}}   

\makeatletter
\newcommand{\subalign}[1]{%
  \vcenter{%
    \Let@ \restore@math@cr \default@tag
    \baselineskip\fontdimen10 \scriptfont\tw@
    \advance\baselineskip\fontdimen12 \scriptfont\tw@
    \lineskip\thr@@\fontdimen8 \scriptfont\thr@@
    \lineskiplimit\lineskip
    \ialign{\hfil$\m@th\scriptstyle##$&$\m@th\scriptstyle{}##$\hfil\crcr
      #1\crcr
    }%
  }%
}
\makeatother

\def \tu {\tilde{u}}

\def \tm {\tilde{m}}
\def \tw {\tilde{w}}

\def \te {\tilde{\epsilon}}
\def \detp {\mathrm{det}^\prime}
\newcommand{\eps}[2]{\varepsilon^{\dot{#1}\dot{#2}}}

\newcommand{\half}{\frac{\scriptstyle 1}{\scriptstyle 2}}

\newcommand{\C}{\mathbb{C}}

\newcommand{\CP}{\mathbb{CP}}

\newcommand{\R}{\mathbb{R}}

\newcommand{\bbS}{\mathbb{S}}

\newcommand{\F}{\mathscr{F}}

\newcommand{\p}{\partial}

\newcommand{\e}{\mathrm{e}}

\newcommand{\cE}{\mathcal{E}}

\newcommand{\cI}{\mathcal{I}}
\newcommand{\cA}{\mathcal{A}}
\newcommand{\cO}{\mathcal{O}}

\newcommand{\Pf}{\mathrm{Pf}}

\newcommand{\tr}{\mathrm{tr}}
\newcommand{\sgn}{\mathrm{sgn}}

\newcommand{\rd}{\, \mathrm{d}}

\newcommand{\pf}{\mathrm{Pf}\,}
\newcommand{\be}{\begin{equation}\label}
\newcommand{\ee}{\end{equation}}
\newcommand{\bea}{\begin{eqnarray}\label}
\newcommand{\eea}{\end{eqnarray}}
\newcommand{\proof}{ \noindent {\bf Proof:} }
\newcommand{\la}{\langle}
\newcommand{\ra}{\rangle}
\newcommand{\lla}{\left\langle}
\newcommand{\rra}{\right\rangle}

\newtheorem{thm}{Theorem}

\newtheorem{propn}{Proposition}[section]

\newtheorem{corol}{Corollary}[section]
\newtheorem{lemma}{Lemma}[section]

\newcommand{\CHY}{{\scalebox{0.6}{$\mathrm{CHY}$}}}
\newcommand{\LG}{{\scalebox{0.6}{$\mathrm{LG}$}}}

\newcommand{\sA}{{\scalebox{0.6}{$A$}}}
\newcommand{\sB}{{\scalebox{0.6}{$B$}}}
\newcommand{\sC}{{\scalebox{0.6}{$C$}}}
\newcommand{\sD}{{\scalebox{0.6}{$D$}}}

\newcommand{\sL}{{\scalebox{0.6}{$L$}}}
\newcommand{\sR}{{\scalebox{0.6}{$R$}}}
\newcommand{\sI}{{\scalebox{0.6}{$I$}}}
\newcommand{\sJ}{{\scalebox{0.6}{$J$}}}
\newcommand{\sdotI}{{\scalebox{0.6}{$\dot I$}}}
\newcommand{\sdotJ}{{\scalebox{0.6}{$\dot J$}}}
\newcommand{\sN}{{\scalebox{0.6}{$N$}}}
\newcommand{\scN}{{\scalebox{0.6}{$\mathcal{N}$}}}
\newcommand{\stN}{{\scalebox{0.6}{$\tilde N$}}}
\newcommand{\scI}{{\scalebox{0.6}{$\mathcal{I}$}}}
\newcommand{\scJ}{{\scalebox{0.6}{$\mathcal{J}$}}}
\newcommand{\ssL}{{\scalebox{0.5}{$L$}}}
\newcommand{\ssR}{{\scalebox{0.5}{$R$}}}
\newcommand{\szr}{{\scalebox{0.8}{$\zeta_r$}}}
\newcommand{\sz}[1]{{\scalebox{0.8}{$\zeta_{#1}$}}}
\newcommand{\ssz}[1]{{\scalebox{0.5}{$\zeta_{#1}$}}}
\newcommand{\ete}{\epsilon\tilde\epsilon}

\newcommand{\mLabel}[1]{\mbox{$\scriptstyle{#1}$}}

\definecolor{light-gray}{gray}{0.95}
\definecolor{light-grayII}{gray}{0.85}

\title{\Large Recursion and worldsheet  formulae for 6d superamplitudes
}
\author[a]{Giulia Albonico,}
\author[b,c]{Yvonne Geyer}
\author[a]{\& Lionel Mason}

\affiliation[a]{The Mathematical Institute, University of Oxford\\24-29 St.~Giles, Oxford OX1 3LP, United Kingdom}

\affiliation[b]{Department of Physics, Faculty of Science, Chulalongkorn University\\
Thanon Phayathai, Pathumwan, Bangkok 10330, Thailand}

\affiliation[c]{School of Natural Sciences, Institute for Advanced Study \\Einstein Drive, Princeton, NJ 08540, USA}

\emailAdd{giulia.albonico@maths.ox.ac.uk}
\emailAdd{yjgeyer@gmail.com}
\emailAdd{lmason@maths.ox.ac.uk}

\abstract{
Recently two of the authors presented  a spinorial extension of the scattering equations, the \emph{polarized scattering equations} that incorporates spinor polarization data.  These led to new worldsheet amplitude formulae for a variety of gauge, gravity and brane theories in six dimensions that naturally incorporate fermions and directly extend to maximal supersymmetry.  This paper provides a number of improvements to the original formulae, together with extended details of the construction, examples and full proofs of some of the formulae by BCFW recursion and factorization.  We show how our formulae reduce to corresponding formulae for maximally supersymmetric gauge, gravity and brane theories in five and four dimensions.  In four dimensions our framework naturally gives the twistorial version of the 4d ambitwistor string, giving new insights into the nature of the refined and polarized scattering equations they give rise to, and on the relations between its measure and the CHY measure. Our formulae exhibit a natural double-copy structure being built from `half-integrands'. We give further discussion of the  matrix of theories and formulae to which our half-integrands give rise, including controversial formulae for amplitudes involving Gerbes.}

\begin{document}

\maketitle

\section{Introduction}
Worldsheet approaches to scattering amplitudes  generate perhaps the most compact and mathematically structured formulae for tree-level S-matrices and loop integrands available.  These formulations cannot at this stage be obtained from space-time action formulations. The first such formulae for field theory amplitudes (as opposed to conventional string theory amplitudes) arose from 
the twistor strings of Witten  \cite{Witten:2003nn}, Berkovits \cite{Berkovits:2004hg} and Skinner \cite{Skinner:2013xp}.  These give rise to remarkable worldsheet formulae for tree-level super Yang-Mills \cite{Roiban:2004yf, Roiban:2004ka} and gravity \cite{Cachazo:2012kg} in four dimensions.  These formulae were extended by Cachazo, He and Yuan (CHY) \cite{Cachazo:2013hca} to tree formulae for gravity and Yang-Mills amplitudes in all dimensions together with a variety of further theories \cite{Cachazo:2014dia} including D-branes and Born-Infeld theories, but without fermions or  supersymmetry. 

The CHY formulae are based on the \emph{scattering equations}.  These are equations for $n$ points on the Riemann sphere arising from the $n$ null momenta taking part in a scattering process. They were first discovered in conventional string theory as a semi-classical approximation in \cite{Fairlie:1972zz} and at high energy \cite{Gross:1987ar}.  They   were then seen to underpin the twistor string \cite{Witten:2004cp} and to naturally arise from string theories in the space of complex null geodesics, ambitwistor space \cite{Mason:2013sva} in an RNS formulation. These RNS ambitwistor models provide the worldsheet theories underpinning the CHY formulae and extend straightforwardly \cite{Casali:2015vta} to  incorporate the later CHY formulae \cite{Cachazo:2014dia}.
 The RNS ambitwistor model was followed by a fully supersymmetric pure spinor formulation  in 10 dimensions \cite{Berkovits:2013xba} but which does not lead to such explicit formulae for amplitudes.  Although the original RNS forms of ambitwistor string theories contain supersymmetry and fermions in their Ramond sectors, as do the pure spinor formulations more directly, it has been difficult to obtain explicit formulae for such amplitudes with  arbitrary numbers of fermions.  As such they don't directly make contact with the original twistor-string formulae by dimensional reduction.  

A framework was subsequently developed in six dimensions \cite{Heydeman:2017yww,Cachazo:2018hqa} that allowed the supersymmetric extension of the original CHY formulae and those for brane theories.
These models had some features of the original RSVW formulae \cite{Roiban:2004yf, Roiban:2004ka} in that moduli of  maps from the worldsheet to chiral spin space in six-dimensions  are integrated out against delta functions.    Although these authors were able to obtain amplitude formulae for a variety of supersymmetric theories in this way, there were a number of issues.  In particular the formulae distinguish between even and odd numbers of particles, and become quite awkward for odd numbers of particles in gauge and gravity theories where  such distinctions are not natural. Although a number of persuasive checks were made, there has been no attempt at a systematic proof of factorization or recursion for these formulae.  Their possible origins from worldsheet models remain obscure.

Subsequently the last two named authors of this paper introduced a distinct approach \cite{Geyer:2018xgb} based on extending the scattering equations to incorporate polarization data.  These \emph{polarized scattering equations}  have a geometric origin in string theories in six-dimensional ambitwistor space expressed in twistorial coordinates (although complete worldsheet theories that give rise to the full supersymmetric worldsheet formulae remain lacking). They were used to obtain compact formulae for amplitudes for a full range of six-dimensional theories,  now without any awkward distinction between even and odd numbers of particles for gauge and gravity theories. These formulae differed from those of \cite{Cachazo:2018hqa, Heydeman:2017yww} both in the underlying form of the scattering equations, and also provided a number of new integrand structures.  These  included  6-dimensional analogues of the 4d formulae of \cite{Geyer:2014fka} that provided a more efficient and compact version of the RSVW \cite{Roiban:2004yf, Roiban:2004ka}  and Cachazo-Skinner  formulae \cite{Cachazo:2012kg} for gauge and gravity theories, as well as formulae for D5 and M5-branes all expressed naturally in new supersymmetry representations. There were also more controversial formulae for \emph{Gerbe} multiplets with $(2,0)$ supersymmetry that were analogous to gauge theory amplitudes and with $(3,1)$ and $(4,0)$ supersymmetry that have some analogy with Gerbe-like gravity amplitudes.

In this article we give an improved and more detailed analysis of the formulae of \cite{Geyer:2018xgb}. We  shift the  supersymmetry representation in such a way as  to maintain the same simple exponential structure but so that it no longer depends on the solutions to the polarized scattering equations.   We present manifestly permutation invariant  expressions for the brane integrands, as well as direct computations for  three and four point amplitudes, which we compare to known answers previously obtained by recursion \cite{Cheung:2009dc,Huang:2010rn}.   For the polarized scattering equations we give a deeper analysis, showing that generically there is a unique solution for each solution to the conventional scattering equations: we prove that, although they are superficially expressed as nonlinear equations, the solutions can be obtained by normalizing solutions to a system of linear equations.  As a further check on the formulae, we derive the symmetry reductions to five  dimensions giving formulae for the same variety of theories there with maximal supersymmetry. We also show that the controversial $(0,2)$-PT, $(3,1)$ and $(4,0)$ formulae for interacting gerbes reduce to standard gauge and gravity formulae in 5d.  Reducing further to 4d we  land directly on the 4d ambitwistor string formulae of \cite{Geyer:2014fka}.  Our treatment  gives new insights there, giving an interpretation of the 4d refined  scattering equations introduced there as also being polarized scattering equations.  We also give a  proof via 6d of the relation between  the CHY measure in 4d with the 4d refined/polarized scattering equations measure.

Our main result consists of a proof of  factorization for all our gauge, gravity and brane formulae.  We also introduce a new spinorial realization of BCFW recursion adapted to 6d for gauge and gravity that therefore leads to a full proof of our formulae. Somewhat surprisingly, despite their poor power counting at large momenta, our brane formulae have no boundary contribution for large BCFW shifts.    

The paper is structured as follows.  In \S\ref{sec:review} we  give an extended introduction.  This contains a review of the formulae of CHY and the original scattering equations, the four dimensional formulae of \cite{Geyer:2014fka}.  We structure this four-dimensional discussion to highlight that these formulae were also based on 4d polarized scattering equations (as are the closely related RSVW formulae \cite{Roiban:2004yf} based on the original twistor-string).  The review  goes on to define the ingredients and  details of the six-dimensional formulae of \cite{Geyer:2018xgb} with some improvements and updates to include for example $(2,0)$-supergravities and statements of the main results. 
In \S\ref{sec:PSE} the polarized scattering equations and measure are studied in more detail.  It is shown that given a solution to the original scattering equations, there exists generically a unique solution to the polarized scattering equations which can be obtained essentially by solving linear equations and then normalizing. The associated measures are also shown to reduce to the CHY measure. Section \ref{sec:integrands} goes on to prove basic properties of the integrands we use, permutation invariance (see also appendix \ref{sec:perm}), invariance under supersymmetry and compatibility of the supersymmetry factors with the reduced determinants.  In \S\ref{sec:low-point} the three  and four point amplitudes are computed from the new formulae and shown to agree with the standard answers for the corresponding theories.  Section \ref{sec:dim-red} gives the symmetry reductions to give new formulae in five dimension, and then to the standard known formulae of \cite{Geyer:2014fka} in four dimensions, giving new insights into the relations between CHY and 4d refined/polarized scattering equations measures there.

The  full proof of the gauge and gravity formulae by  BCFW recursion is given  in section \ref{sec:BCFW}.  Along the way we  prove factorization  for all non-controversial formulae. Our BCFW shifts are different from those of other authors so we give a brief comparison in Appendix \ref{sec:BCFW_Cliff+Donal}. To give a practical example we use our BCFW shift to derive the four point formulae in appendix \ref{sec:3-4points}.  

Finally in \S\ref{sec:discussion} we discuss further issues and directions. These include a brief discussion of the Grassmannian approach of \cite{Cachazo:2018hqa} and its use in \cite{Schwarz:2019aat} to obtain a correspondence between the formulae studied in this paper and those of \cite{Cachazo:2018hqa}.  This leads to some brief remarks concerning  analogues of the momentum amplituhedron of \cite{Damgaard:2019ztj} in 6d.  There is also some discussion of ambitwistor worldsheet models and  the controversial formulae for Gerbe theories with $(2,0)$, $(3,1)$ and $(4,0)$ supersymmetry.


\section{Review and extended summary of results}\label{sec:review}
 We start with a review of the CHY formulae \cite{Cachazo:2013hca} for gauge and gravity theories with a brief mention of those for other theories \cite{Cachazo:2014nsa}.  We further give an introduction to the 4d refined/polarized scattering equation formulae of \cite{Geyer:2014fka} in such as a way as to bring out the analogy with the formulae that come later in 6d as the scattering equations there were extended to include an extra scaling per point that incorporates the polarization data.\footnote{ 
 In that paper, the equations were referred to as the refined scattering equations as the extra data and measures  distinguish the different MHV sectors so they were refined by MHV degree.}  This extended introduction then introduces the six-dimensional spinor-helicity formalism \cite{Cheung:2009dc}, polarized scattering equations, measures and integrands that underlie the formulae for the various different theories, and then summarizes the amplitude formulae and other main results of the paper.

\subsection{Review of CHY} 
For a scattering process involving $n$ null momenta $k_i$, the scattering equations arise from a meromorphic vector-valued function 
\begin{equation}
P(\sigma)_\mu=\sum_{i=1}^n\frac{k_{i\mu}}{\sigma-\sigma_i}\, ,\label{P-def}
\end{equation}
where $\sigma\in\C$ is a coordinate on the Riemann sphere $\CP^1$. When momentum is conserved, $P(\sigma)_\mu$ naturally transforms as a 1-form on $\CP^1$ under M\"obius transforms. Equivalently,  $P(\sigma)$ has weight $-2$ in homogeneous coordinates and  is a section of the line bundle $\cO(-2)$ on $\CP^1$. 
The scattering equations are then 
\begin{equation}
\mathrm{Res}_{\sigma_i} \frac{P^2(\sigma)}2=k_i\cdot P(\sigma_i)=\sum_j\frac{k_i\cdot k_j}{\sigma_{ij}}=0\, , \qquad  \sigma_{ij}=\sigma_i-\sigma_j. \label{SE}
\end{equation}
The scattering equations imply that $P^2(\sigma)$ is global and holomorphic, but it must then vanish as there are no global one-forms squared on $\CP^1$,  so $P(\sigma)_\mu$ is therefore null for all $\sigma$.   

The scattering equations then underpin the CHY formulae for massless scattering amplitudes in the form
\begin{equation}
\mathcal{M}_n=\int_{\mathfrak{M}_{0,n}}\!\! d\mu_n^{\CHY}\,\mathcal{I}\, ,
 \end{equation}
 where the CHY measure is defined by
 \begin{align}\label{eq:CHY-measure}
\int \cI  \; d\mu_n^{\CHY}&=\delta^d\left(\sum_{i=1}^n k_i\right)\int \cI \;\frac{\prod_{i=1}^n \delta(k_i\cdot P(\sigma_i)) d\sigma_i}{\mathrm{Vol}(\mathrm{SL}(2,\mathbb{C})\times \C^3)}\nonumber \\&
    =\delta^d\left(\sum_i k_i\right) \int\cI\;|lmn||pqr|\prod_{i\neq p,q,r} \bar \delta (k_i\cdot P(\sigma_i)) \prod_{j\neq l,m,n} d\sigma_j 
 \\&
    =\delta^d\left(\sum_i k_i\right) \sum\cI \; \frac{|lmn||pqr|}{\det \Phi_{lmn}^{pqr}}\, .\nonumber
 \end{align}
 Here, the Jacobians for the gauge-fixing and solving the scattering equations are given by
 \begin{equation}
 |pqr|:=\sigma_{pq} \sigma_{qr} \sigma_{rp}\, , \qquad \Phi_{ij}:= \frac{ \p k_i\cdot P(\sigma_i)}{\p \sigma_j},   
 \end{equation}
 and the superscript $pqr$ denotes the removal of the corresponding rows and subscript $lmn$ the corresponding  columns.  It is standard  that \eqref{eq:CHY-measure} is permutation invariant \cite{Cachazo:2013hca}.
The integration is over $\mathfrak{M}_{0,n}$, the space of $n$ marked points on the Riemann sphere, having divided by the volume of the M\"obius transformations SL$(2,\mathbb{C})$ in the Faddeev-Popov sense. (The second $\C^3$ factor is removed by removing the $pqr$ delta functions in the product and  replacing them with a further factor of $|pqr|$).
 The delta functions are understood as complex delta functions that localize the integral to a sum over the $(n-3)!$ solutions to the scattering equations of residues given by the integrand $\cI$ divided by the given Jacobian.
 
The integrands denoted $\cI$ vary from theory to theory.  They are usually a product of two factors $\cI=\cI_\sL^{\mathrm{h}}\cI_\sR^\mathrm{h}$ with each ``half-integrand''  $\mathcal{I}_{\sL,\sR}^{\mathrm{h}}$ transforming under M\"obius transformations as a 1-form in each $\sigma_i$.  In the original CHY formulae, two possibilities for these half-intgrands were discussed.  The first was  a Parke-Taylor factor that depends on a permutation $\rho$
\begin{equation}
\mathrm{PT}(\rho)=\prod_{i=1}^n\frac{1}{\sigma_{\rho(i)\, \rho(i+1)}}\, .
\end{equation}
The second was the CHY Pfaffian $\Pf'(M)$ where $M$ is the skew matrix that depends on polarization vectors $e_{i\mu}$ associated to each null momenta $k_{i\mu}$
\begin{equation}
M=\begin{pmatrix}
A&C\\-C^T&B
\end{pmatrix}\, , \quad A_{ij}=\frac{k_i\cdot k_j}{\sigma_{ij}}\, , \quad B_{ij}=\frac{e_i\cdot e_j}{\sigma_{ij}}\, , \quad C_{ij}=\begin{cases}  \frac{k_i\cdot e_j}{\sigma_{ij}}\,,\qquad i\neq j\\
\sum_l \frac{k_i\cdot e_l}{\sigma_{li}}\, , \quad i=j\, .\end{cases}
\end{equation}
On the support of the scattering equations, the matrices $M$ have a two-dimensional kernel, and so the Pfaffian $\pf M$ vanishes. One can however define a non-trivial  reduced Pfaffian by deleting two rows and columns, say $i$ and $j$, and quotienting by the corresponding generators of the kernel,
\begin{equation}
\Pf'(M):=\frac{1}{\sigma_{ij}}\Pf(M_{[ij]}). 
\end{equation}
This reduced Pfaffian is invariant under which rows and columns are removed.
We then obtain
\begin{subequations}\label{eq:CHY_ampl}
 \begin{align}
  & \text{Yang-Mills:} && \int\mathrm{PT}(\alpha)\;\Pf'(M) \,d\mu_{n}^{\mathrm{CHY}}
 \\
  & \text{Gravity:} && \int\Pf'(M)\Pf'(\tilde M)\,d\mu_{n}^{\mathrm{CHY}}\,,
 \end{align}
\end{subequations}
There are many related formulae.  Biadjoint scalar amplitudes are constructed from a product of two Parke-Taylors and further integrands  for Einstein-Yang-Mills, DBI, and other massless theories in \cite{Cachazo:2014nsa,Casali:2015vta}.

\subsection{The refined/polarized scattering equations in 4d}\label{sec:intro-4d}
In four dimensions, polarization data can be  presented in terms of spinor-helicity variables.  A null momentum $k_\mu$, $\mu=1,\ldots, d$,  is  expressed  for $d=4$ in terms of two-component spinors $k_{\alpha\dot \alpha}=\kappa_\alpha\tilde \kappa_{\dot \alpha}$, $\alpha=1,2$, $\dot \alpha= \dot 1, \dot 2$.   We will use  the conventional angle and square bracket notation to denote undotted and dotted spinor contractions
\begin{equation}
\langle\epsilon_i\epsilon_j\rangle:=\varepsilon_{\alpha\beta}\epsilon_i^\alpha\epsilon_j^\beta\, , \qquad [\tilde\epsilon_i\tilde \epsilon_j]= \varepsilon_{\dot \alpha\dot \beta}\tilde{\epsilon}_i^{\dot \alpha}\tilde \epsilon^{\dot \beta}_j\, .
\end{equation} 
We will, for the most part use complexified polarization data as we will take our Maxwell 2-forms to be simple and null, although momenta can be taken to be real.  So the little group is the $\C^*$ subgroup of the complexified Lorentz group that preserves the momentum and acts by rescaling $\kappa_\alpha$ and $\tilde \kappa_{\dot \alpha}$.   We take polarization data for a Maxwell field or gluon to be a null vector $e_\mu$ that is null and orthogonal to $k_\mu$. Null simple 2-forms are then either self-dual or anti-self-dual given by $F_{\mu\nu}=e_{[\mu}k_{\nu]}$ with $F_{\alpha\dot\alpha \beta \dot \beta}=\epsilon_{\alpha}\epsilon_{\beta}\varepsilon_{\dot \alpha \dot\beta}$ or its conjugate in terms of spinor-helicity data
$\epsilon_\alpha=\epsilon\kappa_\alpha$ or $\tilde\epsilon_{\dot\alpha}=\tilde \epsilon \tilde{\kappa}_{\dot \alpha}$  respectively. Thus, polarization simply associates a scale to either $\kappa_\alpha$ or $\tilde \kappa_{\dot \alpha}$.

In order to \emph{polarize} the scattering equations, we can  seek global meromorphic $\lambda(\sigma)_\alpha$ and $\tilde\lambda(\sigma)_{\dot \alpha}$ such that 
\begin{equation}
P(\sigma)_{\alpha\dot\alpha}=\lambda(\sigma)_\alpha\tilde\lambda(\sigma)_{\dot \alpha}\, .\label{P-fact}
\end{equation}
The weights of $\lambda(\sigma)_\alpha$ and $\tilde\lambda(\sigma)_{\dot \alpha}$ must add up to $-2$ to give $P$ and we will take them each to take values in $\cO(-1)$.  In 4d we have the freedom to let them take values in different line bundles $\lambda_\alpha\in\Omega^0(\Sigma,\mathcal{L})$, $\tilde \lambda_{\dot \alpha}\in\Omega^0(\Sigma, \tilde{\mathcal{L}})$ such that $\mathcal{L}\otimes \tilde{\mathcal{L}}\cong K_\Sigma$. While this set-up emerges naturally from  the original twistor-string and related models \cite{Cachazo:2013gna,Skinner:2013xp,Adamo:2014yya}, the higher dimensional analogues of \eqref{P-fact} will only make sense when both spinors take values in $\mathcal{O}(-1)$, and so the 4d ambitwistor-string model \cite{Geyer:2014fka} provides the more natural starting point.

Amplitudes in the 4d ambitwistor string are localized on  scattering equations that are refined by MHV degree as follows. Take $k$ gluons  $i=1,\ldots,k$ to have negative  helicity polarization $\epsilon_{i\alpha}=\epsilon_i \kappa_{i\alpha}$ and $p=k+1,\ldots ,n$ positive with polarization data $\tilde\epsilon_{i\dot \alpha}=\tilde\epsilon_i\tilde{\kappa}_{i\dot\alpha}$.  The equations then  incorporate the polarization data via the following ans\"atze for  $\lambda(\sigma)_\alpha$ and $\tilde\lambda(\sigma)_{\dot \alpha}$; 
\begin{equation}
\lambda(\sigma)_\alpha=\sum_{i=1}^k\frac{u_i\epsilon_{i\alpha}}{\sigma-\sigma_i}\, , \qquad \tilde\lambda(\sigma)_{\dot \alpha}=\sum_{p=k+1}^n\frac{u_i\tilde\epsilon_{i\dot\alpha}}{\sigma-\sigma_i}\, ,\label{lambda-def-4d}
\end{equation}
where the $\sigma_i$ and $u_i$ are together determined by the polarized scattering equations
\begin{equation}
u_p\lambda(\sigma_p)_\alpha=\frac{\kappa_{p\alpha}}{\tilde\epsilon_p}\, ,\quad p=k+1,\ldots n-k, \qquad u_i\tilde\lambda(\sigma_i)_{\dot \alpha}= \frac{\tilde{\kappa}_{i\dot\alpha}}{\epsilon_i}\,, \quad i=1,\ldots,k\, . 
\label{polscatt-4d}
\end{equation}
It is easy to see that the $\sigma_i$ satisfy the original scattering equations. In \cite{Geyer:2014fka}, these equations were incorporated into a measure 
\begin{equation}
d\mu_{n,k}^{4d}= 
\prod_{i=1}^k \delta^2
\left(u_i\tilde\lambda(\sigma_i)_{\dot\alpha}-\frac{\tilde\kappa_{i\dot\alpha}}{\epsilon_i}\right)
\prod_{p=k+1}^n \delta^2\left(u_p\lambda(\sigma_p)_\alpha-\frac{\kappa_{p\alpha}}{\tilde\epsilon_p}\right)
\frac{\prod_{j=1}^n d  \sigma_j d u_j /u_j}{\mathrm{Vol}(\mathrm{GL}(2,\mathbb{C}))}
\end{equation}
where the GL$(2,\mathbb{C})$ extends the SL$(2,\mathbb{C})$ M\"obius invariance to include the little group $\C^*=\mathrm{GL}(1)$ generated by 
\begin{equation}
\sum_{i\leq k} u_i\p/\p u_i -\sum_{p>k} u_p\p/\p u_p.
\end{equation} 
The quotient by GL$(2,\C)$ removes the first three $d\sigma_i$ and one $du_i$ whilst introducing a factor of $\sigma_{12}\sigma_{23}\sigma_{13}$ but no delta functions are removed.  The  four-momentum conserving delta functions, do not need to be inserted manually, as they are  implied by the delta functions.    This measure is related to the CHY measure by
\begin{equation}
\prod_{i,p}\epsilon_i\tilde\epsilon_p \int d \mu_{n,k}^{4d}\,\,\cI^{4d} =\int  d \mu_n^{\CHY}\, \det{}'H^k\,\cI^{4d}\, .\label{4d-measures}
\end{equation}
Although this is clear from an indirect general argument as described in \S\ref{sec:4d-massless}, we also give a detailed proof there via 6d. Here the symmetric matrix $H^k$ is defined on each MHV sector by
\begin{equation}\label{eq:Hodges}
H_{ij}^k=\begin{cases}  \frac{\langle\epsilon_i  \epsilon_j\rangle}{\sigma_{ij}} , \quad i,j\leq k\\
\frac{[\tilde\epsilon_i \tilde \epsilon_j]}{\sigma_{ij}} , \quad i,j >k ,
\end{cases} \quad \mbox{for $i\neq j$,   }\quad H_{ii}^k=\begin{cases} -\frac{\langle \epsilon_i \lambda(\sigma_i)\rangle }{u_i}\, , \quad i\leq k \\  -\frac{[\tilde \epsilon_i \tilde\lambda(\sigma_i)]}{u_i}\, , \quad i>k\, ,\end{cases} 
\end{equation}
with vanishing entries otherwise.  It follows straightforwardly from \eqref{lambda-def-4d} that $H$ has a two-dimensional kernel spanned by  the vectors $(u_1, \ldots ,u_k,0,\ldots ,0)$ and $( 0,\ldots, 0,u_{k+1},\ldots ,u_n)$.  Its reduced determinant is defined  by 
\begin{equation}
\det{}'H^k:=\frac{\det H^k{}_{[ij]}^{[lm]}}{u_iu_ju_lu_m}
\end{equation}
where $H_{[ij]}^{[lm]}$ is the matrix with rows  $i,j$  and columns $l,m$ removed with $l\leq k< m $, $i\leq k <j$.
We remark that $\det{}'H^k$ is supported on the sectors appropriate to N$^{k-2}$MHV degree\footnote{This can be seen from the ranks $k-1$ and $n-k-1$ respectively of the $H$ and $\tilde H$ matrices of the Cachazo-Skiner formulae \cite{Cachazo:2012kg,Cachazo:2012pz} and their relationships to $H^k$ \cite{Geyer:2016nsh}.} \cite{Zhang:2016rzb}.  The full $(n-3)!$ set of solutions to the scattering equations break up into the N$^{k-2}$MHV sectors with $k=2,\ldots ,n-2$ with Eulerian number\footnote{$A(n,m)$ is the number of permutations of $n$ elements in which $m$ elements are greater than their predecessors after the permutation.} $A(n-3,k-2)$ in each sector.  

This reduced determinant plays a dual role in that it agrees with the CHY Pfaffian $\Pf'(M)$ when the polarization data is restricted to the appropriate MHV degree. Thus, because \eqref{4d-measures} already essentially contains one CHY Pfaffian, the integrand for Yang-Mills formula is simply the Parke-Taylor factor and the one for gravity contains one additional copy of $\det{}'(H)$.

These formulae directly extend to incorporate supersymmetry either by using chiral or anti-chiral supermomenta.  For super-Yang-Mills with $\mathcal{N}= 4$ supersymmetries, our  supermultiplets will be either chiral  or antichiral with the supermultiplet given by
\begin{align}\label{eq:4d-multiplet}
(F_{\alpha\beta},\psi_{\alpha \scI},\Phi_{\scI\scJ},\psi^\scI_{\dot \alpha},\tilde F_{\dot\alpha\dot\beta})&=\left(\epsilon_\alpha\epsilon_\beta,\epsilon_\alpha q_\scI, q_\scI q_\scJ, \frac{\tilde\kappa_{\dot\alpha}}{\epsilon}q^{3\scI}, \frac{\tilde\kappa_{\dot\alpha}\tilde\kappa_{\dot\beta}}{\epsilon^2}q^4\right)\e^{ik\cdot x}\nonumber
\\
&=\left(\tilde q^4\frac{\kappa_\alpha\kappa_\beta}{\tilde\varepsilon^2},\frac{\kappa_\alpha}{\tilde\epsilon} \tilde q^3_\scI,\half\varepsilon_{\scalebox{0.6}{$\mathcal{IJKL}$}} \tilde q^{\scalebox{0.6}{$\mathcal{K}$}}\tilde q^{\scalebox{0.6}{$\mathcal{L}$}}, \tilde\epsilon_{\dot\alpha}\tilde q^\scI, \tilde\epsilon_{\dot\alpha}\tilde\epsilon_{\dot\beta}\right)\e^{ik\cdot x}
\end{align}
respectively where  $q^{3\scI}=\varepsilon^{\scalebox{0.6}{$\mathcal{IJKL}$}}q_\scJ q_{\scalebox{0.6}{$\mathcal{K}$}}q_{\scalebox{0.6}{$\mathcal{L}$}}/6$ and $q^4=q^{3\scI}q_\scI/4$ etc..   These are obtained from each other by $\epsilon=1/\tilde \epsilon$ and fermionic Fourier transform from $q_\scI$ to $\tilde q^\scI$.  At $\mathcal{N}=4$ these multiplets are the same. For $\mathcal{N}<4$ we can define them in an obvious way so as to be  complementary. 

To obtain supersymmetric formulae at N$^{k-2}$MHV, we  partition $ \{1,\ldots,n\}=Y\cup \bar Y$ with $|Y|=k$ and particles $i\in Y$  in the first representation and $i\in \bar Y$ in the second and introduce the supersymmetry factor $\e^{F_\scN^k}$  with
\begin{equation}
F_\scN^k=\sum_{i\in Y , j\in \bar Y}\frac{u_iu_j}{\sigma_i-\sigma_j}q_{i\scI} \tilde{q}^\scI_j\, .
\end{equation}
We now obtain the following supersymmetric  4d amplitude formulae
\begin{subequations}\label{eq:4d_ampl}
 \begin{align}
  & \text{Super Yang-Mills:} && \int\mathrm{PT}(\alpha)\;e^{F_\scN^k}\,d\mu_{n,k}^{4d}
 \\
  & \text{Supergravity:} && \int\det{}'H^k\;e^{F_\scN^k}\,d\mu_{n,k}^{4d}\,,
 \end{align}
\end{subequations}
with $\mathcal{N}\leq 4$ for Yang-Mills theory and $\mathcal{N}\leq 8$ for gravity.

\subsection{Polarized scattering equations framework in 6 dimensions}
We here recall basic definitions from \cite{Geyer:2018xgb}.

\paragraph{Spinor  helicity in 6d:}  In six dimensions, vectors transform in the antisymmetric representation of $\mathrm{SL}(4,\C)$, the spin group of the Lorentz group $\mathrm{Spin}(6,\C)$. Thus a 6-momentum can be expressed as  $k^{\sA\sB}=k^{[\sA\sB]}=\gamma_\mu^{\sA\sB}k^\mu$, where $A, B=0,\ldots , 3$ are spinor indices and $\gamma_\mu^{\sA\sB}$ are antisymmetric $4\times4$ Pauli matrices, the chiral constituents of the $\gamma$-matrices satisfying the Clifford algebra. The inner product of vectors is defined via the totally skew, $\mathrm{SL}(4)$-invariant tensor $\half\varepsilon_{\sA\sB\sC\sD}$, which is also used to raise and lower skew pairs of spinor indices. 

For massless particles, the little group is given by $\mathrm{Spin}(4,\C)\cong\mathrm{SL}(2)\times\mathrm{SL}(2)$. Since null momenta $k^{\sA\sB}$ with $k^2=k^{\sA\sB}k^{\sC\sD}\varepsilon_{\sA\sB\sC\sD}=0$ are of rank two due to the antisymmetry of the spinor indices, the on-shell condition can  be solved by chiral (or antichiral) spinors \cite{Cheung:2009dc},
\begin{equation}
k^{\sA\sB}= \varepsilon ^{\dot a\dot b} \kappa_{\dot a}^\sA\kappa_{\dot b}^\sB\equiv\left[\kappa^\sA\kappa^\sB\right]\,,\qquad k_{\sA\sB}= \kappa^{ a}_\sA \kappa^{b}_\sB \varepsilon_{ a b}\equiv\left\langle\kappa_\sA\kappa_\sB\right\rangle\,.\label{eq:momenta}
\end{equation}
Here, $a=0,1$, $\dot a=\dot 0, \dot 1$ are the corresponding $\mathrm{SL}(2,\C)$ little group spinor indices, and we have introduced the four-dimensional notation $\langle\cdot,\cdot \rangle$ and $[\cdot,\cdot]$ brackets now used to denote little group contractions.  

Polarization data is made up of representations of the little group. A Dirac particle has polarization data $\epsilon_\sA=\epsilon_a\kappa^a_\sA$.  A Maxwell field strength is represented by $F^\sA_\sB$, with $F_\sA^\sA=0$ because the Lie algebra  of the Lorentz group is $\mathfrak{sl}(4)$. For a momentum eigenstate, with a null polarization vector orthogonal to $k$, we find 
\begin{equation}
F^\sA_\sB=\epsilon^\sA\epsilon_\sB\,.\label{eq:F}
\end{equation}
The Maxwell equations require $k_{\sA\sB}\epsilon^\sA=0=k^{\sA\sB}\epsilon_\sB$, so that all polarization data is encoded in little group spinors $\epsilon_a$ and $\epsilon_{\dot a}$ 
with\footnote{Note that $\epsilon_a$ and $\epsilon_{\dot a}$ cannot be taken to be real in Lorentz signature.}
\begin{equation}\label{eq:pol}
\epsilon^\sA=\epsilon_{\dot a}\kappa^{\sA\dot a}\, , \qquad \epsilon_\sA=\epsilon_a\kappa_{\sA}^a\, .
\end{equation}

\paragraph{6D polarized scattering equations:}
Now in 6d, we can seek a spinor-helicity
factorization for  $P(\sigma)$ over $\CP^1$
\begin{equation}
P_{\sA\sB}=\lambda_{\sA a}\lambda_{\sB}^a= \half\varepsilon_{\sA\sB\sC\sD}\lambda^\sC_{\dot a}\lambda^{\sD\dot a}\, .
\end{equation}  
The scattering equation $k_i\cdot P(\sigma_i)=0$ implies  $k_i\cdot P=\mathrm{det} (\kappa^a_{i\sA},\lambda^b_\sA)=0$.  This determinant  vanishes iff there exists non zero $(u^a_i,v^a_i)$ defined up to scale so that 
\begin{equation}
\cE_{i\sA}:=u_{ia}\lambda^a_\sA(\sigma_i) -v_{ia}\kappa_{i\sA}^a=0\, .
 \label{6dSE+}
\end{equation}
This is scale invariant in $u$ and $v$, so we can  normalize
\begin{equation}
\langle v_i\epsilon_i\rangle=1\, .\label{SE-norm}
\end{equation} 
We introduce an analogue of \eqref{P-def} for $\lambda_{\sA a}(\sigma)$
\begin{equation}
\lambda_{\sA a}(\sigma)=\sum_{i=1}^n\frac{u_{ia}\epsilon_{i\sA}}{\sigma-\sigma_i}\, .\label{lambda-def}
\end{equation}
Together, \eqref{6dSE+}, \eqref{SE-norm} and \eqref{lambda-def} will constitute the \emph{polarized scattering equations}.  One motivation for this latter formula arises  from a heuristic twistorial ambitwistor-string model that was presented in \cite{Geyer:2018xgb}.

These provide our 6D polarized version of the 4d polarized scattering equation \eqref{polscatt-4d}  as equations on the $(\sigma_i, u_{ia}, v_{ia})$ that determine the $(u_{ia},v_{ia})$ on the support of a solution $\sigma_i$ to the ordinary scattering scattering equations. 
More explicitly we can write
\begin{equation}
\cE_{i\sA}:=\sum_j \frac{\langle u_i u_j\rangle \epsilon_{j\sA}}{\sigma_{ij}}-\langle v_i \kappa_{i\sA} \rangle=0\, .\label{polscatt}
\end{equation}
We can eliminate the $v_i$ from these equations by skewing with $\epsilon_{i\sA}$ to get
\begin{equation}
\epsilon_{i[\sA}\cE_{\sB]i}:=\sum_j \frac{\langle u_i u_j\rangle \epsilon_{j[\sB}\epsilon_{\sA]i}}{\sigma_{ij}}-k_{i\sA\sB}=0\, ,\label{polscatt-k}
\end{equation}
which follows from the normalization condition on $v_i$.  Although these are 6 equations, skewing with $\epsilon_{i\sC}$ vanishes identically by construction and there are only three independent equations per point that serve to determine the $u_{ia}$ and $\sigma_i$.
Summing this version of the equations over $i$, the first double sum vanishes being antisymmetric over $i,j$, leaving the sum of momentum showing that these equations imply momentum conservation.

 Although as presented, the equations for $u_{ia}$ appear nonlinear, later we will see that they are underpinned by linear equations, and, in proposition \ref{unique}, that there exists a unique solution to these equations for each solution $\sigma_i$ to the unpolarized scattering equation.\footnote{Unique up to an SL$(2,\C)$-transformation on the global $a$ index.}

\paragraph{Integral formulae:} Our integral formulae for amplitudes  all take the form
\begin{equation}
\cA_n=\int \mathcal{I}_n \rd \mu_n^{\mathrm{pol}}\label{eq:form_amplitude}
\end{equation}
where the integrands $\mathcal{I}_n$ are theory specific and will be specified in due course.  We define the measure based on the chiral 6D polarized scattering equations by
\begin{equation}
 \rd\mu_n^{\mathrm{pol}}= \frac{\prod_{i=1}^n  \delta^4\Big(\cE_{i\sA}\Big) \delta\big(\langle v_i \varepsilon_i \rangle -1 \big)\; \rd \sigma_i\,\rd^2 u_i\,\rd^2 v_i}{\mathrm{vol}\; \mathrm{SL}(2,\mathbb{C})_\sigma \times \mathrm{SL}(2,\mathbb{C})_+}\,. \label{measure6d}
\end{equation}
Here the  two copies of SL(2) are the M\"obius transformations on $\sigma$ and the little group on the little $a$ index and the division by their volumes are   understood in the usual Faddeev-Popov sense.  We will however see that this measure is equal to the CHY measure in \S\ref{CHY=pol}. 

\subsection{Supersymmetry in 6d}\label{sec:susy}
Here we review supersymmetry representations in 6d, in particular that in \cite{Geyer:2018xgb}.  That representation depends on individual solutions to the scattering equations, so we introduce a variant that maintains the same simple structure, but that is global.   

Supersymmetry representations in 6 dimensions have been explored in the context of scattering ampitudes by a number of authors \cite{Cheung:2009dc, Dennen:2009vk, Huang:2010rn}.
In six dimensions, $(N,\tilde N)$-supersymmetry  possesses an $\mathrm{Sp}(N)\times\mathrm{Sp}(\tilde N)$ R-symmetry group for which we introduce indices $I=1,\ldots , 2N,$ and $\dot I=\dot 1, \ldots,\dot {2\tilde N}$.   
On momentum eigenstates with momentum $k_{AB}$, the supersymmetry generators $Q_{\sA\sI}$ and $Q^\sA_{\sdotI}$ satisfy, temporarily  suppressing the particle index $i$ for readability,
\begin{equation}
\{Q_{\sA \sI},Q_{\sB \sJ}\}=k_{\sA\sB} \, \Omega_{\sI\sJ},\quad\{Q^\sA_{\dot \sI},Q^\sB_{\dot \sJ}\}=k^{\sA\sB} \, \Omega_{\dot \sI\dot \sJ}
\end{equation} where $\Omega_{\sI\sJ}$ and $\Omega_{\sdotI \sdotJ}$ are the R-symmetry symplectic metrics.
The supersymmetry generators thus reduce to the little group as 
\begin{equation}
 Q_{A I}=\kappa_A^a Q_{aI}\,,\qquad Q^{A}_{\dot I}=\kappa^A_{\dot a} Q_{\dot I}^{\dot a}
\end{equation}
where we now have
\begin{equation}
\{Q_{aI},Q_{bJ}\}=\varepsilon_{ab}\Omega_{IJ}\, , \qquad 
\{Q_{\dot a\dot I},Q_{\dot b\dot J}\}=\varepsilon_{\dot a\dot b}\Omega_{\dot I\dot J}\,. \label{LG-Q}
\end{equation}

\paragraph{Super Yang-Mills.} A key example is $(1,1)$ super Yang-Mills theory.  The linearized `super-Maxwell' multiplet is
\begin{equation}
\F:=(F_\sA^\sB, \,\psi^\sA_{\sI},\,\tilde\psi_{\sA \sdotI},\, \phi_{\sI\sdotI})\, ,\label{eq:supermaxwell-mult}
\end{equation}
consisting of a 2-form curvature $F^\sB_\sA$, spinors of each chirality $\psi^\sA_\sI$ and $\tilde \psi_{\sA\sdotI}$ and four scalars $\phi_{\sI\sdotI}$. On momentum eigenstates with null momentum $k_{\sA\sB}$,  $Q_{\sC \sJ}$ acts on this multiplet by
\begin{eqnarray}
Q_{\sC \sJ}\F&=&(k_{\sA\sC}\psi^\sB_{\sJ},\, \Omega_{\sJ\sI}F_\sC^\sA, \,k_{\sA\sC}\phi_{\sJ\sdotI}, \,\Omega_{\sJ\sI} \tilde \psi_{\sdotI \sC})\, .
\end{eqnarray}
To construct a supersymmetry representation, we need to choose half of the $Q_{a\sI}$ as anticommuting supermomenta.  The possibilities  discussed in the literature  \cite{Cheung:2009dc,Dennen:2009vk, Huang:2010rn} focus on halving either the $I$ or the $a$-indices manifesting only  full little-group or only R-symmetry respectively. The former was used successfully implemented in recent work on 6d scattering amplitudes for a variety of theories \cite{Cachazo:2018hqa, Heydeman:2017yww}.  However, the latter is more natural from the perspective of the ambitwistor string \cite{Bandos:2014lja}, and will be the formulation we work with here. The two approaches are of course related by appropriate Grassmann Fourier transforms and we discuss the details of the R-symmetry breaking approach and its correspondence  with the little group breaking approach used in this section in \S\ref{sec:susy-factor}.

For amplitudes in the representation \eqref{eq:form_amplitude} based on the polarized scattering equations, there is a natural choice of supermomenta that manifests the full R-symmetry, because the polarized scattering equations provide  a natural basis $(\epsilon_a, v_a)$ of the little group space for each particle so that $\epsilon^aQ_{a\sI}^\mathrm{pol}$ anti-commute. They can therefore be represented as Fermionic variables
 \begin{equation}
 q_\sI :=\epsilon^aQ_{a\sI}^\mathrm{pol}.\label{eq:superm}
\end{equation}  
This allows us to write the supersymmetry generators as
\begin{equation}
Q_{a\sI}^\mathrm{pol}=\left( v_a q_\sI +\epsilon_a \Omega_{\sI\sJ}\frac{\partial}{\partial q_\sJ}\right)\,,\qquad \tilde Q_{\sdotI}^{\mathrm{pol}\,\dot a}=\left( v^{\dot a} \tilde q_{\sdotI} +\epsilon^{\dot a} \tilde\Omega_{\sdotI\sdotJ}\frac{\partial}{\partial \tilde{q}_{\sdotJ}}\right)\, .\label{eq:def_susy-gen_v1}
\end{equation}
The full super Yang-Mills  multiplet is then obtained from the pure gluon state $\F(0,0)=(\epsilon_\sA\epsilon^\sB,0,0,0)$ as
\begin{equation}
\F_{\mathrm{pol}}(q_\sI,\tilde q_{\sdotI})=\left((\epsilon_\sA+ q^2\langle v \kappa_\sA\rangle)(\epsilon^\sB+ \tilde q^2 \langle v \kappa^\sB\rangle),\right.  \left. q_\sI(\epsilon^\sA+\tilde q^2 \langle v \kappa^\sA\rangle),\, \tilde q_{\sdotI}(\epsilon_\sA+q^2 \langle v \kappa_\sA\rangle),\, q_\sI\tilde q_{\sdotI}\right)\, .
\end{equation}
  This gives a representation of the anti-commutation relations \eqref{LG-Q} such that the $(1,1)$-super-Yang-Mills superfield becomes  
\begin{equation}
 \Phi^{\mathrm{R}}_{\mathrm{pol}} =g^{\ete}+q_\sI\,\psi^{\sI\tilde\epsilon}+\tilde q_{\sdotJ} \,\tilde\psi^{\epsilon \sdotJ}+q^2 g^{v\tilde\epsilon}+\tilde q^2 g^{\epsilon\tilde v}+q_\sI\tilde q_{\sdotJ} \,\phi^{\sI \sdotJ}+\dots +q^2\tilde q^2\,g^{v\tilde v}\,.\label{fieldR_v}
\end{equation}
where $g^{\ete}= \epsilon_a\tilde \epsilon_{\dot a} \,g^{a\dot a}$ is the gluon with polarization $\epsilon_a\tilde \epsilon_{\dot a} $ etc. This explicit form of the multiplet highlights one of the peculiar features of this supersymmetry representation: Since the supersymmetry generators depend via $v$ on the individual solutions to the polarized scattering equations, so do all states in the bottom half of the multiplet,  e.g. $g^{v\tilde v}$ or $g^{v\tilde\epsilon}$. The supersymmetry representation is thus dynamic, not just particle-specific, and varies with the solution to the scattering equations, i.e., $v_{ia}$ is not specified in advance, but depends on the momenta and polarization data and an individual solution to the scattering equations. 
While any issues associated to this peculiarity can be easily avoided by only calculating amplitudes with external states at the top of the multiplet,\footnote{ i.e. taking all gluons as $g^{\ete}$, fermions as $\psi^{\sI\tilde\epsilon}$ or $\tilde\psi^{\epsilon \sdotJ}$ and scalars as $\phi^{\sI\sdotJ}$. This can always be achieved by a choice of polarization. Note in this context that the supermomenta themselves only depend on the $\epsilon_{ia}$  from \eqref{eq:superm}.} we prefer to work with a global supersymmetry representation that can be introduced as follows.

\paragraph{The new representation.}Instead of using the basis $(\epsilon_a, v_a)$ of the little group introduced by the polarized scattering equations (which depends on the solutions to the scattering equations), let us choose a global basis for each particle
\begin{equation}
 \left(\epsilon_{ia},\xi_{ia}\right)\,,\qquad\qquad\text{ with } \;\la\xi_i \epsilon_i\ra=1\,.
\end{equation}
Using this basis,  $\epsilon^aQ_{a\sI}$ again anti-commute, and can be represented by Grassmann viariables $q_\sI=\epsilon^aQ_{a\sI}$. However, the supersymmetry generators are now globally defined,
\begin{equation}
Q_{a\sI}=\left( \xi_a q_\sI +\epsilon_a \Omega_{\sI\sJ}\frac{\partial}{\partial q_\sJ}\right)\,,\qquad \tilde Q_{\sdotI}^{\dot a}=\left( \xi^{\dot a} \tilde q_{\sdotI} +\epsilon^{\dot a} \tilde\Omega_{\sdotI\sdotJ}\frac{\partial}{\partial \tilde{q}_{\sdotJ}}\right)\, .\label{eq:def_susy-gen}
\end{equation}
Note that due to the normalization condition $\la v\epsilon\ra=1$, we know that $v_a$ and $\xi_a$ are related by
\begin{equation}\label{eq:v_xi}
 v_a = \xi_a +\la \xi v\ra \epsilon_a\,.
\end{equation}
This implies that the supersymmetry generators $Q_{a\sI}^\mathrm{pol}$ and $Q_{a\sI}$ are \emph{not} related by a linear transformation of the respective supermomenta $q_\sI$. Returning to the example of super Yang-Mills, the multiplet now takes the form 
\begin{equation}
\F(q_\sI,\tilde q_{\sdotI})=\left((\epsilon_\sA+ q^2\langle \xi \kappa_\sA\rangle)(\epsilon^\sB+ \tilde q^2 \langle \xi \kappa^\sB\rangle),\right.  \left. q_\sI(\epsilon^\sA+\tilde q^2 \langle \xi \kappa^\sA\rangle),\, \tilde q_{\sdotI}(\epsilon_\sA+q^2 \langle \xi \kappa_\sA\rangle),\, q_\sI\tilde q_{\sdotI}\right)\, ,
\end{equation}
and the $(1,1)$-super-Yang-Mills superfield becomes  
\begin{equation}
 \Phi^{\mathrm{R}} =g^{\ete}+q_\sI\,\psi^{\sI\tilde\epsilon}+\tilde q_{\sdotJ} \,\tilde\psi^{\epsilon \sdotJ}+q^2 g^{\xi\tilde\epsilon}+\tilde q^2 g^{\epsilon\tilde \xi}+q_\sI\tilde q_{\sdotJ} \,\phi^{\sI \sdotJ}+\dots +q^2\tilde q^2\,g^{\xi\tilde \xi}\,.\label{fieldR}
\end{equation}
where as above $g^{\ete}= \epsilon_a\tilde \epsilon_{\dot a} \,g^{a\dot a}$ denotes the gluon with polarization $\epsilon_a\tilde \epsilon_{\dot a} $. By construction, this representation is now global and independent of the solution to the polarized scattering equations. Of course, this global definition comes at the expense of having to introduce an additional  reference  spinor $\xi_a$, whereas the dynamic representation $\Phi^{\mathrm{R}}_\mathrm{pol}$ only depends on a single choice of polarization spinor.
\\

For the most part hereon, we will work in the global R-symmetry preserving representation $\Phi^{\mathrm{R}}$. However, it is easy to  convert our formulae to  the little-group preserving  representation: for this we break up $Q_{aI}=(Q^l_a,Q_{al},)$ with $l=1,\ldots,N$ so that $\Omega_{IJ}=\begin{pmatrix}
0&\delta^l_m\\-\delta^m_l&0
\end{pmatrix}$ and introduce supermomenta $\eta_{al}$ so that
\begin{equation}
Q_{aI}=\left(\frac{\p}{\p \eta^a_l},\eta_{al}\right)\, .\label{eq:little-gp-pres}
\end{equation}
We explain the correspondence in more detail in \ref{sec:susy-factor} and give the alternative formulae below.

\subsection{Integrands}  \label{sec:int_intro}
Supersymmetry determines the full super-amplitude from the amplitudes involving only the top of the multiplet.   We will see in \S \ref{sec:susy-factor} that superysmmetry implies that the total dependence on the supermomenta is encoded in the exponential  factor $\e^{F}$, with $F=F_{\scalebox{0.7}{$N$}}+\tilde{F}_{\scalebox{0.7}{$\tilde N$}}$ where\footnote{Here we decompose our factors for the new fixed SUSY representation in terms of  the $ F_{\scalebox{0.7}{$N$}}^\mathrm{pol}$ factors used in \cite{Geyer:2018xgb}. }
\begin{subequations} \label{SUSY-factor}
\begin{align}
 &F_{\scalebox{0.7}{$N$}}=F_{\scalebox{0.7}{$N$}}^\mathrm{pol}-\frac{1}{2}\sum_{i=1}^n \la \xi_i v_i\ra \,q_i^2 \,,&&
 F_{\scalebox{0.7}{$N$}}^\mathrm{pol} = \sum_{i<j}  
 \frac{ \langle u_i u_j\rangle}{\sigma_{ij}} q_{i\sI}q_j^\sI
  \,,\\
  &\tilde{F}_{\scalebox{0.7}{$\tilde N$}}=\tilde{F}_{\scalebox{0.7}{$\tilde N$}}^\mathrm{pol}- \frac{1}{2}\sum_{i=1}^n \left[ \xi_i v_i\right]\, \tilde q_i^2 \,,&&
  \tilde{F}_{\scalebox{0.7}{$\tilde N$}}^\mathrm{pol}=\sum_{i<j}   \frac{ [ \tilde u_i \tilde u_j] }{\sigma_{ij}}
  \tilde{q}_{i\sdotI}\tilde{q}_j^{\sdotI} &&
  \,.
  \end{align}
  \end{subequations}
 For example  for $\mathcal{N}=(1,1)$ super Yang-Mills  we take the exponential factor $\exp F^{\mathrm{YM}}=\exp (F_1+\tilde{F}_1)$.  In the dynamic R-symmetry preserving representations \eqref{fieldR_v} as used in \cite{Geyer:2018xgb}, we only keep the $F_{\scalebox{0.7}{$N$}}^\mathrm{pol}$ terms in the exponential, $e^{F^\mathrm{pol}}$ with   $F=F_{\scalebox{0.7}{$N$}}^\mathrm{pol}+\tilde{F}_{\scalebox{0.7}{$\tilde N$}}^\mathrm{pol}$. Alternatively, we can  Fourier transform in half the fermionic variables to make contact with the little-group-preserving representation of Refs.~\cite{Heydeman:2017yww,Cachazo:2018hqa} as given in \eqref{eq:little-gp-pres}. To do so, we choose an explicit off-diagonal representation for the R-symmetry metric, decompose the fermionic variables $q_\sI = \left(q^l,\la \epsilon \eta_l\ra \right)$ according to this representation, and Fourier transform one of these half-dimensional fermionic subspaces,
\begin{equation}\label{eq:delta-rep}
   \int \prod_{i=1}^n d^Nq_i^l  \; \prod_je^{-q_j^l\la \xi_j\eta_{j l}\ra}\;e^{F_N}=\prod_{i}\delta^{0|N}\left( \sum_{j}\frac{\la u_i u_j\ra}{\sigma_{ij}}\la \epsilon_j \eta_{jl}\ra -\la v_i\eta_{il}\ra\right)\,.
\end{equation}
On the right, we have relabeled $q_{l}=\eta_{\epsilon l}:=\la\epsilon \eta_l\ra$, and grouped the fermionic variables into a little-group spinor $\eta_a^l$. In this representation, the fermionic delta-functions take the same form as the polarized scattering equations with $\eta_{al}$ replacing $\kappa_{aA}$, and we define $\rd\mu_n^{\mathrm{pol}|N+\tilde N}$ to be the measure obtained by   combining the fermionic delta functions \eqref{eq:delta-rep} into $\rd\mu^\mathrm{pol}_n$. 

\smallskip

In general,  given a scattering amplitude of the form \eqref{eq:form_amplitude} for the top states of the multiplet of an $\mathcal{N}=(N,\tilde N)$ theory, the fully supersymmetric amplitude is given by
\begin{subequations}\label{eq:ampl_susy}
 \begin{align}
 &\mathcal{A}_n=\int \rd\mu_n^{\mathrm{pol}}\,\mathcal{I}_n\;e^{F_N+\tilde F_{\tilde N}} && \text{R-symmetry}\\
 & \mathcal{A}_n=\int \rd\mu_n^{\mathrm{pol}|N+\tilde N}\,\mathcal{I}_n  && \text{little-group symmetry}\,.
 \end{align}
\end{subequations}
This gives our formulae for superamplitudes from the formulae for the top states of the supermultiplets. We show in \S\ref{sec:susy-factor} that these are correctly  supersymmetric.

For the ambidextrous spin one contribution, define an $n\times n$  matrix $H$ by 
\begin{equation}
H_{ij} =\begin{cases}\frac{\epsilon_{i\sA}\epsilon^\sA_j}{\sigma_{ij}} \qquad i\neq j\\
e_i\cdot P(\sigma_i) \, , \qquad i=j\label{eq:def1_H_ii}
 \end{cases} 
\end{equation}
where $e_i$ is the null polarization vector and $P(\sigma
)$ is as defined in \eqref{P-def}.  We can define $H_{ii}$ equivalently by 
\begin{equation}\label{eq:def2_H_ii}
\lambda_{a\sA}(\sigma_i)\epsilon^\sA_i=-u_{ia} H_{ii} \, , \quad \lambda^{\dot a\sA}(\sigma_i) \epsilon_{i\sA}=-u_i^{\dot a} H_{ii} \,.
\end{equation}
See \S\ref{YM-int} for details. 

On the polarized scattering equations, the determinant $\det H$ vanishes because  $H$ has co-rank 2 due to
\begin{equation}\label{eq:co-rank}
\sum_i u_{ia} H_{ij}=\lambda_{a \sA} (\sigma_j)\epsilon^\sA_j +u_{ja} H_{jj}=0\, .
\end{equation}
The first term follows from the definition \eqref{lambda-def} of $\lambda_{a \sA} $ and the second equality from \eqref{eq:def2_H_ii}. Similarly,  $\sum_j H_{ij} u_{j\dot a}=0$. These identities nevertheless imply that $H$ has a well defined  reduced determinant 
\begin{align}
\det{}'H: =\frac{\det(H^{[i_1i_2]}_{[j_1j_2]})}{\langle u_{i_1} u_{i_2}\rangle [u_{j_1} u_{j_2}]}\,.\label{gendet2}
\end{align}
Here $H^{[i_1j_1]}_{[i_2j_3]}$ denotes the matrix $H$ with the rows $i_1,\,i_2$ and columns $j_1,\,j_2$ deleted, and $\det{}'H$ is well-defined in the sense that the \eqref{gendet2} is invariant under permutations of particle labels, and thus independent of the choice of $i_{1,2},\,j_{1,2}$, see \S\ref{YM-int} for the proof. 

The reduced determinant $\det{}'H$ is manifestly gauge invariant in all particles, carries $\mathrm{SL}(2,\mathbb{C})_\sigma$ weight $-2$, as expected for a half-integrand $\mathcal{I}^{\mathrm{spin-1}}$ and is equally valid for even and odd numbers of external particles. On the support of the polarized scattering equations, it is verified using factorization in \S \ref{sec:factorization} that $\det{}'H$ is equal to the CHY half-integrand $\pf{}'M$.
\smallskip

Another important building block, relevant for the D5 and M5 theory, is the skew matrix $A$, familiar from the CHY formulae \cite{Cachazo:2013hca, Cachazo:2014xea}, with
\begin{equation}
 A_{ij}=\frac{k_i\cdot k_j}{\sigma_{ij}}\,.
\end{equation}
Again, the Pfaffian $\mathrm{Pf} A$ vanishes on the scattering equations \eqref{SE}, but the reduced Pfaffian $\mathrm{Pf}'A=\frac{(-1)^{i+j}}{\sigma_{ij}}\mathrm{Pf} A^{ij}_{ij}$ is well-defined and non-zero for even numbers of particles  \cite{ Cachazo:2013hca,Cachazo:2014xea}.

\smallskip

The final ingredients are constructed from  $(\sigma_i, u_{ia}, \tilde u_{i\dot a})$, and are only needed for  M5-branes.  These only lead to amplitudes with even  numbers of particles. We present a formulation pointed out by \cite{Schwarz:2019aat} using \cite{Roehrig:2017wvh}, giving  a useful alternative formulation to that in \cite{Geyer:2018xgb}, the connections to which we discuss in \S\ref{sec:M5-integrands}.  Define the family of matrices $U^{(a,b)}$ by
\begin{equation}
 U^{(a,b)}_{ij}=\frac{\la u_i u_j\ra ^a\left[ \tilde u_i\tilde u_j\right]^b}{\sigma_{ij}}\,.\label{Uab}
\end{equation}
In fact we will only need $U^{(2,0)}$ and $U^{(0,2)}$ although for even numbers of particles we have the identity
\begin{equation}\label{eq:detH=pfA}
\det{}'H= \frac{\Pf'\,A}{\Pf\, U^{(1,1)}}\, ,
\end{equation}
allowing for the use of $U^{(1,1)}$ according to taste.

With these ingredients, we have the following integrands of various supersymmetric theories  as follows
\begin{subequations}\label{eq:integrands}
 \begin{align}
  & \text{(1,1)-Super Yang-Mills:} && \mathrm{PT}(\alpha)\;\det{}'H\;e^{F_1+\tilde{F}_1} \label{eq:int_sYM}\\
  & \text{(2,2)-Supergravity:} && \det{}'H\;\det{}'\tilde H\;e^{F_2+\tilde{F}_2}\\
  & \text{(1,1)-D5-branes:} && \det{}'A\;\det{}'H\;e^{F_1+\tilde{F}_1} \\
  & \text{(2,0)-M5-branes:} && \det{}'A\;\frac{\pf{}'A}{\pf U^{(2,0)}}\;e^{F_2}
 \end{align}
\end{subequations}
The resulting superamplitudes are $\mathrm{SL}(2,\mathbb{C})_\sigma\times \mathrm{SL}(2,\mathbb{C})_\pm$ invariant, the super Yang-Mills and supergravity amplitudes are gauge invariant, and the supergravity amplitudes are permutation invariant. We also see colour-kinematics duality expressed in the form of the super Yang-Mills and supergravity amplitudes.  The M5 amplitudes are manifestly chiral. 

The integrands used here improve the formulae in \cite{Geyer:2018xgb} by having a static, fixed once and for all supersymmetry representation.  We  have furthermore replaced the determinants of $n/2 \times n/2$ blocks of $U$-matrices with manifestly permutation invariant Pfaffians. These integrands are quite different from those of \cite{Cachazo:2018hqa}, not only in the supersymmetry representation, but also in the Pfaffians of our $U$ matrices and our spinorially constructed $\det{}'H$ replaces the CHY Pfaffians. 

\smallskip

The main result of this paper, expressed and proved in detail in \S\ref{sec:BCFW}, is:
\begin{thm}\label{thm:BCFW}
The amplitude formulae \eqref{eq:form_amplitude} with integrands \eqref{eq:integrands} all factorize correctly.  There exists good BCFW shifts for the gauge and gravity formulae so that their equivalence with the corresponding tree-level S-matrices is guaranteed by recursion and the three-point examples of \S\ref{sec:low-point}. 
\end{thm}
We will see later explicitly that these formulae all correctly reproduce the known three- and four- point amplitudes.  We will see further that the supergravity and super Yang-Mills amplitudes reduce to the four-dimensional expressions  given in terms of the four-dimensional polarized scattering equations above.  

\subsection{The double copy and Gerbe-theories.}  As remarked in \cite{Geyer:2018xgb}, our half-integrands provide a double-copy matrix of theories given in terms of the improved half-integrands of this paper in the first four columns of \cref{tabletheories} below.  This table is analogous to  those obtained in \cite{Cachazo:2014xea,Casali:2015vta} in the CHY and RNS ambitwistor-string
 framework and   
the entries provide nodes in the web of theories of \cite{Bern:2019prr}. The table contains  the amplitude formulae for  the theories described above, but the last column also gives three expressions that may not correspond to an amplitude in a well-defined theory. Analogous formulae were also found in the framework of \cite{Cachazo:2018hqa}.

A key feature of this last column is that is provides   amplitude-like formulae of the type that might arise for theories that  contain \emph{Gerbes} in their linear multiplets.  Gerbes are  closed self-dual 3-forms and correspond to fields $B_{AB}=B_{(AB)}$ in spinors.  The spin-2 analogues have spinors $\psi^A_{BCD}$ for $(3,1)$  and $\psi_{ABCD}$ for $(4,0)$ in their linear multiplets (whereas the spinor corresponding to the Weyl tensor of a genuine gravitational field is the trace-free symmetric spinor $\Psi_{AB}^{CD}$).   See \cite{Hull:2000zn, Borsten:2017jpt,Henneaux:2017xsb,Henneaux:2018rub} for further discussion of  these spin-2 linear fields and their possible links with interesting interacting theories.

\begin{table}[!h]
\begin{center}
\begin{tabular}{|l|llll|}
\hline
 & PT & $\det{}'A$ & $\det{}'H \;e^{F_1+\tilde{F}_1}$ & $\displaystyle \frac{\pf{}'A}{\Pf\, U^{(2,0)}}\; e^{F_2}$\Bbstrut\Ttstrut\\\hline
 PT & Bi-adjoint scalar &  NLSM & $\mathcal{N}=(1,1)$ sYM & $\mathcal{N}=(2,0)$-PT \Tstrut\\
 $\det{}'A$ &&Galileon& $\mathcal{N}=(1,1)$ D5 & $\mathcal{N}=(2,0)$-M5 \\
 $\det{}'H \;e^{F_1+\tilde{F}_1}$ &&&$\mathcal{N}=(2,2)$ sugra& $\mathcal{N}=(3,1)$\\
$\displaystyle \frac{\pf{}'A}{\Pf\, U^{(2,0)}}\; e^{F_2}$ &&&& $\mathcal{N}=(4,0)$\Bbstrut\Tstrut\\\hline
\end{tabular}
\caption{All integrands constructed from the building blocks discussed above.}
\label{tabletheories}
\end{center}
\end{table}

The example that may be of most interest in this column is the `$(2,0)$-PT formula', obtained from combining the M5 half-integrand $\pf{}'A\,\det X\, e^{F_2}/\Pf\, U^{(2,0)}$ with a Parke-Taylor factor, i.e. replacing the $\det{}'A$ of the M5-integrand by a Parke-Taylor half-integrand.  This leads to an expression with a non-abelian structure and $\mathcal{N}=(2,0)$ supersymmetry. While this formula may seem suggestive of amplitudes for the famous $(2,0)$-theory arising from coincident M5-branes, this is certainly too simplistic, because that theory lacks a perturbative parameter and thus has no S-matrix.\footnote{See also the no-go theorems of \cite{Huang:2010rn, Dennen:2010dh} for the existence of a 3-point amplitude.} Ref.~\cite{Cachazo:2018hqa} have further shown that the equivalent four-particle expression in their framework factorizes into non-local three-particle formulae that are not even well-defined, \footnote{The three-particle kinematics carries a special redundancy, under which amplitudes must be invariant --- but these three-particle formulas are not.} and thus cannot be interpreted as amplitudes.
Moreover, the  formulae in the right-hand column are not obviously defined for an odd number of particles. The $(2,0)$-M5 theory is not expected to have amplitudes for odd numbers of particles, but that is already guaranteed by the additional $\det{}' A$ factor which, being the determinant of a skew matrix, automatically vanishes for odd  $n$.  However, for the other factor we have no analogue of \eqref{eq:detH=pfA} to provide a meaning for odd $n$.  This issue may well be connected to the difficulties in defining three-particle extensions for the $(2,0)$-PT mentioned above and discussed in \cite{Cachazo:2018hqa}. 

Despite these difficulties in identifying underlying theories for these formulae, they are all well-defined and manifestly chiral and supersymmetric, and we discuss them  further in \S\ref{sec:discussion}.


\paragraph{Further theories, (2,0) supergravity.}
Our matrix in table \ref{tabletheories} can be extended further using the half-integrands from \cite{Cachazo:2014xea,Casali:2015vta} to give potentially supersymmetric 6d versions of the theories discussed there. Further half-integrands in \cite{Azevedo:2017lkz,Azevedo:2018dgo} will give further potentially supersymmetric  formulae for the higher order theories treated there.

This larger matrix will  by no means be exhaustive and many further theories can be constructed by stripping out some of the supersymmetry and adjoining fewer or more fields than are present in the maximally supersymmetric multiplet.   This yields further half-integrands and theories.  In many settings the correct couplings will then be ensured from the original  supersymmetric theory.  We give an example that follows the analysis of Heydeman et al. \cite{Heydeman:2018dje}.  In the context of their 6d framework, they extract all chiral $\mathcal{N}=(2,0)$ 6d supergravity amplitudes together with the abelian $(2,0)$  tensor multiplets from the known formulae for $\mathcal{N}=(2,2)$ supergravity.  The number of tensor multiplets can then be changed with impunity.  If there are 21 of them, this  leads to anomaly cancellation and a correspondence with a K3 reduction of type IIB string theory.

The $(2,2)$ supergravity multiplet can be regarded as the tensor product of the $(2,0)$ multiplet with the $(0,2)$ multiplet.  The latter consists of fields $(B^{\sA\sB},\Psi^\sA_{\sdotI},\phi_{\sdotI\sdotJ})$ with $\phi_{\sdotI\sdotJ}\Omega^{\sdotI\sdotJ}=0$ so that there are only 5 scalars.  This can be  truncated to throw out the $\Psi^\sA_{\sdotI}$ and the number of scalars can be reduced or increased.  In the tensor product with the $(2,0)$ multiplet, the scalars correspond to $(2,0)$ abelian tensor multiplets.
With just one flavour of $(2,0)$ abelian tensor multiplet embedded into the $(2,2)$ multiplet (together with the $(2,0)$ gravity), integrating out the $(0,2)$ part of the supersymmetry from the $(2,2)$ formula yields, with $m$ abelian multiplets and $n$ graviton multiplets
\begin{equation}
 \mathcal{M}_{n+m}^{(2,0)}=\int \rd\mu_{n+m}^{\mathrm{pol}}\;\det{}'H\det{}'\tilde H\; \det U^{(0,1)}_m \;e^{F_2}\,. \label{(2,0)-grav}
\end{equation}
where $U^{(0,1)}_m$ is the $m\times m$ matrix  of \eqref{Uab} whose particle indices are those for the $m$ abelian tensor multiplets.  If we now wish to have an arbitrary number of flavours of abelian tensor multiplets,  we can extend $U_m^{(0,1)}$ to 
\begin{equation}
\mathcal{U}_{ij}^{(0,1)}=\frac{[ \tilde u_i\tilde u_j] \delta_{f_if_j}}{\sigma_{ij}}
\end{equation}
into which the flavour vectors of the $m$ abelian tensor multiplets can be contracted before taking the  determinant in \eqref{(2,0)-grav}.

We remark that this formula superficially contains more polarization data than expected for the $m$ abelian tensor multiplets as it contains an $\epsilon^A$ in addition to the $\epsilon_A$ for each tensor multiplet, coming from the $(n+m)\times (n+m)$ reduced determinant $\det'H$.  However, it will be seen  in \S\ref{sec:multiplet} that these expressions are independent of the spurious $\epsilon^A$ as they should be.

\section{Polarized scattering equations and measure}\label{sec:PSE}
 
In this section we prove various statements made in the introduction.  We first give an alternative form of the scattering equations that manifests that the scattering equations imply momentum conservation.  In \S\ref{sec:unique} we prove the existence and uniqueness for  solutions to the polarized scattering equations given an initial solution to the scattering equations.  Underlying this is a linear formulation of the polarized  scattering equations that we make explicit in \S\ref{lin-SE}. This is not used explicitly in what follows and can be omitted by a casual reader. The final subsection \S\ref{sec:reltoCHYmeasure} proves that the polarized scattering equations measure is equivalent to the standard CHY measure.

We first recall the form of the polarized scattering equations in which 
we eliminate the $v_{ia}$  by skew-symmetrizing the $i$th polarized  scattering  equation with $\epsilon_{i\sA}$ to obtain
\begin{equation}
\epsilon_{i[\sA]}\cE_{\sB]i}= \epsilon_{i[\sA}\langle u_i\lambda_{\sB]}(\sigma_i)\rangle+k_{i\sA\sB}=
\sum_j \frac{\langle u_{i},u_j\rangle \epsilon_{i[\sA}\epsilon_{\sB]j}}{\sigma_{i}-\sigma_j}-k_{i\sA\sB}\, . 
\label{polscatt-k0}
\end{equation}
These leads to 
\begin{lemma} We have the identity
\begin{equation}
K_{\sA\sB}:=\sum_i k_{i\sA\sB}=\sum_i\epsilon_{i[\sA]}\cE_{\sB]i}\, .
\label{total-mom}
\end{equation}
Thus
if $\cE_{i\sA}=0$ then momentum conservation $K_{\sA\sB}=\sum_i k_i=0$ follows.
\end{lemma}

\noindent
{\bf Proof:} This follows from  
\begin{equation}
 \sum_{i,j} \frac{u_{ij} \epsilon_{j[\sA}\epsilon_{\sB]i}}{\sigma_{ij}}=0\, ,
\end{equation}
as the argument of the double sum is skew symmetric in $i,j$.\hfill $\Box$\\

We also wish to know that $\lambda_{a\sA}$ provides a  spinor-helicity decomposition of $P(\sigma)$.

\begin{propn}
\label{6dCons}
On the support of the polarized scattering equations
\begin{equation}
\lambda_{\sA a}(\sigma)\lambda_{\sB}^a(\sigma)=P_{\sA\sB}(\sigma):=\sum_i\frac{k_{i\sA\sB}}{\sigma-\sigma_i}
\end{equation}
\end{propn}

\noindent
{\bf Proof:} We have 
\begin{equation}
\lambda_\sA^a(\sigma)\lambda_{\sB}^a(\sigma)=\sum_{ij}\frac{ u_{ia}u_{j}^a \epsilon_{i\sA} \epsilon_{j\sB}}{(\sigma\,\sigma_i)(\sigma\sigma_j)}\, .
\end{equation}
There are no double poles because $u_{ia}u_{i}^a=0$. The residue of the LHS  at $\sigma_i$ is
$$
\mathrm{Res}_{\sigma_i} \lambda_{\sA a}(\sigma)\lambda_{\sB}^a (\sigma)=\epsilon_{i[\sA}\sum_{j}\frac{u_{ia}u_{j}^a  \epsilon_{j|\sB]}}{(\sigma_i \sigma_j)}=\epsilon_{i[\sA} u_{ia} \lambda(\sigma_i)_{\sB]}^a\, .
$$
The polarized scattering equations reduces the RHS of this to
$$
\mathrm{Res}_{\sigma_i} \lambda_{\sA a}(\sigma)\lambda_{\sB}^a (\sigma)=\epsilon_{ic}\kappa_{i[\sA}^c\kappa_{\sB]i}^b v_{ib}= \langle v_i \epsilon_i\rangle\kappa_{i[\sA|a}\kappa_{\sB]i}^a=:k_{i\sA\sB} \,,
$$
as desired.\hfill $\Box$\\

When the scattering equations are not imposed, although the residue of Res$_{\sigma_i}P(\sigma)$ is no longer $k_i$, there is nevertheless an alpha-plane that contains both $P(\sigma_i)$ and $k_i$.

\subsection{Linear form of equations, and existence and uniqueness 
of solutions
}\label{sec:unique}
In this subsection we prove existence and uniqueness using algebreo-geometric arguments.  We define the bundle over $\CP^1$ in which $\lambda_{aA}$, $a=0,1$, takes its values to show that it is a rank-two  bundle with canonically defined skew form, and so generically has a pair of sections that can be normalized. 

We work with bundles on $\CP^1$ which will be direct sums of line bundles $\cO(n)$ whose sections can be represented in terms of homogeneous functions of degree $n$ in terms of homogeneous coordinates $\sigma_{\alpha}$, $\alpha=0,1$ on $\CP^1$ with skew inner product $(\sigma_i \sigma_j):=\sigma_{i0}\sigma_{j1}-\sigma_{i1}\sigma_{j0}$.   
We  prove:

\begin{propn}\label{unique}
For each solution $\{\sigma_i\}$ to the scattering equations  and compatible polarization data in general position, there exists a unique  solution to the polarized scattering equations \eqref{6dSE+}, \eqref{SE-norm} and \eqref{lambda-def} up to a global action of $SL(2,\C)$ on the little-group index.
\end{propn}

\noindent {\bf Proof:} 
Let  $P^{AB}(\sigma)$ arise from the given solution to the scattering equations as the spinor form of \eqref{P-def}.  To remove the poles, define $\Pi(\sigma)^{AB}:=P^{AB}\prod (\sigma\sigma_i)$ which is now holomorphic object of weight $n-2$ on $\CP^1$ and is a null 6-vector so as a skew matrix has rank 2 on $\CP^1$ (for momentum and $\sigma_i$ in general position it will be vanishing on $\CP^1$). 

We require  $\lambda_{aA}P^{AB}=0$ for $a=0,1$ so to study solutions to this equation, define the rank-2 bundle  $E= \ker P \subset \bbS_A$ on $\CP^1$ where $\bbS_A$ is the rank four trivial bundle of spinors over $\CP^1$. To calculate the number of sections we wish to compute the degree of this bundle.  To do so consider the short exact sequence 
\begin{equation}
0\longrightarrow E\longrightarrow \bbS_A \longrightarrow E^0(n-2)\longrightarrow 0\,,
\end{equation}
where the second map is multiplication by $\Pi(\sigma)^{AB}$ and $E^0(n-2)\subset \bbS^A(n-2)$ is the annihilator of $E$ twisted by $\cO(n-2)$, that being the weight of $\Pi^{AB}$.  In such a short exact sequence the degree of  $\S_A$ is the sum of that of $E$ and $E^0(n-2)$ since the degree is the winding number of the determinant of the patching function, and the maps of the exact sequence determine these up to upper triangular terms that dont contribute to the determinant.   Since $\bbS_A$ is trivial, it has degree 0, so we find 
\begin{equation}
\deg E + \deg E^0 + 2(n-2)=0\,.
\end{equation}
Because $E^0=(\bbS/E)^*$ and $\bbS$ is trivial, we have $\deg E^0=\deg E$ so this gives $\deg E=2-n$.

Now $\Lambda_{aA}:=\lambda_{aA}\prod(\sigma\sigma_i)$ is a section of $E(n-1)$ which by the above has degree $n$.  Our $\Lambda_{aA}$ is subject to the $n$ conditions, one at each  marked point, as we impose $\Lambda_{aA}|_{\sigma=\sigma_j}\propto\epsilon_{jA}$. This has the effect of defining a subbundle with a reduction of degree by 1 at each marked point, so the total degree is now zero.  Thus this subbundle therefore has degree zero.  For data in general position, it will therefore be trivial  with a two-dimensional  family of sections spanned by $\Lambda_{aA}$, $a=0,1$.  These can be normalized because $\Lambda_{0[A}\Lambda_{1B]}=f\Pi_{AB}$ where $f$ is a holomorphic function of the sphere of weight $n$.  The conditions on $\Lambda_{aA}$ at $\sigma_i$ imply that $f$ vanishes at each $\sigma_i$ so $f=c\prod_i(\sigma\, \sigma_i)$ and we can normalize our sections so that $c=1$ reducing the freedom in the choiced of frame $\Lambda_{aA}$ to SL$(2)$.   On dividing through by $\prod_i(\sigma\, \sigma_i)^2$ we obtain
$P_{AB}=\lambda_{aA}\lambda^a_B\, .$
\hfill$\Box$\\

  For the non-chiral theories that we are considering, we will need both chiralities of spinors satisfying polarized scattering equations i.e, we can also define
\begin{equation}
\lambda_{\dot a}^\sA(\sigma):=\sum_i \frac{u_{i\dot a}\epsilon^\sA_i}{\sigma-\sigma_i}\, , \qquad u_{i\dot a}\lambda^{\dot a\sA}(\sigma_i) =v_{i\dot a}\kappa_i^{\dot a \sA}\, .
 \label{6dSE-}
\end{equation}

\subsubsection{An explicit linear version  of the polarized scattering equations}\label{lin-SE}
This is not essential to the logical structure of the paper and can be omitted by the casual reader.  However, the above argument is rather abstract and it is helpful to see explicitly at least the underlying linearity of the problem of solving the polarized scattering equations. However we have not been able to give explicit versions of all the  algebreo geometric proofs above.

According to the above, we are trying to find a pair of solutions $\lambda_{a\sA}$, $a=1,2$ to the equations
\begin{equation}
P(\sigma)^{\sA\sB}\lambda_\sB(\sigma)=0 \,,   \label{pol-lam}
\end{equation}
where $\lambda_\sA(\sigma)$ has projective weight $-1$ in $\sigma$ and $P$ weight $-2$.  The argument above gives $\lambda_{\sA}\prod(\sigma\sigma_i)$ as a section of $E(n-1)$ which has degree $n$ and rank 2 so generically has $n+2$ global sections.  To make this more explicit, make the ansatz\footnote{We attach the additional $i$-index to $a_i$ here to distinguish this $u_{a_ii}$ from the $u_{ia}$ in the original ansatz for $\lambda_{\sA a}$; the $a_i$ is a little group index associated to momentum $k_i$ rather than the global one associated to $\lambda_{Aa}$. We will drop these sub-indices when the equations are unambiguous.}
\begin{equation}
\lambda_\sA=\sum_i\frac{u_{ia_i}\kappa_\sA^{a_i}}{(\sigma\sigma_i)}\, ,
\end{equation}
which removes double poles from \eqref{pol-lam}. Given that the total weight of \eqref{pol-lam} is negative, it will be satisfied if the residues at its poles vanish.  The vanishing of the residue at $\sigma_i$ yields
\begin{equation}
k_i^{\sA\sB}\sum_j \frac{\kappa_{j\sB}^{a_j}}{\sigma_{ij}} u_{a_jj}+ P(\sigma_i)^{\sA\sB}\kappa_{i\sB}^{a_i}u_{a_ii}=0 \,.  \label{pol-lam-res}
\end{equation}
Now define  $p_i^{a\dot a}$  after solving the CHY scattering equations \eqref{SE} by 
\begin{equation}
P^{\sA\sB}(\sigma_i)\kappa^a_{i\sA} = \kappa_{i\dot a}^\sB p_i^{a\dot a}\, . \label{piaa}
\end{equation}
This makes sense at $\sigma_i$ as $\kappa_{i\sA}^a$ annihilates the pole, and  a second contraction with $\kappa^{b}_{i\sB}$ leads to zero as it gives $k_i\cdot P$, so it must be a multiple of $\kappa_{i\dot a}^\sB$.   
We can understand this also by considering the 2-form $P(\sigma_i)\wedge k_i$ which in spinors gives, using the above,
\begin{equation}
P(\sigma_i)_{\sA\sC}k^{\sB\sC}_i=P(\sigma_i)^{\sB\sC}k_{i\sA\sC}=p^\sB_{i\sA}\, , \qquad p_{i\sA}^\sB=\kappa_{i\sA a}\kappa_{\dot a}^\sB p_i^{a\dot a}\, .
\end{equation}
We can now  see for example that
\begin{equation}
e_i\cdot P(\sigma_i)=[ \epsilon_i |p_i | \epsilon_i \rangle\, ,\label{edotp}
\end{equation}
using $e_{i\sA\sB}=\epsilon_{i[\sA}\tilde\epsilon_{\sB]i}$ where $\tilde\epsilon_{\sA}\kappa^{\sA\sB}_i=\epsilon_i^\sB$.
Following Cheung and O'Connell \cite{Cheung:2009dc}, we further define 
\begin{equation}
\kappa_{ij}^{\dot a a}:=\kappa^{\sA\dot a}_i\kappa_{j\sA}^a\,,
\end{equation}
that relate the $ij$-particles little group indices.

With this notation we see that \eqref{pol-lam-res} can be written as $\kappa_{i\dot a}^\sA$ multiplied by
\begin{equation}
    \sum_{a,j}H^{\dot{a}a_j}_{ij}u_{a_jj}=0,\qquad \qquad H^{\dot{a}a}_{ij}=
    \begin{cases}
    \frac{\kappa_{ij}^{\dot{a} a}}{\sigma_{ij}}&\quad i\neq j\\
     p_i^{a\dot{a}}&\quad i=j\,.
    \end{cases} \label{new-u-eq}
\end{equation}
The discussion of the previous subsection implies that generically these equations have $n+2$ solutions.
These equations reduce to the original polarized scattering equations if we supplement them with $n$ further equations $\la\epsilon_ju_{j}\ra=0$, since we will then have $u_{a_jj}=\epsilon_{ja_j}u_j$ as in the original ansatz \eqref{lambda-def}. We then expect to find a pair of linearly independent solutions $u_{ia}$, with $a=1,2$ now global little group indices,  so that we now have 
\begin{equation}
u^a_{a_ii}=\epsilon_{ia_i}u_i^a.\label{uaiai}
\end{equation}
In order to normalize these solutions, observe that for a pair of solutions $\lambda_\sA^1, \lambda^2_\sA$ to \eqref{pol-lam}, we must have that 
\begin{equation}
\lambda_{[\sA}^1 \lambda^2_{\sB]}=f P_{\sA\sB} \label{P-lambda}
\end{equation}
 for some meromorphic function $f$ on $\CP^1$ with poles at the $\sigma_i$.  However, when we impose \eqref{uaiai}, the double poles in \eqref{P-lambda} vanish and $f$ must be constant, so we can normalize the pair of solutions $u_i^a$ so that the coefficient is $1$.  The full $n+2$-dimensional space of solutions also has a volume form determined by \eqref{P-lambda}.

In general \eqref{new-u-eq} are $2n$-equations on $2n$-unknowns, so we must have $n+2$ relations to agree with the discussion of the previous subsection and to allow us to impose these extra $n$ conditions.  The relations follow from the original equation \eqref{pol-lam} and the nilpotency $P^{\sA\sB}P_{\sB\sC}=0$ that follows from the original scattering equations.  
This leads to the nilpotency
\begin{equation}
\sum_{ja} H_{ji}^{a\dot a} H_{jk}{}_a^{\dot b}=0\, .
\end{equation}
 This can be checked explicitly using a Schouten identity. We can use this nilpotency to generate solutions
\begin{equation}
\lambda_\sA(\sigma)= P(\sigma)_{\sA\sB} W^\sB(\sigma)\, , \qquad W(\sigma)^\sA=\sum_i \frac{\kappa_{i\dot a}^\sA w^{\dot a }_i}{(\sigma\sigma_i)}
\end{equation}
where the $W^\sB$ has weight 1 in $\sigma$ so $w_{\dot ai}$ has weight 1 in $\sigma_i$ and 2 in $\sigma$.  The ansatz guarantees no double poles in $\lambda_\sA$ and by taking residues we obtain\footnote{ We also have the special solutions when $W(\sigma)^\sA$ has no poles that leads to the 8 solutions
\begin{equation}
u_{ai}=\kappa_{ia\sA}(W_0^\sA+\sigma_iW_1^\sA)\, .
\end{equation}
}
\begin{equation}
u^a_{i}=\sum_{\dot a,j}H^{a\dot a}_{ij}w_{\dot a j}\, .
\end{equation}

\subsection{The equivalence of measures}\label{sec:reltoCHYmeasure}\label{CHY=pol}
 We first show that
\begin{equation}
\bar\delta(k\cdot P)=\int d^{2}u\, d^{2}v \,\delta^{4}( \cE_\sA)\delta(\langle\epsilon v\rangle-1)\, , \qquad \text{with }\,\cE_\sA:=\langle u\lambda_\sA\rangle -\langle v\kappa_{\sA}\rangle\, .\label{6dSE-delta+}
\end{equation}
After integrating out the four components of $(u_a,v_b)$, we are left with a single delta-function on both sides of the equation.  It is easy to see that they have the same support as the latter delta function on the left implies that $v_a\neq0$, but this can only be true when $(\lambda_\sA^a,\kappa_\sA^b)$ have rank less than four, which happens iff $\varepsilon^{\sA\sB\sC\sD}\lambda_\sA^0\lambda_\sB^1\kappa_\sC^0\kappa_\sD^1:=k\cdot P=0$. Furthermore the weights in $\lambda_\sA^a$ and $\kappa_\sA^a$ are $-2$ on both sides.  A  systematic proof uses a basis with $\epsilon_a=(0,1)$, $\kappa^0_{3}=\kappa^1_{4}=1$ and all other components zero.  This allows us to integrate out the $v^a$ directly  against the delta functions reducing the right side to
\begin{equation}
\int d^{2}u \,\delta( u_{a}\lambda^a_0)\,\delta( u_{a}\lambda^a_1)\,\delta(u_a\lambda^a_3- 1)=\delta(\langle\lambda_0\, \lambda_1\rangle) \, ,
\end{equation}
where the latter equality follows by direct calculation integrating out the $u_a$; this gives  \eqref{6dSE-delta+} in this basis.


The CHY measure is defined to be
\begin{equation}
 d\mu_n^{\CHY}:= \delta^6\left(K\right) \frac{\prod_{i=1}^n \bar \delta (k_i\cdot P(\sigma_i)) d\sigma_i}{\mathrm{Vol} (\mathrm{SL}(2,\C)_\sigma \times  \C^3)}= \delta^6\left(K\right) (\sigma_{12} \sigma_{23}\sigma_{31})^2\prod_{i=4}^n \bar \delta (k_i\cdot P(\sigma_i)) d\sigma_i\, ,
\end{equation}
where $K=\sum_i k_i$, the volume of SL$(2,\C)_\sigma$ quotients by the M\"obius invariance of $\sigma$, and the $\C^3$ is a symmetry of the ambitwistor string whose quotient removes the linearly dependent scattering equations delta functions. 

\begin{propn} \label{prop:measures} We have 
\begin{equation}
 d\mu_n^{\mathrm{pol}}:= \int \frac{\prod_{i=1}^n d^{2}u_i\, d^{2}v_i \, d \sigma_i \,\delta^{4}( \cE_{iA} )\delta(\langle\epsilon_i v_i\rangle-1)}{\mathrm{Vol} ( \mathrm{SL}(2,\C)_\sigma\times \mathrm{SL}(2,\C)_u)}= d\mu_n^{\CHY}\, ,
\end{equation}
where SL$(2,\C)_\sigma$ denotes M\"obius invariance of $\sigma$ as above  in the CHY measure,  the SL$(2,\C)_u$ is acting  on the little group index of $u_a$, and the integrals are over the $(u_i,v_i)$ variables. 
\end{propn}

\noindent
{\bf Proof:} We first reduce the SL$(2,\C)_\sigma$ factor fixing $(\sigma_1,\sigma_2,\sigma_3)$ to be constant with the standard 
\begin{equation}
\frac{\prod_i d\sigma_i}{\mathrm{Vol}\; \mathrm{SL}(2,\C)_\sigma}=\sigma_{12}\sigma_{13}\sigma_{23} \prod_{i\geq 4} d\sigma_i\, .
\end{equation} 
Similarly Faddeev-Popov gauge fixing\footnote{This entails contracting a normalized basis of the Lie algebra of SL$(2,\C)_u$ into the form $\prod_id^2u_i$ and restricting to the given slice.} SL$(2,\C)_u$ by 
\begin{equation}\label{eq:u_gauge}
u^a_1=(1,0),\qquad u^a_2=(0,u_{12}), \qquad u_3^a=\left(-\frac{u_{23}}{u_{12}},u_{13}\right)\, ,
\end{equation}
so that $u_{ij}=\langle u_iu_j\rangle $ for $i<j\leq 3$  yields
\begin{equation}
\frac{\prod_id^2u_i}{\mathrm{Vol}\; SL(2)_u}=du_{12} d u_{13} du_{23} \prod_{i=3}^n d^2u_i\, ,
\end{equation}
On the support of the delta functions $\prod_{i>3}\delta^4(\cE_{i\sA})$  we can write, using \eqref{total-mom},
\begin{equation}
K_{\sA\sB}=\left(\sum_{i=1}^3 \epsilon_{i[\sA}\cE_{i\sB]}\right)\, .
\label{total-mom-pol-scatt}
\end{equation}
We can trivially perform one of each of the $v_i$ integrals against the $\delta(\langle v_i\epsilon_i\rangle-1)$ delta functions by choosing a basis of the little group spin space for each $i$ so that $\epsilon_{ia}=(1,0)$ fixing $v_i^a=(v_i,1)$.

Choosing a basis of spin space consisting of $\{\epsilon_{i\sA}, \epsilon_{0\sA}\}$ with $i=1,2,3$ and $\epsilon_{0\sA}$ chosen so that $\langle 0123\rangle=1$, and dual basis $\tilde\epsilon_i^\sA$, $i=,0,\ldots ,3$ we find via \eqref{total-mom-pol-scatt}
\begin{equation}
K_{0i}=\cE_{i0}\, , \qquad K_{ij}=\cE_{[ij]}\, , 
\end{equation}
so that these polarized scattering equations can be replaced by $\delta^6(K)$.  The remaining scattering equations  in $\prod_{i=1}^3 \delta^4(\cE_{i\sA})$ are, for $i,j=1,\ldots,3$,
\begin{equation}
\cE_{(ij)}=\begin{cases}
\frac{u_{ij}}{\sigma_{ij}} + \ldots\,  \qquad i\neq j\,\\
v_i+\ldots \, , \qquad \; \, i=j\, 
\end{cases}
\end{equation}
where the $\ldots$ denotes terms involving $i,j> 3$.  Thus we can integrate out $du_{ij}$ and $dv_i$ against these remaining polarized scattering equation delta functions $\delta(\cE_{(ij)})$ for $i,j \leq 3$ yielding an extra numerator factor of $\sigma_{12}\sigma_{23}\sigma_{13}$.

Finally we can use \eqref{6dSE-delta+} to replace the remaining polarized scattering equations delta functions by standard ones thus yielding the desired formula.\hfill$\Box$

\section{Integrands}\label{sec:integrands}
In this section, we discuss the integrands $\mathcal{I}_n$ and  the supersymmetry representation in more detail. 
We first show that the spin-one contribution $\det{}'H$ is permutation invariant, and that it is equivalent to the CHY pfaffian $\Pf'M$ in providing the correct dependence on the spin-one polarization data.  We move on to giving further details of the supersymmetry factors and of the ingredients required for  brane theories. 
Finally, we prove crucial properties such as linearity of the spin-one contribution in the polarization data, and the compatibility of the reduced determinant with the supersymmetry representation.

\subsection{The kinematic reduced determinant \texorpdfstring{$\det{}'H$}{det'H}.}\label{YM-int}
 For our ambidextrous spin one contribution, recall that we defined an $n\times n$  matrix $H$ by 
\begin{equation}
H_{ij} =\begin{cases}\frac{\epsilon_{iA}\epsilon^A_j}{\sigma_{ij}} & i\neq j\\
e_i\cdot P(\sigma_i) \, , & i=j\label{eq:def1_H_ii1}
 \end{cases} \,,
\end{equation}
where $e_i$ is the null polarization vector above and $P(\sigma
)$ is as defined in \eqref{P-def}.  We first prove the equivalence between this definition of $H_{ii}$ and that in \eqref{eq:def2_H_ii}.    
In order  to use the vector representation of the polarization vector, we  introduce a spinor $\tilde \epsilon_A$ so that $\epsilon^A=k^{AB}\tilde\epsilon_B$.  Then the polarization vector is $e_{AB}=\epsilon_{[A}\tilde\epsilon_{B]}$.
 The equivalent  definition of $H_{ii}$ \eqref{eq:def2_H_ii} is
\begin{equation}\label{eq:def2_H_ii1}
\lambda_{aA}(\sigma_i)\epsilon^A_i=-u_{ia} H_{ii} \, , \quad \lambda^{\dot aA}(\sigma_i) \epsilon_{iA}=-u_i^{\dot a} H_{ii} \,.
\end{equation}
The left side  is a multiple of $u_{ia}$ (or $u_i^{\dot a} $) due to the scattering equation  and the identity $k^{AB} \kappa_{A}^a=0$. 
Starting from the second last formula we obtain the first from
\begin{equation}
e_i\cdot P(\sigma_i)= \epsilon^{[A}\tilde\epsilon^{B]}\lambda_{aA}(\sigma_i)\lambda^a_B(\sigma_i)=-H_{ii} \tilde \epsilon^{B}u_a\lambda^a_B(\sigma_i)=-H_{ii} \tilde \epsilon^{B}v_a\kappa^a_B=-H_{ii}\, . \label{Hii-id}
\end{equation}
This then, being neither chiral nor antichiral justifies the equivalence.\\

The matrix $H_{ij}$ is not full rank because 
\begin{equation}
\sum_i u_{ia} H_{ij}=\lambda_{a A} (\sigma_j)\epsilon^A_j +u_{ja} H_{jj}=0\, ,
\end{equation}
 and so, as above, we define the generalized determinant 
\begin{align}
\det{}'(H):&=\frac{\det(H^{[ij]})}{\langle u_i u_j\rangle [u_i u_j]}
\label{gendet1}
=\frac{\det(H^{[i_1i_2]}_{[j_1j_2]})}{\langle u_{i_1} u_{i_2}\rangle [u_{j_1} u_{j_2}]}
\end{align}
where $H^{[ij]}$ denotes the matrix $H$ with the $ij$ rows and columns deleted and $H^{[i_1i_2]}_{[j_1j_2]}$ the matrix with the with rows $i_1,i_2$ and columns $j_1,j_2$ removed.  These are well-defined as 
\begin{lemma}
The generalized determinant defined above is permutation invariant.  \label{lemma:perm-inv}
\end{lemma}
{\bf Proof:} We can extend the argument of appendix A of \cite{Cachazo:2012pz}  on such generalized determinants  as follows.   

 Consider an $n\times n$ matrix $H_{i}^{j}$ with a $p$-dimensional kernel and cokernel, i.e., that satisfies $\sum_i w^{i}_aH_{i}^{j}=0$ and $\sum_j H_{i}^{j}\tilde w_{j}^b=0$ where $a,b=1,\ldots,p$. We must also assume that there are  volume $p$-forms on these kernels, $\langle w_1\ldots w_p\rangle$ and $[\tilde w_1,\ldots\tilde w_p]$. 
Our  reduced determinant can be understood as the determinant of the exact sequence
\begin{equation}
0\rightarrow \C^p\stackrel{\tilde w}{ \rightarrow }\C^n\stackrel{H}{\rightarrow} \C^n\stackrel{w}{\rightarrow} \C^p\rightarrow 0\, .
\end{equation}
To make this explicit, note that we have 
 \begin{equation}
\varepsilon_{j_{1}\ldots j_n}\varepsilon^{i_1\ldots i_n}H_{i_{p+1}}^{j_{p+1}}\ldots H_{i_{n}}^{j_{n}}\langle w_1\ldots w_p\rangle\langle \tilde w^1\ldots \tilde w^p\rangle=\det{}'(H)   w^{[i_{1}}_{1}\ldots w_{p}^{i_p]} \tilde{w}^{1}_{[j_1}\ldots \tilde{w}^{p}_{j_p]} \label{red-det}
 \end{equation}
 for some $\det' (H)$.  This formula follows  because skew symmetrizing a free index on the left with a  $w_r$ or  $\tilde w_r$ vanishes as it dualizes via the $\varepsilon$ to contraction with $H_i^j$.  Thus it must be a multiple of the right hand side as defined.  The definitions
\eqref{gendet1}, \eqref{gendet2} then follow by taking components  of this definition in the case $p=2$ on the $i_{1}, i_2, j_{1}, j_2$ indices.  In our context the natural volume form on the kernel is defined on the 2-dimensional space of $u_{ia_i}=u_i \epsilon_{a_i}$ by the $f$ on the right hand side of \eqref{P-lambda} but for our polarized scattering equation framework, the normalizations are such that this is  1 so the bracketed terms on the left of \eqref{red-det} reduce to unity in \eqref{gendet1}.\hfill $\Box$
\\

Note  that the first term on the left side of \eqref{red-det} is simply the $p^{\mathrm{th}}$ derivative of $\det{H}$ where we have to relax the scattering equations and momentum conservation to make the determinant not identically zero.  The CHY matrix is also non-degenerate away from the support of the scattering equations and momentum conservation.  We have

\begin{propn} The determinant is related to the full CHY Paffian  by
$\det(H)=\mathrm{Pf} \,M$. 
\end{propn}

\noindent
{\bf Proof:} We use the form of the CHY Pfaffian due to Lam \& Yao \cite{Lam:2016tlk}.  They show that the full Pfaffian of $M$ can be expanded into a sum over the permutations $\rho\in S_n$ of the particle labels, 
\begin{equation}
 \text{Pf}\big( M\big) =\sum_{\rho\in S_n}\sgn(\rho)M_I ... M_J\,,\label{full-Pfaff}
\end{equation}
where each term has been decomposed into the disjoint cycles $I=(i_1\dots i_I)$, $J=(j_1\dots j_J)$ of the permutation $\rho$.  The terms in this cycle expansion are given by 
\begin{align}
 M_I =\begin{cases} \frac{ \tr(F_{i_1}...F_{i_I})}{\sigma_I} & \text{if }|I|>1\,, \\ C_{ii} & \text{if } I=\{i\}\,,\end{cases}
\end{align}
and $\sigma_I=\big(\sigma_{i_1i_2}\dots\sigma_{i_Ii_1}\big)^{-1}$ denotes the Parke-Taylor factor associated to the cycle.

Euler's formula for the determinant of $H$ similarly gives
\begin{align}
 \det(H)&=\sum_{\rho \in S_{n}}\sgn(\rho)H_I ... H_J
\end{align}
where the terms $H_I$ are given by
\begin{align}
 H_I = H_{i_1i_2}...H_{i_Ii_1}=\begin{cases} \frac{ \tr(F_{i_1}...F_{i_I})}{\sigma_I} & \text{if }|I|>1\,, \\ H_{ii} & \text{if } I=\{i\}\,,\end{cases}
\,.
\end{align}
Here the trace over the $F$s is taken in the spin representation and we have $C_{ii}=H_{ii}$ hence the equivalence.\hfill $\Box$
\\

This result provides some circumstantial evidence that $\Pf'M=\det'H$ on the support of the scattering equations, but we do not have a direct proof.  We prove this only indirectly via factorization in \S\ref{sec:factorization}.
Our $\det'H$ can therefore be used as a half-integrand in place of $\Pf'(M)$ in the theories as described in \cite{Cachazo:2014xea} to give full integrands
\begin{subequations}\label{eq:integrands_2}
 \begin{align}
  & \text{Yang-Mills:} && \mathrm{PT}(\alpha)\;\det{}'H\\
  & \text{Gravity:} && \det{}'H\;\det{}'\tilde H\\
  & \text{D5-branes:} && \det{}'A\;\det{}'H\, .
 \end{align}
\end{subequations}

\subsection{The supersymmetry factors and transform to little-group preserving representation}\label{sec:susy-factor}
Here we show that the supersymmetry factors $\e^{F_N}$, with
\begin{subequations} 
\begin{align}
 &F_{\scalebox{0.7}{$N$}}=F_{\scalebox{0.7}{$N$}}^\mathrm{pol}-\frac{1}{2}\sum_{i=1}^n \la \xi_i v_i\ra \,q_i^2 \,,&&
 F_{\scalebox{0.7}{$N$}}^\mathrm{pol} = \sum_{i<j}  
 \frac{ \langle u_i u_j\rangle}{\sigma_{ij}} q_{i\sI}q_j^\sI
  \,,\\
  &\tilde{F}_{\scalebox{0.7}{$\tilde N$}}=\tilde{F}_{\scalebox{0.7}{$\tilde N$}}^\mathrm{pol}- \frac{1}{2}\sum_{i=1}^n \left[ \xi_i v_i\right]\, \tilde q_i^2 \,,&&
  \tilde{F}_{\scalebox{0.7}{$\tilde N$}}^\mathrm{pol}=\sum_{i<j}   \frac{ [ \tilde u_i \tilde u_j] }{\sigma_{ij}}
  \tilde{q}_{i\sdotI}\tilde{q}_j^{\sdotI} &&
  \,.
  \end{align}
  \end{subequations}
 are invariant under supersymmetry. 
The full supersymmetry generator for $n$ particles is defined by the sum  $Q_{\sA\sI}=\sum_{i=1}^n Q_i{}_{\sA\sI}$ for each particle as defined by \eqref{eq:def_susy-gen},
 \begin{equation}
Q_{i\sA\sI}= \la \xi_i \kappa_{i\sA}\ra q_{i\sI} +\epsilon_{i\sA}\,  \Omega_{\sI\sJ}\frac{\partial}{\partial q_{i\sJ}}\,,\qquad \tilde Q_{i\sdotI}^{\sA}= \left[\xi_i \kappa^{\sA}_i\right] \tilde q_{i\sdotI} +\epsilon_i^{\sA}\, \tilde\Omega_{\sdotI\sdotJ}\frac{\partial}{\partial \tilde{q}_{i\sdotJ}}\, .
\end{equation}
Superamplitudes must be supersymetrically invariant and so are annihilated by  the total $Q_{\sA\sI}$ and indeed this determines the amplitude for the whole multiplet from the amplitudes involving only the top of the multiplets.

 It is easily verified that the supersymmetry factors give an amplitude  that is supersymetrically invariant, since
\begin{align}
 Q_{\sA\sI} \,e^{F_N}&=\left(\sum_i \Big(\langle \xi_i\kappa_{i\sA}\rangle+\la \xi_i v_i\ra \epsilon_{i\sA}\Big) \,q_{i\sI}-\sum_{i,j}\frac{\langle u_i u_j\rangle\,\epsilon_{i\sA}}{\sigma_{ij}}q_{j\sI}\right)e^{F_N}\nonumber\\
 &=\left(\sum_i \langle v_i\kappa_{i\sA}\rangle \,q_{i\sI}-\sum_{i,j}\frac{\langle u_i u_j\rangle\,\epsilon_{i\sA}}{\sigma_{ij}}q_{j\sI}\right)e^{F_N}=0\,,
\end{align}
 and similarly $Q^\sA_{\sdotI} \,e^{F}=0$.  Here, the second equality follows from $v_i = \xi_i +\la \xi_i v_i\ra \epsilon_i$, and the sum vanishes on the support of the polarized scattering equations. 
 Conversely, given an integrand $\mathcal{I}_n$ for the top states of a multiplet,  \eqref{eq:ampl_susy} is the unique supersymmetric completion using the supersymmetry representation \eqref{eq:def_susy-gen}, as can be verified using supersymmetric Ward identities.

\paragraph{The little-group preserving supersymmetry representation.} In six dimensions, amplitudes can alternatively be written in a supersymmetry representation that breaks R-symmetry, but preserves little group symmetry. We construct this representation by choosing an $\mathcal{N}$-dimensional subspace on which $\Omega_{\sI\sJ}$ vanishes indexed by $l,m=1\ldots \mathcal{N}$ so that $a^\sI=(a^l,a_l)$ with $\Omega_{\sI\sJ}a^\sI b^\sJ=a^lb_l-b^la_l$. Then
\begin{equation} 
Q_\sA^\sI=(Q_\sA^l,Q_{\sA l})=\kappa_\sA^a(Q_a^l,Q_{al})
\end{equation}  
satisfying
\begin{equation}
\{Q_a^l,Q_b^m\}=0=\{Q_{al},Q_{bm}\}\, , \qquad \{Q_{al},Q_b^m\}=\epsilon_{ab}\delta^m_l\, , \label{Q_LG}
\end{equation} 
with similar relations for $Q^{\sA\sdotI}=(Q^{\sA l}, Q^\sA_{l}) =\kappa^\sA_{\dot a}(Q^{\dot a l},Q^{\dot a}_l)$.
Thus we can introduce  supermomenta $\eta_{la}$ as fermionic eigenvalues of $Q_{la}$ so that our supermomentum eigenstates satisfy 
\begin{equation}
Q_{la}\phi=\eta_{la}\phi\, , \qquad Q^{la}\phi =\frac{\p\phi}{\p \eta_{la}}\,,\qquad\tilde Q_{l\dot a}\phi=\tilde \eta_{l\dot a}\phi\, , \qquad \tilde Q^{l\dot a}\phi =\frac{\p\phi}{\p \tilde \eta_{l\dot a}}\,.
\end{equation}  
This clearly gives a representation of \eqref{Q_LG}.
For $\mathcal{N}=(1,1)$ super Yang-Mills, we can replace the $l$-index by `1' when $l=1$ is an upper index  and `2' when $l$ is a lower index to
 find
\begin{equation}\label{eq:multiplet_LG}
 \Phi^{\mathrm{LG}} = \phi^{1\dot 1} + \eta_a\psi^{a\dot 1}+\tilde\eta_{\dot a}\tilde\psi^{1\dot a}+\eta_a\tilde\eta_{\dot a} A^{a\dot a}+\eta^2 \phi^{2\dot 1}+\tilde\eta^2\phi^{1\dot 2}+\dots +\eta^2\tilde\eta^2\phi^{2\dot 2}\,,
\end{equation}
for the R-symmetry breaking representation.\footnote{The indices are chosen to agree with the conventions in \cite{Heydeman:2017yww,Cachazo:2018hqa}.}

\paragraph{Fermionic Fourier transform.} The sets of supermomenta from the $R$-symmetry preserving representing are related to those above by decomposing $q_I=(q_l,q^l)$ and observing that the definitions allow us to identify
\begin{equation}
\eta_{1l}:=\eta_{\epsilon l}=\epsilon^a\eta_{al}=q_l\, , \qquad \eta_{2l}:=\eta_{\xi l}=\xi^a\eta_{al}= \frac{\p}{\p q^l}\, .
\end{equation}
The latter relation implies a fermionic half-Fourier transform on the supermultiplets written for general $(N,\tilde N)$ as
\begin{equation}
 \Phi^{\mathrm{R}} = \int d^{\sN}\!\eta_2 \,d^{\stN}\!\tilde\eta_{\dot 2}\,e^{q^l \eta_{2l}}e^{\tilde q^l\tilde\eta_{\dot2 l}}\,
  \Phi^{\mathrm{LG}}\Bigg|_{\substack{\eta_{1l}= q_l \\ \tilde\eta_{\dot 1l}=\tilde q_l}}\,,\qquad 
   \Phi^{\mathrm{LG}} = \int d^{\sN}\!q\, d^{\stN}\!\tilde q\,e^{-q^l \eta_{\xi l}}e^{-\tilde q^l\tilde\eta_{\tilde \xi l}}\,
  \Phi^{\mathrm{R}}\Bigg|_{\substack{q_l=\eta_{\epsilon l}  \\ \tilde q_l=\tilde\eta_{\tilde \epsilon l}}}\,. \label{Ferm-half-Fourier}
\end{equation}
As discussed in  \cref{sec:int_intro}, we can implement the fermionic half-Fourier transform at the level of the amplitudes. Starting from the exponential (R-symmetry preserving) representation, we find that the supersymmetry factors turn into delta-functions that mimic the polarized scattering equations,
\begin{equation}\label{eq:Delta_susy}
   \int \prod_{i=1}^n d^Nq_i^l  \; \prod_je^{-q_j^l\eta_{\xi l}}\;e^{F_N}=\prod_{i}\delta^{0|N}\left( \sum_{j}\frac{\la u_i u_j\ra}{\sigma_{ij}}\la \epsilon_j \eta_{jl}\ra -\la v_i\eta_{il}\ra\right)=:\Delta^{0|N}_n\,.
\end{equation}
In this representation, it is convenient to include the fermionic delta-functions in the definition of the measure, $d\mu_n^{\mathrm{pol}|N+\tilde N}= d\mu_n^{\mathrm{pol}}\Delta_n^{0|N}\tilde{\Delta}_n^{0|\tilde N}$. We remark that in this delta-function representation of the superamplitude, all components are monomials in the Grassmann variables $\eta$, and the all-gluon amplitude sits in the middle of the multiplet \eqref{eq:multiplet_LG}. It is straightforward to check that we recover the integrand $\det{}'H$ of the gluon amplitude in the top state by extracting the component proportional to $\prod_i \la v_i\eta_i\ra[\tilde v_i\tilde \eta_i]$.\\

We can also verify directly that the supersymmetry factors $\Delta_n^{0|\sN}$ are invariant under supersymmetry, and that superamplitudes in the delta-function representation are annihilated by the supersymmetry generator $Q_{\sA\sI}$, defined as before  by the sum $Q_{\sA\sI}=\sum_{i=1}^n Q_{i\,\sA\sI}$. This is particularly easy to see for the multiplicative operator $Q_{\sA l}$, which vanishes on the support of the polarized scattering equations,
\begin{equation}
 Q_{\sA l}\, \Delta_n^{0|N} = \sum_{i=1}^n \kappa_{i\sA}^a\eta_{ila}\, \Delta^{0|N}_n = \sum_{i,j}\frac{\la u_i u_j\ra}{\sigma_{ij}}\Big(-\la \epsilon_i\eta_{il}\ra \epsilon_{j\sA}+\la \epsilon_j\eta_{jl}\ra \epsilon_{i\sA}\Big)\Delta^{0|N}_n=0\,.
\end{equation}
Here we have used both the support of the polarized scattering equations and their fermionic analogues, and the last equality holds because the argument of the sum in $i$ and $j$ is skew symmetric. The remaining supersymmetry generators annihilate the superamplitude by a similar argument,
\begin{equation}
 Q^l_\sA\, \Delta_n^{0|N} = \sum_{i=1}^n \kappa_{i\sA}^a\frac{\partial}{\partial \eta_{il}^a}\Delta^{0|N}_n = \sum_{i,j}\frac{\la u_i u_j\ra}{\sigma_{ij}}\left(-\epsilon_{j\sA}\mathcal{E}_j^F+\epsilon_{i\sA}\mathcal{E}_i^F\right)\Delta_{n\,[i_lj_l]}^{0|N}=0\,,
\end{equation}
where $\mathcal{E}_i^F$ denote the fermionic delta-functions, and $\Delta_{n\,[i_lj_l]}^{0|N}$ is the usual product \eqref{eq:Delta_susy}, but with the delta-functions $\mathcal{E}_{il}^F,\mathcal{E}_{jl}^F$ removed. The sum vanishes again by the skew-symmetry of its argument.

\subsection{M5 and D5 theories}\label{sec:M5-integrands}

We first recall the ingredients for D5 and M5-branes.  
These are supersymmetric theories that share a scalar sector with Lagrangian of the form $L\sim \sqrt{-\det(\eta_{\mu\nu} +k\sum_r\p_\mu\phi^r\p_\nu\phi^r)}$.  For  D5 branes $r=1,\ldots ,4$  and for M5 branes $r=1,\ldots ,5$ thought of as transverse coordinates to 6d worldvolumes in 10d or 11d respectively. D5-branes are naturally completed with $(1,1)$-supersymmetry, and M5 with $(2,0)$-supersymmetry.  In the case of D5-branes, the linearised multiplet then coincides with the $(1,1)$ super-Maxwell multiplet \eqref{eq:supermaxwell-mult}.  The Lagrangian for the bosonic parts of the multiplet extends the Born-Infeld action to give 
$$
L\sim \sqrt{-\det(\eta_{\mu\nu} +k\sum_3 \p_\mu\phi^r\p_\nu\phi^r +\kappa F_{\mu\nu})}.
$$  
For M5 branes, the $(2,0)$ supermultiplet is $(G_{AB},\psi_{IA}, \phi_{IJ})$ with $\phi_{IJ}=\phi_{[IJ]}$ and $\phi_{IJ}\Omega^{IJ}=0$.  Here the spinor $G_{AB}=G_{(AB)}$ corresponds to a self-dual 3-form whose linearized equations are that it should be closed (and hence co-closed by self-duality). Such a field is known as a Gerbe, often thought of as a curvature associated to a 2-form potential $B_A^B$.  See \cite{Schwarz:2020emu} for a modern review.  

There are CHY formulae \cite{Cachazo:2014xea} for the bosonic brane theories with any number of scalars, and further including the Born-Infeld contribution.
As in \cite{Geyer:2018xgb}, we follow the strategy in \cite{Heydeman:2017yww} that obtains full superamplitudes for D5 and M5 theories by incorporating supersymmetry factors on top of these CHY formulae for scalar amplitudes. This makes use of the fact that both theories share  an SU$(2)$ subsector of the scalars. The full supersymmetric amplitudes can then be reconstructed from the known scalar amplitudes 
in this sector by applying supersymmetry. 
We go on to explain their relationship with the half-integrands \eqref{Uab} given in the introduction.

\paragraph{The D5 integrand.} The bosonic part of this is well-known from  \cite{Cachazo:2014xea} in the original CHY-format, where it takes the form $\mathcal{I}_{\mathrm{D5}}=\det{}'A\,\pf{}'M$. Substituting the spin-one half-integrand in the 6d spinor-helicity formalism, and inserting the correct supersymmetry factors immediately gives the 6d integrand
\begin{equation}
 \mathcal{I}_{\mathrm{D5}}=\det{}'A\;\det{}'H\; e^{F_1+\tilde F_1} \,,
\end{equation}
of the full superamplitudes. We can now extract the shared subsector of scalar amplitudes from this  D5 integrand
by a suitable integration over the super momenta $q_\sI$, $\tilde q_{\sdotI}$. For an all-scalar amplitude where we scatter generic scalars $\phi_i^{\sJ_i\sdotJ_i}$, the integrand takes the form
\begin{align}\label{eq:D5_scalar_generic}
 \mathcal{I}_{\mathrm{D5}}^{\sJ_1\sdotJ_1\dots \sJ_n\sdotJ_n}&=\int \prod_{i=1}^n d^2\! q_i\,d^2\!\tilde{q}_i \,q_i^{\sJ_i}\,\tilde{q}_i^{\sdotJ_i}\;\mathcal{I}_{\mathrm{D5}}=\det{}'A\;\det{}'H\;\left(\pf \mathcal{U}\right)^{\sJ_1\dots \sJ_n}\;(\pf \tilde{\mathcal{U}})^{\sdotJ_1\dots \sdotJ_n}\,.
\end{align}
Here, $\mathcal{U}$ and $\tilde{\mathcal{U}}$ are $n\times n$ matrices carrying the R-symmetry indices of the scalars, with entries
\begin{equation}\label{eq:def_Ufull}
 \mathcal{U}_{ij}^{\sJ_i\sJ_j}=\frac{\langle u_iu_j\rangle}{\sigma_{ij}}\Omega^{\sJ_i\sJ_j}\,,\qquad  \tilde{\mathcal{U}}_{ij}^{\sdotJ_i\sdotJ_j}=\frac{[\tilde{u}_i\tilde{u}_j]}{\sigma_{ij}} \Omega^{\sdotJ_i\sdotJ_j}\,,
\end{equation}
and $\pf \mathcal{U}$ and $\pf \tilde{\mathcal{U}}$ are defined by specifying the R-symmetry indices, and then taking the Pfaffian as usual.  To construct the M5 integrand, we further have to restrict this amplitude to the shared SU$(2)$ scalar subsector between M5 and D5 theory, which is the subspace of non-self-interacting scalars of the respective theories. This sector can be constructed along similar lines to the discussion in \S\ref{sec:susy-factor}. Let us choose again an $\mathcal{N}$-dimensional subspace of the supersymmetry generators on which the metric $\Omega_{\sI\sJ}$ vanishes, indexed by $a^\sI=(a^l,a_l)$ with $\Omega_{\sI\sJ}a^\sI b^\sJ=a^lb_l-b^la_l$. From this we can directly construct two  non-self-interacting scalar subsectors, $Y=\{\phi^{l\dot l}\}$ and  $\overline{Y}=\{\phi_{l\dot l}\}$ for D5, and  $Y=\{\phi^{lm}\}$ and  $\overline{Y}=\{\phi_{lm}\}$ for M5. Any other non-self-interacting subsector is related to $Y$ and $\overline{Y}$ by an SU$(2)$ transformation. Note that each of the non-self-interacting subsectors contains exactly one scalar state; this is obvious for D5, where $\phi^{l\dot l}=\phi^{1\dot 1}_{\LG}$ and $\phi_{l\dot l}=\phi^{2\dot 2}_{\LG}$ in the notation of the last section, and for M5 theory this follows from the antisymmetry constraint on the scalar indices, $\phi^{\sI\sJ}=-\phi^{\sJ\sI}$. Moreover, amplitudes in this SU$(2)$ subsector are non-trivial, as long as $n/2$ of the scalars are in $Y$, and the other $n/2$ in $\overline{Y}$. This is most easily seen in the R-symmetry breaking representation, where the multiplets take the form
\begin{subequations}
 \begin{align}
  \Phi^{\LG}_{\mathrm{D5}} &= \phi^{1\dot 1}_{\LG} + \eta_a\psi^{a\dot 1}+\tilde\eta_{\dot a}\tilde\psi^{1\dot a}+\eta_a\tilde\eta_{\dot a} A^{a\dot a}+\eta^2 \phi^{2\dot 1}_{\LG}+\tilde\eta^2\phi^{1\dot 2}_{\LG}+\dots +\eta^2\tilde\eta^2\phi^{2\dot 2}_{\LG}\,,\\
  \Phi^{\LG}_{\mathrm{M5}} &= \phi_{\LG} + \eta_{al}\psi^{al}+\varepsilon^{lm}\eta_{al}\eta_{bm}\; B^{ab}+\eta_{al}\eta^{a}_m\, \phi^{lm}_\LG+(\eta^3)^{al}\,\tilde\psi_{al} +\eta^4\,\tilde\phi_\LG\,,
 \end{align}
\end{subequations}
with $\phi_\LG=\phi^{lm}$, $\tilde\phi_\LG=\phi_{lm}$ and  $\phi_\LG^{lm}=\varepsilon^{mn}\phi^l{}_n$ in the M5 multiplet. In this representation, amplitudes are monomials of degree $2n$ in the fermionic variables, so scalar amplitudes from the SU$(2)$ subsector are generically non-trivial when $n/2$ particles are in $\overline{Y}$, as claimed above. 

Using this construction, we can restrict the generic scalar amplitudes of \eqref{eq:D5_scalar_generic}  to the SU$(2)$ subsector with  $|Y|=|\overline{Y}|=n/2$. The matrices $\mathcal{U}$ and $\tilde{\mathcal{U}}$ then take the form
\begin{equation}\label{eq:reduced_U}
 \mathcal{U}=\begin{pmatrix} 0 & U_Y\\ -U_Y^T & 0\end{pmatrix}\,,\qquad  \tilde{\mathcal{U}}=\begin{pmatrix} 0 & \tilde{U}_Y\\ -\tilde{U}_Y^T & 0\end{pmatrix}\,,
\end{equation}
where $U_Y$ and $\tilde{U}_Y$ are $n/2\times n/2$ matrices with entries $U_Y{}_{ip}=U^{(1,0)}_{ip}$ and $\tilde U_Y{}_{ip}=U^{(0,1)}_{ip}$ for for $i\in Y$ and $p\in\overline{Y}$.
In this SU$(2)$ scalar subsector, the scalar D5 amplitudes are thus given by
\begin{equation}\label{eq:D5_scalar}
 \mathcal{I}_{\mathrm{D5}}^{\mathrm{SU}(2)}=\det{}'A\;\det{}'H\;\det U_Y\;\det\tilde{U}_Y\,.
\end{equation}
We can compare this to the same scalar subsector in the CHY formalism \cite{Cachazo:2014xea}, where the integrand is given by $\mathcal{I}_{\mathrm{D5}}^{\mathrm{SU}(2)}=(\pf{}'A)^3 \det X_Y$. Here,  $X_Y$ is 
an $n/2\times n/2$ matrix with entries $X_Y{}_{ip}=\sigma_{ip}^{-1}$, again with
for $i\in Y$ and $p\in\overline{Y}$. This gives the identity
\begin{equation}\label{eq:pfA_vs_detH}
 \frac{\det X_Y}{\det U_Y\,\det \tilde U_Y}\;\pf{}'A=\det{}'H\, .
\end{equation}

\paragraph{The M5 integrand.} 
As discussed above, the scalar amplitudes \eqref{eq:D5_scalar} are the same in both the M5 and D5 theory. Supersymmetry then uniquely determines the M5 integrand $\mathcal{I}_n^{\mathrm{M5}}$ from this SU$(2)$ scalar subsector as follows. Consider the following generic ansatz for the M5 superamplitude,
\begin{equation}
 \mathcal{A}_n^{\mathrm{M5}}=\int d\mu_n^{\mathrm{pol}} \;\mathcal{I}_{\mathrm{M5}}\,e^{F_2}\,.
\end{equation}
By integrating over suitable supermomenta $q_\sI$, we can again extract the SU$(2)$ scalar sector, and a similar calculation to the above D5 case gives
\begin{equation}\label{eq:SU2_M5}
 \mathcal{I}_{\mathrm{M5}}^{\mathrm{SU}(2)}=\mathcal{I}_{\mathrm{M5}}\;\det{}^2 U_Y\,.
\end{equation}
There is no contribution of the local terms $-\frac{1}{2}\sum_i\la \xi_iv_i\ra q_i^2\subset F_2$ in the exponential because the scalars obey $\Omega_{\sI\sJ}\phi^{\sI\sJ}=0$.
As discussed above, the amplitudes \eqref{eq:SU2_M5} in the SU$(2)$ scalar subsector have to agree with the D5 case \eqref{eq:D5_scalar}, which uniquely determines the M5 integrand to be
\begin{equation}
 \mathcal{I}_{\mathrm{M5}}=\det{}'A\,\det{}'H \, \frac{\det \tilde U_Y}{\det U_Y}=\left(\pf{}'A\right)^3\frac{\det X_Y}{\det{}^2 U_Y}\,,
\end{equation}
where the second equality follows from \eqref{eq:pfA_vs_detH}.\\

While this gives a valid formula for the M5 integrand, it obscures the permutation invariance of the Gerbe amplitudes, because the integrand superficially seems to depend on $Y$. However, it turns out that all of the combinations
\begin{equation}\label{eq:ratios}
 \frac{\det \tilde U_Y}{\det U_Y}\,,  \qquad \frac{\det X_Y}{\det{}^2 U_Y}\,,\qquad \frac{\det X_Y}{\det U_Y\, \det \tilde U_Y}\,,
\end{equation}
are in fact permutation invariant, and in particular independent of the choice of $Y$. This can be made manifest, as pointed out in \cite{Schwarz:2019aat}, by using results first derived in \cite{Roehrig:2017wvh} relating the above ratios to Pfaffians of a family of matrices $U^{(a,b)}$, defined as before by
\begin{equation}
 U^{(a,b)}_{ij}=\frac{\la u_i u_j\ra ^a\left[ \tilde u_i\tilde u_j\right]^b}{\sigma_{ij}}\,.
\end{equation}
The main theorem we will need here, derived in \cite{Roehrig:2017wvh}, gives a fundamental identity for the splitting of the Pfaffian $\pf U^{(a,b)}$ into two determinants,
\begin{equation}\label{eq:lemma_Kai}
 \pf U^{(a,b)} = \frac{\det U_{Y_1}^{(a_1,b_1)}}{V_{Y_1}V_{\overline{Y}_1}}\frac{\det U_{Y_2}^{(a_2,b_2)}}{V_{Y_2}V_{\overline{Y}_2}}V\,,\qquad \text{with }\,a=a_1+a_2\,,\;b=b_1+b_2\,.
\end{equation}
Here, $V$ denotes the Vandermonde determinant, and $V_{Y_{1,2}}$ are the Vandermonde determinants for the subsets $Y_{1,2}$ etc. Ref.~ \cite{Roehrig:2017wvh} further proves that each of the factors $\det U_{Y_1}^{(a_1,b_1)}/V_{Y_1}V_{\overline{Y}_1}$ are invariant under the full $S_{n}$ permutation group,  despite only manifesting permutation invariance on the subgroup $S_{n/2}\times S_{n/2}\times \mathbb{Z}_2$. The only further identity we will need is for $\det X_Y$, which can be expressed as 
\begin{equation}
 \det X_Y = \frac{V_Y^2 V_{\overline Y}^2}{V}\,.
\end{equation}
If we choose $Y_1=Y_2=Y$ in \eqref{eq:lemma_Kai}, we thus find that 
\begin{equation}\label{eq:lemma_kai_v2}
  \pf U^{(a,b)} = \frac{\det U_{Y}^{(a_1,b_1)}\det U_{Y}^{(a_2,b_2)}}{\det X_Y}\,.
\end{equation}
This gives manifestly permutation invariant formulae for all of the ratios in \eqref{eq:ratios},
\begin{equation}
 \pf U^{(2,0)}=\frac{\det{}^2 U_Y}{\det X_Y}\,,\qquad  \pf U^{(1,1)}=\frac{\det U_Y\, \det \tilde U_Y}{\det X_Y}\,,
\end{equation}
from which we deduce the following manifestly permutation invariant representation fo the M5 half-integrand, as well as the following relation between the reduced determinant $\det{}'H$ and $\det{}'A$,
\begin{equation}\label{eq:M5_half-int}
 \mathcal{I}^{\mathrm{h}}_{\mathrm{M5}}=\frac{\pf{}'A}{\pf U^{(2,0)}}\,,\qquad \det{}'H=\frac{\pf{}'A}{\pf U^{(1,1)}}\,.
\end{equation}
In particular, the full M5 superamplitude takes the form
\begin{equation}
 \mathcal{A}_n^{\mathrm{M5}}=\int d\mu_n^{\mathrm{pol}} \;\det{}'A\;\frac{\pf{}'A}{\pf U^{(2,0)}}\,e^{F_2}\,.
\end{equation}
This integrand now manifests $\mathcal{N}=(2,0)$ supersymmetry and is manifestly chiral and permutation invariant. We note that all dependence on the polarization data is encoded by the Pfaffian $\pf{}U^{(2,0)}$, an argument similar to the one presented in \S\ref{sec:poldata} guarantees that the amplitude is indeed linear. While the integrand is guaranteed to be correct by construction (supersymmetry and agreement with the SU$(2)$ scalar subsector of D5 theory), we  verify in \S\ref{sec:dim-red} that both M5 and D5 amplitudes agree upon dimensional reduction to five dimensions as an additional check.

\subsection{Consistency of the reduced determinant with the supersymmetry representation}\label{sec:multiplet}
Our gauge (and gravity) formulae in effect give two different representations of bosonic amplitudes with gluons coming from different parts of the multiplets.  One comes from simply substituting gluon polarizations from different parts of the multiplet in the kinematic integrand $\det{}'H$ and the other from expanding out the supersymmetry factors. In this subsection we show that these give the same formulae.   

When  a subset $I$ of the particles are in states at the bottom of the (chiral part of the) supersymmetry multipet, the integrals over the supercharges lead to the integrand
\begin{equation}\label{eq:I=UH}
 \mathcal{I}_n^{\mathrm{h}}=\det U^I\det{}'H\;e^{F^{\bar{I}}+\tilde F}\,,
\end{equation}
where  $U^I_{ij} = U^{(1,0)}_{ij}$ and the superscripts indicate the restriction to the subsets $I$ and $\bar I$ respectively. On the other hand, for \emph{any} choice of polarization data, the integrand for gluons (gravitons) takes the form of a reduced determinant, 
\begin{equation}\label{eq:I=hatH}
\mathcal{I}_n^{v_{i_1}\dots v_{i_{\scalebox{0.5}{$|I|$}}}}=\det{}'  H^I\;e^{F^{\bar{I}}+\tilde F}\,,\qquad\qquad\text{with }\;\;
  H^I_{ij} = \begin{cases}
                                  H_{ij} & i\notin I\\
                                  \frac{\la \xi_i\kappa_{i\sA}\ra\epsilon_j^\sA}{\sigma_{ij}} & i\in I\,,
                              \end{cases}
\end{equation}
where $ H^I$ is defined with polarization spinors $\la \xi_i \kappa_{i\sA}\ra$ instead of $\epsilon_{i\sA}$ for $i\in I$. For the supersymmetry to be compatible with the representation of the integrand, the two prescriptions for the amplitude must agree,  $\mathcal{I}_n^{\mathrm{h}}=\mathcal{I}_n^{\xi_{i_1}\dots \xi_{i_{\scalebox{0.5}{$|I|$}}}}$.

\paragraph{A lemma on reduced determinants.}  To prove the equivalence of \eqref{eq:I=UH} and \eqref{eq:I=hatH}, the general strategy will be to first identify the relation between $H$ and $H^I$. To draw conclusions about the behaviour of their reduced determinants though, we will need a few results discussed in appendix A of \cite{Cachazo:2012pz}, which we review here for convenience. 

In contrast to regular determinants, it does not make sense to ask how a reduced determinant behaves under the addition of an arbitrary vector to a row or column of $H$, because this will in general spoil the linearity relations among its rows and columns. On the other hand, we \emph{can} define a new  reduced determinant by multiplication with an invertible $n\times n$ matrix $U$, since this leaves the (full) determinant $\det H=\det \hat H=0$ unaffected,
\begin{equation}\label{eq:HU_transf}
 \hat H{}_i^j:= U_i^k\,H_k^j\,.
\end{equation}
Since the kernel and co-kernel of $H$ are spanned by $w$ and $\tilde w$,\footnote{As discussed above, for super Yang-Mills and supergravity, we take $w^i_a=u_{ia}$, where $a$ denotes the chiral little group index, and similarly for $\tilde w_j^{\dot b}=\tilde u_j^{\dot b}$.} the kernel of $\hat H = UH$ is $\hat w=U^{-1}w$. To be explicit, $\hat H$ and $\hat w$ satisfy relations analogous to \eqref{eq:co-rank},
\begin{equation}
 \sum_i \hat w{}_a^i \hat H{}_i^j=0\,,\qquad
 \sum_j \tilde w_j^b \hat H{}_i^j=0\,,\qquad\qquad\text{for } \hat w_a^i=\left(U^{-1}\right)^i_k\, w_a^k\,. 
\end{equation}
We can thus define a reduced determinant $\det{}' \hat H$ as in \eqref{red-det} by 
\begin{equation}\label{eq:detH'}
 \varepsilon^{i_1 i_2\dots i_n}\varepsilon_{j_1j_2\dots j_n} \hat H{}_{i_{p+1}}^{j_{p+1}}\dots \hat H{}_{i_n}^{j_n}\, \la \hat w_1\dots \hat w_p\ra \left[\tilde w^1\dots \tilde w^p \right]
 =\det{}' \hat H\,\, \hat w^{[i_1}_1\dots  \hat w^{i_p]}_p\,\, \tilde w_{[j_1}^1\dots \tilde w_{j_p]}^p \,.
\end{equation}
Let us multiply this equation by $p$ facors of $U$. On the right-hand-side, this cancels the factors of $U^{-1}$ from the kernel $  \hat w^{[i_1}_1\dots  \hat w_p^{i_p]}$, whereas on the left, it combines with the $(n-p)$ factors from $\hat H=UH$ to $\det U$. Putting this all together, we arrive at the following lemma  \cite{Cachazo:2012pz}:

\begin{lemma}\label{lemma:red-det}
 Under multiplication by an invertible matrix $U$, the reduced determinant of a matrix $\hat H{} := U\,H$ behaves as
 \begin{equation}
  \det{}'\hat H=\det U \det{}'H\,,
 \end{equation}
with the reduced determinant defined using the kernel $\hat w=U^{-1} w$.
\end{lemma}

This implies in particular that the usual row- and column operations leave the reduced determinant unaffected,  $\det{}'\hat H=\det{}'H$, due to $\det U=1$. 

\paragraph{Equivalence of the reduced determinants.}  \Cref{lemma:red-det} now allows us to prove the compatibility of the supersymmetry representation with the reduced determinant. We first note that on the support of the  polarized scattering equations, $H^I$ and $H$ are related via 
 \begin{align}
  H^I_{ij} &= \sum_{k\neq i}\frac{\la u_i u_k\ra}{\sigma_{ik}}\frac{\epsilon_{k\sA}\epsilon_j^\sA}{\sigma_{ij}} -\la \xi_i v_i\ra \frac{\epsilon_{i\sA}\epsilon_j^\sA}{\sigma_{ij}}\nonumber \\
  &= \sum_{k\neq i}\frac{\la u_i u_k\ra}{\sigma_{ik}}H_{kj}-\frac1{\sigma_{ij}}\underbrace{\sum_{k\neq i}\la u_i u_k\ra H_{kj}}_{=0} -\la \xi_i v_i\ra \,H_{ij}=:\sum_k U^{I}_{ik} H_{kj}\,,
 \end{align}
for $i\in I$. In the second equality, the middle term vanishes because $u$ spans the kernel of $H$, and we use the last equality to define $U^{I}$.
Combining the above result with $H_{ij}^I=H_{ij}$ for $i\notin I$, we thus have
\begin{align}
 &H^I=U^IH\,,&& \text{with }\;\;U_{ij}^I=\begin{cases}U_{ij}^{(1,0)} & i\neq j\,,\,i\in I \\
  -\la \xi_i v_i\ra & i=j\in I\\
 \delta_{ij} & i\notin I\,.  \end{cases}
\end{align}
Since $\det U^I$ is generically non-zero, and \cref{lemma:red-det}  gives directly that
\begin{equation}\label{eq:det_hatH=UH}
 \det{}'H^I = \det U^I\det{}'H\,,
\end{equation}
confirming the equivalence of the two prescriptions.

\subsection{Linearity in the polarization data}\label{sec:poldata} 
As another important check on the amplitudes \eqref{eq:integrands}, we verify  that they are multilinear in the polarization data. This is of course a mandatatory requirement for amplitudes, but is not manifest in the integrands for gauge and gravity theories because the reduced determinants depend on the $u$-variables and these can potentially depend in a complicated way on the polarization data via the polarized scattering equations.  We first observe that linearity is manifest for amplitudes with two external scalars and $n-2$ gluons.  Given the supersymmetry of the formulae this provides strong circumstantial evidence.  Then we show explicitly that the reduced determinant is linear on the support of the polarized scattering equations and go on to the full superamplitude.

\subsubsection{Linearity from supersymmetry}\label{sec:lin_from_susy}
Linearity of the gluon states is most easily seen from the mixed amplitudes with two external scalars, e.g. $j=1,2$, and $n-2$ gluons. In this case, we can choose to reduce the determinant $\det{}'H$ on the scalar states, giving
\begin{equation}
 \mathcal{A}^{\phi_1\phi_2\epsilon_3\tilde\epsilon_3\dots}=\int d\mu_n^{\mathrm{pol}}\; \frac{1}{\sigma_{12}^2}\det H^{[12]}_{[12]}\; \mathrm{PT}(\alpha)\,.
\end{equation}
The integrand is then manifestly independent of $\{u_{i},v_{i}\}$ as well as $\epsilon_{1,2}$, and only depends on the punctures $\sigma_i$ and the polarization of the gluons. Due to the invariance of the measure established by \cref{prop:measures}, the `polarization' spinors of the scalars $\epsilon_{1,2}$ are choices of reference spinors. For the gluons on the other hand, the integrand is now manifestly linear in $\epsilon_i$. Supersymmetry then guarantees that linearity extends to the all-gluon amplitude.

The consistency between the supersymmetry representation and the reduced determinant discussed in the last section further guarantees that the argument above holds for gluons both at the top and the bottom of the multiplet; we simply replace $H$ by $H^I$. For gravity and  brane-amplitudes, the argument is completely analogous, and follows again from the multilinearity of the amplitude $\mathcal{M}^{\phi_1\phi_2\epsilon_3\tilde\epsilon_3\dots}$ with two scalars and $n-2$ gravitons. 

\subsubsection{Linearity for non-supersymmetric amplitudes.}
We now study the dependence of the reduced determinant on the polarization data directly by expanding the spinors $\epsilon^a$ in a basis. This gives the desired linearity for  pure Yang-Mills  and gravity directly, where the above supersymmetry argument seems excessive, but can equally be applied to supersymmetric theories. We first discuss (chiral) linearity for gluons, but the proof extends straightforwardly to linearity in the anti-chiral polarization data, as well as (bi-)linearity for gravity amplitudes.

Consider the amplitude $A^{\epsilon_1}$ or the superamplitude $\mathcal{A}^{ \epsilon_1}$, where one of the particles is a gluon with polarization $ \epsilon_1$, and all other particles are in arbitrary states. We can expand $\epsilon_1$ in an (arbitrarily chosen) polarization basis $\zeta^a_1,\zeta^a_2$ via
\begin{equation}
 \epsilon^a_1 = \alpha_1 \zeta^a_1 +\alpha_2\zeta^a_2\,,\qquad \qquad \text{with }\,\la \zeta_2\zeta_1\ra=1\,.
\end{equation}
It will be helpful to think of this new basis $(\zeta_1,\zeta_2=:\xi_1^{\sz{1}})$ as playing a similar role to $(\epsilon_1,\xi_1)$, both in the polarized scattering equations and in the integrands. 
To prove linearity of the (super-) amplitudes in the polarization, we then have to show that amplitudes in the two different bases are related via
\begin{equation}\label{eq:ampl_lin_pol}
 A^{\epsilon_1} = \alpha_1\, A^{\zeta_1}+\alpha_2\,A^{\zeta_2}\,,
\end{equation}
where the amplitudes $A^{\epsilon_1}$ and $A^{\szr}$ are respectively given by
\begin{equation}
 A^{\epsilon_1}= \int d\mu_n^{\mathrm{pol}}\det{}' H\; \mathrm{PT}(\alpha)\,,\quad A^{\szr}=\int d\mu_{n}^{\mathrm{pol},\szr} \det{}' H^{\szr}\; \mathrm{PT}(\alpha)\,,
\end{equation}
and the superscripts $\zeta_r$ indicate that the respective quantities are defined using the polarization $\zeta_r$. For the measure, \cref{prop:measures} guarantees that $ d\mu_n^{\mathrm{pol}}= d\mu_n^{\mathrm{pol},\szr}$, but the integration variables $ u_i^{\szr}=u_i(\zeta_r)$ defined by  $ d\mu_n^{\mathrm{pol},\szr}$ enter into  the definition of the reduced determinant $\det{}'H^{\szr}$. Since the measure and the Parke-Taylor factors are invariant under changes of polarization, the linearity relation \eqref{eq:ampl_lin_pol} for the amplitude is equivalent to linearity of the spin-one contribution;
\begin{equation}
 \det{}'H = \alpha_1\det{}'H^{\sz{1}}+\alpha_2\det{}'H^{\sz{2}}\,,
\end{equation}
where the (implicit) map  between $\{u_i,v_i\}$  on the left-hand side and $\{u_i^\szr,v_i^\szr\}$ on the right hand side is determined by the polarized scattering equations.

\begin{propn}\label{prop:cov}
For $\epsilon^a_1 = \alpha_1 \zeta^a_1 +\alpha_2\zeta^a_2$ expand also $ v_1^a=\beta_1\zeta_1^a+\beta_2\zeta_2^a$ so that $\la \epsilon_1 v_1\ra=1$  gives $\alpha_1\beta_2-\alpha_2\beta_1=1$.  Then we have that 
$\{u_i,v_i\}$   and $\{u_i^\szr,v_i^\szr\}$ are related by
\begin{subequations}\label{eq:uv_full_change}
 \begin{align}
  &v_1^a=\beta_2\, v_1^{\sz{1}\,a} && u_1^a=\beta_2\, u_1^{\sz{1}\,a} \label{eq:uv_change_1}\\
  & v_i^a = v_i^{\sz{1}\,a} +\alpha_2\beta_2\,  \frac{\la u_1^\sz{1} u_i^\sz{1}\ra^2}{\sigma_{1i}^2}\epsilon_i^a && 
  u_i^a = u_i^{\sz{1}\,a} -\alpha_2\beta_2  \frac{\la  u_1^\sz{1} u_i^\sz{1}\ra}{\sigma_{1i}} u_1^{\sz{1}\,a}\,,
 \end{align}
\end{subequations}
with  identical  expressions for  $\{u_i,v_i\}$ in terms of  $\{u_i^\sz{2},v_i^\sz{2}\}$.
\end{propn}

\proof  First note that the punctures $\sigma_i$ are unaffected so we omit the superscripts here. First write $\epsilon_1^a=( \zeta_1^a+  \alpha_2v_1^a)/\beta_2 $. Using this,  the polarized scattering equations $\mathcal{E}_i$ can be written in the form
 \begin{align}
  \mathcal{E}_{1\sA}&= \sum_{j\neq 1}\frac{\la u_1u_j\ra}{\sigma_{1j}}\epsilon_{j\sA}-\la v_1\kappa_{1\sA}\ra\\
  \mathcal{E}_{i\sA}&= \sum_{j\neq 1,i}\underbrace{\left(\frac{\la u_i u_j\ra}{\sigma_{ij}}+\frac{\alpha_2}{\beta_2} \frac{\la u_1 u_i\ra}{\sigma_{1i}}\frac{\la u_1 u_j\ra}{\sigma_{1j}}\right)}_{\stackrel{!}{=}\frac{\la u_i^{\scalebox{0.5}{$\zeta_1$}} u_j^{\scalebox{0.5}{$\zeta_1$}} \ra}{\sigma_{ij}}}\epsilon_{j\sA}+\frac{1}{\beta_2}\frac{\la u_1 u_i\ra}{\sigma_{1i}}\la \zeta_1\kappa_{1\sA}\ra -\underbrace{\left(\la v_i\kappa_{i\sA}\ra-\frac{\alpha_2}{\beta_2} \frac{\la u_1 u_i\ra^2}{\sigma_{1i}^2}\epsilon_{i\sA}\right)}_{\stackrel{!}{=}\la v_i^{\scalebox{0.5}{$\zeta_1$}} \kappa_{i\sA} \ra}\,. \nonumber
 \end{align}
It is now simple to map this to the polarized scattering equations  $\mathcal{E}_i^{\sz{1}}$ via the change of variables \eqref{eq:uv_change_1}.\hfill $\Box$\\

As an aside, although Proposition \ref{prop:measures} implies that the measures are unchanged, it is easily checked directly that $d\mu_n^{\mathrm{pol}}=d\mu_n^{\mathrm{pol},\sz{1}}$: the rescaling \eqref{eq:uv_change_1} gives an overall  factor of $\beta_2^{-4}$ coming from the scattering equation $\delta( \mathcal{E}_1) =\beta_2^{-4}\delta(\mathcal{E}_1^\sz{1})$, which exactly compensates the factor from $d^2u_1 d^2 v_1 = \beta_2^{4}\, d^2 u_1^\sz{1} d^2 v_1^\sz{1}$. The remaining part of the measure is  invariant under the linear shift in $\alpha_2\beta_2$, and thus the polarized measure is invariant under the choice of polarization data.

\begin{thm} With the above definitions
\begin{equation}
 \det{}'H =  \alpha_1\det{}'H^{\sz{1}}+\alpha_2\det{}'H^{\sz{2}}\,.
\end{equation}
\end{thm}

\proof  For each solution to the scattering equations, the above correspondence \eqref{eq:uv_full_change} maps the reduced determinant by
\begin{equation}\label{eq:det'H=binvdet'H1}
 \det{}'H =  \frac{1}{\la  u_1  u_i\ra\,\left[ \tilde u_1 \tilde u_i\right]}\det H^{[1i]}_{[1i]} =\frac{1}{\beta_2}\det{}'H^{\sz{1}}\,.
\end{equation}
Here, we have reduced on particle 1 for convenience, and used the fact that the diagonal entries $H_{ii}$ for $i\neq 1$ are independent of the polarization $\epsilon_1$ by \eqref{Hii-id}. Similarly, the map from $\{u_i,v_i\}$ to  $\{u_i^\sz{2},v_i^\sz{2}\}$ induced by the polarized scattering equations gives 
\begin{equation}\label{eq:det'H=binvdet'H2}
 \det{}'H =-\frac{1}{\beta_1}\det{}'H^{\sz{2}}\,.
\end{equation}
Note that $\beta_{1,2}$ depend on the solutions to the polarized scattering equations, so the relations \eqref{eq:det'H=binvdet'H1} and \eqref{eq:det'H=binvdet'H2} between the reduced determinants only hold on individual solutions to the scattering equations, and do not lead to an analogous relation for the amplitudes. However, by combining the two expression we get the following linearity relation
\begin{equation}
 \det{}'H = \left(\alpha_1\beta_2-\alpha_2\beta_1\right)\det{}'H = \alpha_1\det{}'H^{\sz{1}}+\alpha_2\det{}'H^{\sz{2}}\,,
\end{equation}
as required.  This is now independent of the solutions to the scattering equations, and thus lifts to the full amplitudes, confirming \eqref{eq:ampl_lin_pol}. \hfill $\Box$

\paragraph{Superamplitudes.} The above analysis extends straightforwardly to superamplitudes to give checks on the supersymmetry factors. As before, we take particle 1 to be a gluon, though we do not restrict its position in the multiplet in the supersymmetric case. In the top state, its polarization is $\epsilon_1=\alpha_1\zeta_1+\alpha_2\zeta_2$ as above, and in the bottom state we choose the polarization \begin{equation}
  \xi_1=\alpha_1^\xi\zeta_1+\alpha_2^\xi\zeta_2\,,                                                                                                                                                                                                                                                                                                                                                                                                                    \end{equation}
with constant $\alpha_{1,2}^\xi$ such that $\alpha_1\alpha_2^\xi - \alpha_2\alpha_1^\xi=1$ due to the normalization condition $\la\epsilon_1 \xi_1\ra=1$.

As indicated above, in the supersymmetric case it will be helpful to treat the basis spinors $(\zeta_1,\zeta_2)$ as the new basis for the multiplet of particle 1. In the explicit change of variables given in \cref{prop:cov}, $\zeta_1$  plays the r\^ole of the original $\epsilon_1$, and $\zeta_2$ provides the additional polarization spinor to parametrize the full mutiplet, i.e. $\xi_1^{\sz{1}}=\zeta_2$.\footnote{Of course, we are free to reverse the roles of $\zeta_1$ and $\zeta_2$ in this discussion, at the expense of a minus sign due to our normalization conventions.} Using this choice, we can verify by expanding out both sides and using the relation between  $\{u_i,v_i\}$ and $\{u_i^\sz{1},v_i^\sz{1}\}$ from \cref{prop:cov} that 
\begin{align}\label{eq:eFhat}
 \int d^2\! q_1\, q_1^2\, e^{F} =
 \int d^2\! q_1^\sz{1}\, \beta_2\,\left(\alpha_1 \big(q_1^\sz{1}\big)^2+\alpha_2\right)\, e^{F^\ssz{1}}  \,.
\end{align}
The superscript $\zeta_1$ again indicates that the supersymmetry factor is defined with the multiplet parametrized by the polarization $\zeta_1$, as well as the variables $u_i^\sz{1}$. Similarly, for gluon states at the bottom of the multiplet, we find 
\begin{equation}
 \int d^2\! q_1\,  e^{F} = \int d^2\! q_1^\sz{1}\, \beta_2\,\left(\alpha_1^\xi\ \big(q_1^\sz{1}\big)^2+\alpha^\xi_2\right)\, e^{F^\ssz{1}}  \,.
\end{equation}
Combining this with the result \eqref{eq:det'H=binvdet'H1} for the reduced determinant $\det{}'H=\beta_2^{-1}\det{}'H^\sz{1}$, we find the expected linearity relations for supersymmetric integrands with one gluon, 
\begin{align}\label{eq:lin_susy-int}
 \det{}'H\,\int d^2\! q_1\, q_1^2\, e^{F} &=\det{}'H^\sz{1}\,\int d^2\! q_1^\sz{1}\, \left(\alpha_1 \big(q_1^\sz{1}\big)^2+\alpha_2\right)\, e^{F^\ssz{1}} \,,
\end{align}
and similarly for the gluon at the bottom of the multiplet with polarization $\xi_1$. The simplicity of this relation is due to our choice of $\xi_1^{\sz{1}}=\zeta_2$: using this, as well as the results from \S\ref{sec:multiplet}, the second term on the right gives indeed the amplitude for a gluon with polarization $\zeta_2$ with a proportionality factor of $\alpha_2$.
As in the bosonic case, the final linearity relation \eqref{eq:lin_susy-int} is independent of the solution to the polarized scattering equations, and thus lifts to the full superamplitude,
\begin{equation}
 \mathcal{A}^{\epsilon_1} = \alpha_1\, \cA^{\zeta_1}+\alpha_2\,\cA^{\zeta_2}\,,\qquad\qquad \mathcal{A}^{\xi_1} = \alpha_1^\xi\, \cA^{\zeta_1}+\alpha_2^\xi\,\cA^{\zeta_2}\,.
\end{equation}

\section{The three and four-point amplitudes}\label{sec:low-point}

In this section, we discuss the three-particle and four-particle amplitudes in our polarized scattering equations formalism \eqref{eq:integrands}, and compare them to previous results available in the literature, e.g. \cite{Cheung:2009dc}. We first focus on the three-particle amplitudes that will serve as the seed amplitudes for the BCFW recursion relation of \cref{sec:BCFW}. Since the configuration of three momenta is highly degenerate, we include a treatment of the four-particle case for further  illustration.\\

For the calculations below, two general observations will be helpful. First, for low numbers of external particles, the most useful formulation of the scattering equations  arise from \eqref{polscatt-k}, obtained by skew-symmetrizing the $i$th polarized  scattering  equation with $\epsilon_{i\sA}$ to give
\begin{equation}
\sum_j \frac{\langle u_iu_{j}\rangle \epsilon_{j[\sA}\epsilon_{\sB]i}}{\sigma_{ij}}=K_{i\sA\sB}\, .\label{SE-k}
\end{equation}
This can be skewed with further polarization spinors to obtain formulae for $u_{ij}:=\langle u_iu_j\rangle/\sigma_{ij}$. We will use this below to construct explicit solutions to the polarized scattering equations, both for three and four particles.

After solving the polarized scattering equations and simplifying the integrands on these solutions, amplitudes are expressed in the form $A^{\epsilon_1\tilde\epsilon_1\dots\epsilon_n\tilde\epsilon_n}$, with all little group indices contracted linearly into the polarization spinors $\epsilon_i^a$ and $\tilde \epsilon_i^{\dot a}$. To compare our results to the formulae obtained in e.g. \cite{Cheung:2009dc}, we thus have to convert between our polarized formalism and the standard, little-group covariant spinor-helicity formalism, where amplitudes $A_n^{a_1\dot a_1\dots a_n\dot a_n}$ carry the little group indices of the scattered particles. Using that the amplitudes \eqref{eq:integrands} are linear in the polarization spinors $\epsilon_i^a$ and $\tilde \epsilon_i^{\dot a}$ as shown in \S\ref{sec:poldata}, the two formalisms are related via
\begin{equation}\label{eq:pol_vs_sh}
 A^{\epsilon_1\tilde\epsilon_1\dots\epsilon_n\tilde\epsilon_n} = \prod_i \epsilon_{i a_i}\tilde\epsilon_{i\dot a_i}\dots A_n^{a_1\dot a_1\dots a_n\dot a_n}\,.
\end{equation}

\subsection{Three-point amplitudes}\label{sec:3pt}
We now compute the three particle case to compare to the Yang-Mills result given in \cite{Cheung:2009dc}.
This case is somewhat degenerate as momentum conservation implies that the three null momenta are also mutually orthogonal.  In Lorentz signature they would of necessity be proportional, which would be too degenerate to calculate with.  We therefore allow complex momenta so that they span a null two-plane. This can be expressed by the non-vanishing $2-$form that is given in spinors by
\begin{equation}
\kappa_\sB\kappa^\sA:=(k_1\wedge k_2)_\sB^\sA=-(k_1\wedge k_3)_\sB^\sA=(k_2\wedge k_3)_\sB^\sA\,.
\end{equation}
The spinors $\kappa_\sA$ and $\kappa^\sA$ are defined up to an overall scale and its inverse and are orthogonal to each momentum.  

\begin{figure}[!h]
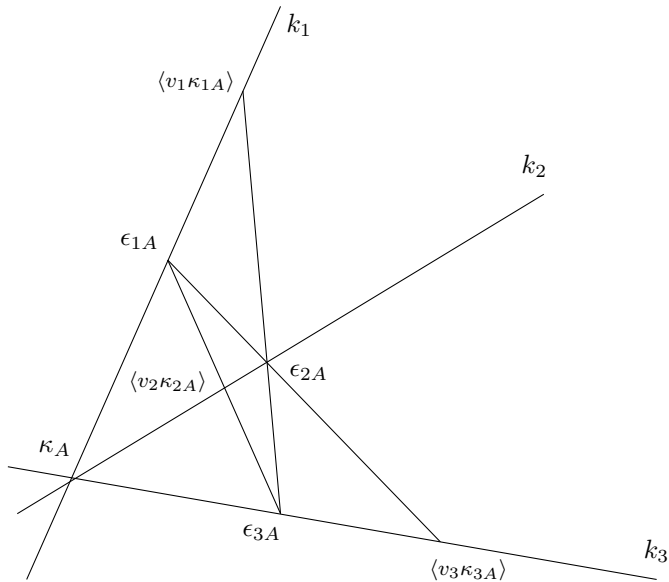

\begin{center}
\ctikzfig{3point}
\label{fig:3pt}
\caption{Each $k_i$  corresponds to a line in the projective spin space spanned by $\kappa_{iaA}$. The 
lines lie in a common two-plane orthogonal to $\kappa^A$ and are concurrent meeting at $\kappa_A$ defined by \eqref{defkappa}. Thus the line  $k_1$ joins   $\epsilon_{1A}$ and $\kappa_A$ and so on. The polarized scattering equations give 3 further lines,   e.g.\ with $\mathcal{E}_{1A}$ giving the line joining $\epsilon_{2A}$ and $\epsilon_{3A}$ and  intersecting $k_1$ at $\la v_1\kappa_{1A}\ra$.
}
\end{center}
\end{figure}

We can represent each momentum $k_{i\sA\sB}$ as a line in the projective spin space $\CP^3$ through the two spinors  $\kappa_{ia\sA}$ for $a=1,2$.  That each line contains $\kappa_\sA$ means that they are concurrent and that they are orthogonal to $\kappa^\sA$ means that they are co-planar as in the diagram \ref{fig:3pt}.

To compare to the results of \cite{Cheung:2009dc}, we introduce little group spinors $m_i^a$, $\tilde m_i^{\dot a}$ for each $i$
\begin{equation}\label{defkappa}
\kappa_\sA =\, , \qquad \kappa^\sA=\tilde{m}_i^{\dot a}\kappa^\sA_{i\dot a}\, .
\end{equation}
These  are defined in \cite{Cheung:2009dc} equivalently by
\begin{equation}\label{Cheungm}
\kappa_{i\sA a}\kappa_{j\dot{b}}^\sA=m_{ia}\tm _{j\dot{b}}\,.
\end{equation}
As in \cite{Cheung:2009dc}, we further introduce spinors  $w_i, \tw_i$ normalized against $m_i$, $\tilde m_i$ such that
\begin{equation}\label{Cheungw}
m_{ia}w_i^{a}=1\, , \qquad \tm_{i\dot a}\tw^{\dot a}_i=1\,.
\end{equation}
This normalization does not fully fix $w_i, \tw_i$, since we have the further freedom to add on terms proportional to $m_i, \tilde m_i$. We can partially fix this redundancy $w_{ia}\rightarrow w_{ia} + c_i m_{ia}$ by the condition
\begin{equation}\label{momcons}
w_1^a\kappa_{1\sA a}+w_2^a\kappa_{2\sA a}+w_3^a\kappa_{3\sA a}=0\,,
\end{equation}
which imposes co-linearity of the three points  $\la w_i \kappa_{i\sA}\ra$ on the lines $k_i$ and reduces the redundancy to shifts satisfying $c_1+c_2+c_3=0$.\\

In what follows we will compute the three gluon amplitude from the general formula (\ref{eq:ampl_susy}) in Yang Mills theory. For three particles the $\sigma_i$ can be fixed to $(0,1,\infty)$ and the formula reduces to
\begin{equation}\label{amp}
A_3=\detp H|_*=\frac{\epsilon_{1\sA}\epsilon_2^\sA}{U_{23}\tilde U_{13}}\,,
\end{equation}
evaluated on the solution to the polarized scattering equations, as indicated by the star. Note that the Jacobian from solving the polarized scattering equations is trivial due to \cref{prop:measures}. Having gauge fixed three of the $u$ variables as in \S\ref{sec:reltoCHYmeasure}, we only need to solve the polarized scattering equations  for the three $U_{ij}:=U_{ij}^{(1,0)}=\la u_i u_j\ra/\sigma_{ij}$, with $U_{ij}=U_{ji}$ for $i\neq j$,
\begin{equation}\label{3ptSE}
U_{12}\epsilon_{2\sA}+ U_{13}\epsilon_{3\sA}=\la v_1\kappa_{1\sA}\ra \,,\quad \mbox{ and cyclic, } 
\end{equation}
together with the normalization conditions $ \la v_i\epsilon_{i}\ra=1$. These three scattering equations equations define lines in the plane spanned by the three momenta in the projective spin space as in the diagram \ref{fig:3pt}. \\

In order to solve the polarized scattering equations we use the $\epsilon_{i\sA}$ as a basis of the plane in the projective spin space orthogonal to $\kappa^\sA$ to write
\begin{equation}\label{3pk}
\kappa_\sA=\sum_ia_i\epsilon_{i\sA}
\end{equation}
Using the normalization $\langle v_i\epsilon_i \rangle=1$, we can further expand $v_i$ in the polarization basis $\epsilon_i, m_i$;
$$
v_{ia}=\frac{1}{\langle m_i\epsilon_i \rangle} \left(\langle m_i v_i\rangle \epsilon_{ia}+m_{ia}\right)\,,
$$ 
and solve the system \eqref{SE} to obtain
\begin{equation}
 U_{ij}=\dfrac{a_i}{\langle m_j\epsilon_j\rangle}=\dfrac{a_j}{\langle m_i\epsilon_i \rangle}\, , \qquad \langle m_i v_i\rangle=a_i\, .
\end{equation}
To  compare to  \cite{Cheung:2009dc}, we can similarly  decompose 
\begin{equation}
w_i=-\dfrac{1}{\la \epsilon_im_i\ra}\epsilon_i+\dfrac{\la \epsilon_i w_i\ra}{\la \epsilon_im_i\ra}m_i,
\end{equation}
and impose the condition (\ref{momcons}) to obtain:
\begin{equation}
\hspace{15pt}a_i=\dfrac{\prod_{k\neq i}\la \epsilon_km_k\ra}{\la \epsilon_1m_1\ra\la \epsilon_2m_2\ra\la \epsilon_3w_3\ra+\mathrm{cyc.}}
\end{equation}
The scattering equations for spinors in the antifundamental representation are solved entirely analogously and together we obtain from (\ref{amp}) the three point amplitude  as 
\begin{equation}
A_3=\Big(\la \epsilon_1m_1\ra\la \epsilon_2m_2\ra\la \epsilon_3w_3\ra+\mathrm{cyc.}\Big)\Big(\la \te_1\tm_1\ra\la \te_2\tm_2\ra\la \te_3\tw_3\ra+\mathrm{cyc.}\Big)\,, \label{eq:3pt-fin}
\end{equation}
where we have used that $\epsilon_{1\sA}\epsilon_2^\sA=\la \epsilon_1m_1\ra[\epsilon_2 \tm_2]$ from \eqref{Cheungm}.
This is precisely the result in \cite{Cheung:2009dc}, contracted into the polarization spinors as discussed around \eqref{eq:pol_vs_sh}.

\subsection{Four-point Yang-Mills amplitudes}\label{sec:4pt}
To illustrate these techniques in a slightly more generic setting, consider next the four-gluon amplitude in Yang-Mills theory. As before, we can fix three of the marked points on the sphere, e.g.  $\sigma_1,\sigma_2$ and  $\sigma_4$, so that the solution to the scattering equation in homogeneous coordinates is
\begin{equation}\label{solSE}
\sigma_1=[(1,0)]\hspace{5pt}\sigma_2=[(1,1)]\hspace{5pt}\sigma_3=[(1,-\frac{s_{13}}{s_{12}})]\hspace{5pt}\sigma_4=[(0,1)]\,.
\end{equation}
From the measure, we thus pick up the CHY Jacobian $|\Phi|_{\scalebox{0.6}{$[i_1i_2i_3]$}}^{\scalebox{0.6}{$[j_1j_2j_3]$}}:=|\partial \mathcal{E}_i/\partial\sigma_j|_{\scalebox{0.6}{$[i_1i_2i_3]$}}^{\scalebox{0.6}{$[j_1j_2j_3]$}}$ as well as the usual Fadeev-Popov factors $(\sigma_{i_1i_2}\sigma_{i_2i_3}\sigma_{i_3i_1})$ and $(\sigma_{j_1j_2}\sigma_{j_2j_3}\sigma_{j_3j_1})$  due to the equality between the polarized measure and the usual CHY measure established in \cref{prop:measures}. Combining this with the four-particle Yang-Mills integrand \eqref{eq:int_sYM} gives
\begin{align}\label{amp4}
A_4^{\epsilon_1\tilde\epsilon_1\dots\epsilon_4\tilde\epsilon_4}&=\frac{(\sigma_{i_1i_2}\sigma_{i_2i_3}\sigma_{i_3i_1})(\sigma_{j_1j_2}\sigma_{j_2j_3}\sigma_{j_3j_1})}{\det \Phi_{\scalebox{0.6}{$[i_1i_2i_3]$}}^{\scalebox{0.6}{$[j_1j_2j_3]$}}}\; \mathrm{PT}(1234)\,\detp H\bigg|_*&\nonumber \\
&=\frac{\sigma_{12}^2(\sigma_{13}\sigma_{34}\sigma_{41})(\sigma_{23}\sigma_{34}\sigma_{42})}{s_{12}}\;\mathrm{PT}(1234)\,\frac{H_{13}H_{24}-H_{14}H_{23}}{\la u_{3}u_4\ra\,[\tu_{1}\tu_2]}\bigg|_*\\
&=\frac{1}{\la u_{3}u_4\ra\,[\tu_{1}\tu_2]}\frac{\sigma_{12}\sigma_{34}}{s_{12}}\bigg(\epsilon_{1\sA}\epsilon^\sA_3\epsilon_{2\sB}\epsilon^\sB_4-\frac{\sigma_{31}\sigma_{42}}{\sigma_{41}\sigma_{32}}\epsilon_{1\sA}\epsilon^\sA_4\epsilon_{2\sB}\epsilon^\sB_3\bigg)\bigg|_*\,,\nonumber
\end{align}
where $*$ again denotes evaluation on the (single) solution to the polarized scattering equations. Using \eqref{solSE}, the amplitude then becomes
\begin{equation}
A_4^{\epsilon_1\tilde\epsilon_1\dots\epsilon_4\tilde\epsilon_4}=-\frac{1}{s_{12}U_{34}\tilde U_{12}}\bigg(\epsilon_{1\sA}\epsilon^\sA_3\epsilon_{2\sB}\epsilon^\sB_4+\frac{s_{13}}{s_{14}}\epsilon_{1\sA}\epsilon^\sA_3\epsilon_{2\sB}\epsilon^\sB_4\bigg)\bigg|_*,
\end{equation}
evaluated on the solution to the scattering equations. At four points there are $8-3$ independent variables $u_i^a$ and we can take them to be $U_{ij}=\la u_i u_j\ra/\sigma_{ij}=U_{ji}$, $i\neq j$, with the extra relation
\begin{equation}
\la u_{i}u_j\ra \la u_{k}u_l\ra + \big(\mbox{ cyc  } jkl\big) =0,
\end{equation}
given by the Schouten identity. The skewed form \eqref{SE-k} of the scattering equations give
\begin{equation}\label{SE4}
\sum_{j\neq i} U_{ij}\epsilon_{j[\sA}\epsilon_{i\sB]}=k_{i\sA\sB}\, ,
\end{equation}
In order to solve for $U_{34}$ we contract this for $i=3$ with $\varepsilon^{\sA\sB\sC\sD}\epsilon_{1\sC}\epsilon_{2\sD}$ to obtain
\begin{equation}\label{u34}
U_{34}=-\frac{\la k_312\ra}{\la1234\ra}, 
\end{equation}
where we define
\begin{equation}
\la 1234\ra=\varepsilon^{\sA\sB\sC\sD}\epsilon_{1\sA}\epsilon_{2\sB}\epsilon_{3\sC}\epsilon_{4\sD}\, , \qquad \la k_3 12\ra=\varepsilon^{\sA\sB\sC\sD}k_{3\sA\sB}\epsilon_{1\sC}\epsilon_{2\sD}\, .
\end{equation}
Similarly we obtain, using square brackets for 4-brackets of upper-indexed quantities,
\begin{equation}
\tilde U_{12}=-\frac{[k_134]}{[1234]}.
\end{equation}
Using these we can solve for the $v_{ia}$ to give
\begin{equation}\label{eq:sol_v_4pt}
v_{1a}=\frac{\la \kappa_{1a}234\ra}{\la 1234\ra}\, ,
\end{equation}
and so on.

The resulting expression for $A_4$ can be simplified by expanding the product of upper and lower $\varepsilon$ tensors as skew product of Kronecker deltas. Consider the quantity
\begin{equation}
\la k_312\ra[k_134]=4\,\epsilon_{1\sD}\epsilon_3^\sD\, k_{3\sA\sB}\,k_1^{\sA\sC}\epsilon_{4}^\sB\epsilon_{2\sC}+2k_1\cdot k_3(\epsilon_{1\sA}\epsilon^\sA_4\,\epsilon_{2\sB}\epsilon^\sB_3-\epsilon_{1\sA}\epsilon^\sA_3\,\epsilon_{2\sB}\epsilon^\sB_4).
\end{equation}
The first term can be rewritten using  using momentum conservation as 
\begin{align}
    k_{3\sA\sB}\,k_1^{\sA\sC}\kappa_{4\dot{a}}^\sB\kappa_{2\sC a}&= 
-k_{2\sA\sB}k_1^{\sA\sC}\kappa_{4\dot{a}}^\sB\kappa_{2\sC a}=
-\frac{1}{2}\,\kappa_{2\sA a}\kappa_{4\dot a}^\sA\;k_1\cdot k_2\,,
\end{align}
such that $\la k_312\ra[k_134]$ is proportional to the numerator of the amplitude,
\begin{equation}
\la k_312\ra[k_134]=s_{14}\left(\epsilon_{1\sA}\epsilon^\sA_3\,\epsilon_{2\sB}\epsilon^\sB_4+\frac{s_{13}}{s_{14}}\epsilon_{1\sA}\epsilon^\sA_4\epsilon_{2\sB}\epsilon^\sB_3\right)\,.
\end{equation}
The amplitude then agrees with the result of \cite{Cheung:2009dc},
\begin{equation}\label{fourpoint}
A_4^{\epsilon_1\tilde\epsilon_1\dots\epsilon_4\tilde\epsilon_4}=\frac{\la1234\ra[1234]}{s_{12}s_{14}}\,,
\end{equation}
upon the usual identification \eqref{eq:pol_vs_sh}.\\

As discussed in \cref{sec:susy}, the supersymmetry representation we  use breaks little group symmetry so that little group multiplets are spread in different degrees  in the superfield expansion (\ref{fieldR}) in terms of supermomenta. All above expressions are for gluons in the top state $g^{\epsilon\tilde\epsilon}$, but the calculations extend directly to other amplitudes as well. As we have seen in \cref{sec:multiplet}, amplitudes for gluons appearing at order $q^2$ in the multiplet can be calculated either from the supersymmetry representation, or by replacing $\epsilon_i\rightarrow \xi_i$ in the integrand. At four points, this can be seen explicitly: consider first the amplitude $A_4(g^{\epsilon_1\tilde{\epsilon}_1}g^{\epsilon_2\tilde{\epsilon}_2}g^{\xi_3\tilde{\epsilon}_3}g^{\xi_4\tilde{\epsilon}_4})$  obtained from the supersymmetry representation,
\begin{equation}
A_4(g^{\epsilon_1\tilde{\epsilon}_1}g^{\epsilon_2\tilde{\epsilon}_2}g^{\xi_3\tilde{\epsilon}_3}g^{\xi_4\tilde{\epsilon}_4})=A_4^{\epsilon_1\tilde\epsilon_1\dots\epsilon_4\tilde\epsilon_4}
\;\Omega^{\sI\sJ}\Omega^{\scalebox{0.6}{$KL$}}\frac{\partial}{\partial q_3^\sI}\frac{\partial}{\partial q_3^\sJ}\frac{\partial}{\partial q_4^{\scalebox{0.6}{$K$}}}\frac{\partial}{\partial q_4^{\scalebox{0.6}{$K$}}}\;e^{F+\tilde F}\bigg|_*\bigg|_{q_i=0}\,.
\end{equation}
The only non-vanishing term comes from the $F^2$ in the expansion of the exponential, and gives an extra factor of $\det U^{\{34\}} = -U_{34}^2+\la \xi_3 v_3\ra\,\la \xi_4v_4\ra$ in the amplitude. When we evaluate this on the solutions to the polarized scattering equations we obtain, using \eqref{u34} and \eqref{eq:sol_v_4pt},
\begin{equation}
 \det U^{\{34\}}\bigg|_*=\frac{1}{\la 1234\ra^2}\Big(\la \xi_3\, 312\ra\,\la \xi_4\, 412\ra - \la \xi_3\, 124\ra \,\la \xi_4\, 123\ra\Big)=\frac{\la 12\, \xi_3\xi_4 \ra}{\la 1234\ra}\,.
\end{equation}
Here we have used  $k_{i\,\sA\sB} = \xi_{i[\sA}\epsilon_{i|\sB]}$ in the first equality, as well as the notation $\xi_{i\sA}:=\la \xi_i\kappa_{i\sA}\ra$, and the last equality follows from a Schouten identity in the two-dimensional space defined by $\varepsilon^{\sA\sB\sC\sD}\epsilon_{1\sC}\epsilon_{2\sD}$. Using the result \eqref{fourpoint} for the amplitude where all gluons are in the top state, we thus find
\begin{equation}
A_4(g^{\epsilon_1\tilde{\epsilon}_1}g^{\epsilon_2\tilde{\epsilon}_2}g^{\xi_3\tilde{\epsilon}_3}g^{\xi_4\tilde{\epsilon}_4})=\frac{\la12\,\xi_3\xi_4\ra[1234]}{s_{12}s_{14}}.
\end{equation}
This clearly agrees with the result from the integrand $\det{}' H_{I}$ for $I=\{3,4\}$, i.e. by replacing $\epsilon_{ia}$ by $\xi_{ia}$ for $i=3,4$ in  \eqref{fourpoint}.
Similar conjugate formulae  apply for amplitudes with a pair of external particles in the $g^{\epsilon\tilde \xi}$ states.

\subsection{Other theories}\label{Sec:Other}
The Yang-Mills calculations extend directly to the other theories expressed as integrals over the polarized scattering equations. For any theory that admits the representation \eqref{eq:ampl_susy}, the four point amplitude for the top states of the supersymmetry multiplet has the form:
\begin{equation}
    A_4=\frac{1}{\detp{\Phi}\,}\cI^{\mathrm{h}}_\sL \,\cI_\sR^{\mathrm{h}}\bigg|_*\,,
\end{equation}
where the $*$ indicates that the formula is evaluated on the solutions to the polarized scattering equations.
Having solved the polarized scattering equations at four point, \eqref{solSE}, it is now an easy task to evaluate the amplitude for other theories than Yang-Mills \eqref{eq:integrands}. We have already discussed the Jacobian,
\begin{equation}
 \frac{1}{\detp{\Phi}}=\frac{(\sigma_{i_1i_2}\sigma_{i_2i_3}\sigma_{i_3i_1})(\sigma_{j_1j_2}\sigma_{j_2j_3}\sigma_{j_3j_1})}{\det \Phi_{\scalebox{0.6}{$[i_1i_2i_3]$}}^{\scalebox{0.6}{$[j_1j_2j_3]$}}}=-\frac{s_{12}^4}{s_{12}s_{13}s_{14}}
\end{equation}
The main ingredients that appear in the half integrands evaluated on such solutions are as follows:
\begingroup
\addtolength{\jot}{1em}
\begin{subequations}
\begin{align}\label{ingredients}
&\mathrm{PT}(1234)=-\frac{s_{12}}{s_{14}} &
&\detp{H}=\la1234\ra[1234]\frac{s_{12}^2}{s_{12}s_{13}s_{14}} \\
&\pf{}U^{(1,1)} = \frac{s_{13}s_{14}}{\la 1234\ra\,[1234]} &
&\pf{}U^{(2,0)} = \frac{s_{13}s_{14}}{\la 1234\ra^2}\\
&\mathrm{Pf}^\prime A=s_{12}\,. &&
\end{align}
\end{subequations}
\endgroup
It is then straightforward to calculate all four-particle amplitudes for the theories we have discussed. In $(2,2)$ supergravity, for all particles in the top state, we obtain:
\begin{equation}\label{sugra}
M_4^{\mathrm{grav}}=\frac{\la1234\ra^2[1234]^2}{s_{12}s_{13}s_{14}}\,,
\end{equation}
which corresponds to the result in \cite{Dennen:2009vk,Cachazo:2018hqa} and reproduces the KLT relation.
For the brane theories we have
\begingroup
\addtolength{\jot}{1em}
\begin{align}
&A_4^{\mathrm{D5}}=\la1234\ra[1234]\,,\\
&A_4^{\mathrm{M5}}=\la1234\ra^2\,,
\end{align}
\endgroup
agreeing with \cite{Heydeman:2017yww}.  As expected these give the same result on reducing to four or five dimensions where fundamental and anti-fundamental spinors are identified, see \cref{sec:dim-red}. 

\smallskip

The more exotic and controversial formulae in \cref{tabletheories}, obtained by double-copying the above integrands. When combining the M$5$ half integrand with a Parke Taylor factor, we get
\begin{equation}
    A_4^{(2,0)-\mathrm{PT}}= \frac{\la1234\ra^2}{s_{12}s_{14}}\,.\label{4pt-02-PT}
\end{equation}
As expected, the formula is chiral, and has the same reduction to 5d as the Yang-Mills amplitude. 
We can also look at the formulae for other `double copied' theories in table \ref{tabletheories}:
\begingroup
\addtolength{\jot}{1em}
\begin{align}
&A_4^{(3,1)}=\frac{\la1234\ra^3[1234]}{s_{12}s_{13}s_{14}}\label{31}\\
&A_4^{(4,0)}=\frac{\la1234\ra^4}{s_{12}s_{13}s_{14}}\,.\label{40}
\end{align}
\endgroup
We note that \eqref{31}-\eqref{40} give the same result as the gravity amplitudes \eqref{sugra} upon reduction to four and five dimensions.  However, in six dimensions, as remarked in \cite{Huang:2010rn, Cachazo:2018hqa}, the formulae are  more problematic as soft limits (or factorization) to three-point amplitudes are not obviously well-defined.  This is because the three-particle kinematics $\kappa_A=m_i^a\kappa_{ia\sA}$ and $\kappa^A=\tilde m_i^{\dot a}\kappa_{i\dot a}^{\sA}$ of \eqref{defkappa} each have a scaling ambiguity
\begin{align}\label{eq:3pt_scaling}
 & m_i^a\rightarrow \alpha\, m_i^a\,, && \tilde m_i^{\dot a}\rightarrow \alpha^{-1} \tilde m_i^{\dot a}\,,
\end{align}
that cancels in $\kappa_A\kappa^B$.  In our discussion of the Yang-Mills three-particle amplitudes, this was reflected in the  the two factors $\big(\la \epsilon_1m_1\ra\la \epsilon_2m_2\ra\la \epsilon_3w_3\ra+\mathrm{cyc.}\big)\times $ (its tilded version)  not being  individually invariant under the scaling \eqref{eq:3pt_scaling}, although of course this ambiguity cancels in the full amplitude \eqref{eq:3pt-fin}.  In the chiral double-copied amplitudes \eqref{4pt-02-PT} - \eqref{40} however, this scaling ambiguity cannot cancel anymore, so there are no invariant three-point amplitudes for gerbe theories. On reduction to 5d, there is an identification between the chiral and anti-chiral spinors so the scaling in \eqref{eq:3pt_scaling} is fixed up to sign. This is also reflected in the factorization discussion of the related formulae in \cite{Cachazo:2018hqa}, where it was shown that the resulting three-particle formulae are non-local. As discussed there, the non-locality can be made manifest in two different ways. To factorize the four-particle formula into the product of two three-particle objects summed over internal states, we have to either fix a scale $\alpha$ or fix the shift redundancy  $w_{ia}\rightarrow w_{ia} + c_i m_{ia}$ of the dual variables. In both cases, the required `frame choice' depends on the kinematics of all \emph{four} particles, and the three-particle objects are not invariant under the a rescaling of $\alpha$ (in the first case) or a shift in $c_i$ (in the latter case).

Thus it seems unlikely that the formulae \eqref{4pt-02-PT} - \eqref{40}  can be interpreted  as tree-level S-matrices in the normal sense.

\subsection{Fermionic amplitudes}
We can also evaluate amplitudes involving the fermionic sector. We will show here how this works for the scattering of two gluons with two gluini in $(1,1)$ super Yang-Mills, but the results can be adapted easily to supergravity and the brane theories.\\

Consider the four particle amplitude $A_4(g_1^{\epsilon\tilde{\epsilon}},g_2^{\epsilon\tilde{\epsilon}},\psi_3^{\sI\tilde{\epsilon}},\psi_4^{\sJ\tilde{\epsilon}})$ for two gluons and two gluini, obtained in our supersymmetry representation by extracting the fermionic components as follows,
\begin{align}
    A_4(g_1^{\epsilon\tilde{\epsilon}},g_2^{\epsilon\tilde{\epsilon}},\psi_3^{\sI\tilde{\epsilon}},\psi_4^{\sJ\tilde{\epsilon}})&=\frac{\la1234\ra[1234]}{s_{12}s_{14}}\frac{\partial}{\partial q^\sI_3}\frac{\partial}{\partial q^\sJ_4}(1+F_1+\tilde{F}_1+...)\bigg|_{q_i=\tilde{q}_i=0}\nonumber\\
    &=\frac{\la1234\ra[1234]}{s_{12}s_{14}}U_{34}\Omega_{IJ}
\end{align}
Inserting  the solution to the polarized scattering equations (\ref{u34}) we obtain,
\begin{equation}\label{4ptgluini}
    A_4(g_1^{\epsilon\tilde{\epsilon}},g_2^{\epsilon\tilde{\epsilon}},\psi_3^{I\tilde{\epsilon}},\psi_4^{J\tilde{\epsilon}})=\frac{\la 12k_3\ra[1234]}{s_{12}s_{14}}\Omega_{IJ}
\end{equation}
We can compare this to the  amplitude representation of \cite{Dennen:2009vk} in the little-group preserving supersymmetry representation;
\begin{equation}
    A_4^{\mathrm{susy}}=\frac{\delta^4(\sum q)\delta^4(\sum \tilde{q})}{s_{12}s_{14}}\,,
\end{equation}
where the supercharges are $q^{\sA\sI}=\eps{a}{b}\kappa_{\dot{a}}^\sA\tilde{\eta}_{\dot{b}}^\sI$ and $q_{\sA}^{\sI}=\varepsilon_{ab}\kappa_\sA^{a}\eta^{b\sI}$. The amplitude $A_4(g^{a\dot{a}}_1,g^{b\dot{b}}_2,\psi^{\dot c }_3,\psi^{\dot d}_4)$ is now the following coefficient of the Grassmann variables $\eta$ and $\tilde \eta$,
\begin{align}
    A_4(g^{a\dot{a}}_1,g^{b\dot{b}}_2,\psi^{\dot c }_3,\psi^{\dot d}_4)&=\frac{\partial}{\partial \eta^a_1}\frac{\partial}{\partial \tilde{\eta}^{\dot a}_1}\frac{\partial}{\partial \eta^b_2}\frac{\partial}{\partial \tilde{\eta}^{\dot b}_2}\frac{\partial}{\partial \tilde{\eta}^{\dot c}_3}\frac{\partial}{\partial \tilde{\eta}^{\dot d}_4}\frac{\partial}{\partial \eta^{e}_4}\frac{\partial}{\partial \eta^{g}_4}\;\varepsilon^{eg}\;
    \frac{\delta^4(\sum q)\delta^4(\sum \tilde{q})}{s_{12}s_{14}}\bigg|_{\eta_i=\tilde\eta_i=0}\nonumber\\
    &=\frac{\la1_a2_bk_3\ra\, [1_{\dot a}2_{\dot b}3_{\dot c}4_{\dot d}]}{s_{12}s_{14}}
\end{align}
This agrees with our result \eqref{4ptgluini} after contraction into the external polarization states.

\section{Dimensional reduction}\label{sec:dim-red}
As an additional check on our formulae, we examine their behaviour under dimensional reduction. When we reduce  D5 and M5 amplitudes to 5d,  both expressions are expected to agree there.  Similarly when we reduce our (controversial) $(0,2)$ formula with the Parke-Taylor, the formulae agree with those of the reduced $(1,1)$ super Yang-Mills formula.  Similarly the reduced $(3,1)$ and $(0,4)$ formulae also agree with the reduced $(2,2)$ supergravity formulae.  When $(1,1)$ super Yang-Mills and $(2,2)$ super gravity  theories are reduced to 5d, we see that our supersymmetry representation naturally extends the R-symmetry to Sp$(2)$ and Sp$(4)$ respectively.

We further reduce the  super Yang-Mills and supergravity  to the 4d massless case, where we recover the 4d version of the polarized scattering equations reviewed in \cref{sec:intro-4d}. The main new feature of the 4d massless case is the emergence of (MHV) sectors for the amplitude, whereas neither the 4d massive nor the higher dimensional amplitudes split into sectors. We will see below that the dimensional reduction  gives rise to a unified formula for all sectors, with the separation into different MHV sectors appearing naturally from different classes of solutions to the 6d polarized scattering equations. The reduction to massive 4d  kinematics, and in particular the  Coulomb branch in super Yang-Mills, has already been discussed in previous work \cite{Geyer:2018xgb}, and we refer the interested reader to that paper, as well as \cite{Cachazo:2018hqa} for related topics in the little-group preserving supersymmetry representation.

\subsection{Dimensional reduction to 5d} 
On reduction to 5d, the sixth direction is represented as a skew spinor that we will denote $\Omega_{AB}$ so that a five vector $k_{AB}$ must satisfy $k_{AB}\Omega^{AB}=0$. In 5d spinor indices can now be raised and lowered with $\Omega_{AB}$ and its inverse.  This
reduces the spin group from SL$(4,\C)$  to Spin$(5)=\rm{Sp}(2)$. 

\newcommand{\CI}{\mathcal{I}}
\newcommand{\CJ}{\mathcal{J}}

\newcommand{\CN}{\mathcal{N}}

Starting with a theory in 6d with $(N,\tilde N)$-supersymmetry, we can lower the supersymmetry generator spinor index $\tilde Q_{A\dot I}=\Omega_{BA} Q^B_{\dot I}$ so that now in 5d we can write $Q_{A\CI}=(Q_{AI},\tilde Q_{A\dot I})$ where $\CI=1, \ldots, 2\CN$ where $\CN=N+\tilde N$. We can define the skew form $\Omega_{\CI,\CJ}=\Omega_{IJ}\oplus \Omega_{\dot I\dot J}$ and with this the R-symmetry has the possibility of extending from Sp$(N)\times \rm{Sp}(\tilde{N})$ to Sp$(N+\tilde N)$. Thus we see that reduction of theories with $(1,1)$ and $(0,2)$-supersymmetry in 6d can naturally reduce to theories with identical supersymmetry in 5d if there is nothing in the spectrum to break the increased R-symmetry. This is typically the case in the massless sectors of the reduced theories (although differences will generally be seen in Kaluza-Klein massive modes). 

\paragraph{5d spinor helicity and scattering equations  from 6d.} In 5d, the massless little group will be Spin$(3,\C)=\rm{Sl}(2,\C)$ rather then Spin$(4)=\mathrm{Sl}(2,\C)\times \mathrm{Sl}(2,\C)$.  Given a 5d massless momentum $k_{\sA\sB}$, we can  introduce the spinor helicity frame $\kappa_\sA^a$ satisfying 
\begin{equation}
 k_{\sA\sB}=\kappa_\sA^a\kappa_\sB^b\varepsilon_{ab}\,, \qquad k\cdot \Omega=0  \label{eq:kappa_5d}
\end{equation}
But we can now raise the indices with $\Omega^{AB}$ to obtain $\kappa^{\sA a}$ providing also the $\kappa^{\sA\dot a}$ thus  identifying the dotted little group in $6d$ with the undotted one. Now $\kappa_{\sA}^a$  
 transforms in the fundamental representation of  Spin$(5,\mathbb{C})\cong\mathrm{Sp}(4,\mathbb{C})$, and $a$ labels 
the little group for massless particles, Spin$(3,\mathbb{C})\cong\mathrm{SL}(2,\mathbb{C})$. 




Spin one polarization data are 2-forms given in 5d by symmetric spinors $F_{AB}=F_{(AB)}$ satisfying $k^{AB}F_{BC}=0$.  Thus they arise from  little group spinors $\epsilon_{ab}=\epsilon_{(ab)}$ with $F_{AB}=\kappa_A^a\kappa_B^b\epsilon_{ab}$ and we can take $\epsilon_{ab}=\epsilon_a\epsilon_b$.  When reduced from 6d, we therefore identify both the 6d $\epsilon_{\dot a}$ and $\epsilon_a$ with the $5d$ $\epsilon_a$s.  This therefore becomes the same polarization data as one obtained from the symmetry reduction of the 6d Gerbe field.

The chiral polarized scattering equations reduce straightforwardly, with the $u$'s, $\epsilon$'s and $v$'s now all transforming in the 5d little group. However, the same is true for the anti-fundamental scattering equations, where the $\tilde u$'s etc now transform under the \emph{same} SL$(2,\mathbb{C})$, i.e. $\tilde u_i^{\dot a}\rightarrow\tilde u_i^a$. Moreover, we have seen that we should take  $\tilde \epsilon_i=\epsilon_i$ after reduction. Thus the fundamental and anti-fundamental scattering equations are identified
\begin{equation}
 \tilde{\mathcal{E}}_i^{5d \,\sA}=\Omega^{\sA\sB}\mathcal{E}_{i\,\sB}\bigg|_{\substack{u\rightarrow\tilde u\\ v\rightarrow \tilde v}}\,.
\end{equation}
We therefore have the same  equations for both $(u_i,v_i)$ and $(\tilde u_i,\tilde v_i)$.  By the uniqueness of the solution ensured by \cref{unique}, we have
\begin{equation}\label{eq:u=tilde u}
 \tilde u_i^a= u_i^a\,,\qquad \tilde v_i^a=v_i^a\,.
\end{equation}

We can implement the reduction from 6d amplitude formulae to 5d via a projection operator
\begin{equation}
 \Pi_{6\rightarrow5}=\int\prod_{i=1}^n \rd k_{i}\cdot \Omega\prod_{j=1}^{n-1}\delta\left(k_{j}\cdot \Omega\right)\,.
\end{equation}
The second product goes only up to $n-1$ so that the $n$th integral can absorb the sixth component of the momentum-conserving delta-function.  The resulting formula then has the correct count of variables vs symmetries and delta-functions, and leading to the required  $\delta^5$ for momentum conservation. 
We therefore define
\begin{equation}
 d\mu_n^{\mathrm{pol},5d}=\Pi_{6\rightarrow5},d\mu_n^{\mathrm{pol}}\,.
\end{equation}
The polarized measure $d\mu_n^{\mathrm{pol},5d}$ in 5d thus has none of the subtleties of the 6d case, and all constraints are manifestly imposed via delta-functions.

\paragraph{Dimensional reduction of the integrands and formulae.}  Upon reduction, the spin-one matrix $H^{6d}_{ij}\rightarrow H_{ij}^{5d}$ becomes symmetric as $\epsilon_i=\tilde\epsilon_i$ gives 
$$
H_{ij}^{5d}=\frac{\epsilon_i^A\epsilon_j^B\Omega_{AB}}{ \sigma_{ij}}\, , \qquad i\neq j.$$
This is sufficient to give Yang-Mills with integrand $\det'H^{5d} \,\mathrm{PT}$  and gravity with $(\det'H^{5d})^2$.

The dimensional reduction of the supersymmetry factors proceeds along the same lines, driven again by the equality  \eqref{eq:u=tilde u}. We find
\begin{subequations}
\begin{align}
 F_N\Big|_{5d}&= \frac1{2}\sum_{i,j}\frac{\la u_i u_j\ra }{\sigma_{ij}}q_{i\sI}q_{j\sJ}\Omega^{\sI\sJ}
 -\frac{1}{2}\sum_{i=1}^n\la \xi_iv_i\ra q_{i\sI}q_{i\sJ}\Omega^{\sI\sJ}\,,\\
 \tilde F_{\tilde N}\Big|_{5d} &= \frac1{2}\sum_{i,j}\frac{\la u_i u_j\ra }{\sigma_{ij}}\tilde q_{i\dot \sI}\tilde q_{j\dot \sJ}\tilde \Omega^{\dot\sI\dot\sJ}
 -\frac{1}{2}\sum_{i=1}^n\la \xi_iv_i\ra  q_{i\dot \sI}\tilde q_{j\dot \sJ}\tilde \Omega^{\dot\sI\dot\sJ}\,.
\end{align}
\end{subequations}
For $\mathcal{N}=(1,1)$ supersymmetry, we can thus naturally combine the fermionic variables $q_{i\sI} = (q_{il},\tilde q_{i\dot l})$ into $\mathcal{N}=2$ supermomenta, with the  symplectic metric  $\Omega=\mathrm{diag}(\Omega,\tilde\Omega)$ composed of the $N=1$, $\tilde N=1$ metrics in 6d. This manifests that
\begin{equation}
 F_1+\tilde F_1\Big|_{5d}=F_2^{5d}\,,\qquad  F_2\Big|_{5d}=F_2^{5d}\,.
\end{equation}
Thus for maximally supersymmetric Yang-Mills we obtain the integrand $\e^{F_2^{5d}}\det' H^{5d} \,\mathrm{PT}$. 
Similarly, the 6d $(2,2)$-supersymmetry factor reduces to $F_4^{5d}$ giving the maximal supergravity integrand $\e^{F_4^{5d}} \det' (H^{5d})^2$.

Finally, for the brane integrands, we first note that from $u_i=\tilde u_i$, that
 $U^{(a,b)}$ reduces to  $U^{(m)}$ 
  \begin{equation}
     U^{(m)}_{ij}:=\frac{\la u_i u_j\ra^m}{\sigma_{ij}}\,,
\end{equation}
with $a+b=m$. 
Further, from \eqref{eq:detH=pfA}, we find  
\begin{equation}
 \det{}'H^{5d}=\frac{\pf{}'A}{\pf U^{(2)}}\,,
\end{equation}
On the other hand, the M5 integrand reduces to the same expression due to the equality between $u_i$ and $\tilde u_i$, 
\begin{equation}
 \frac{\pf{}' A}{\pf U^{(2,0)}_{5d}}= \frac{\pf{}'A}{\pf U^{(2)}}=\det{}'H\,,\label{Pf-det}
\end{equation}
This in particular gives a nontrivial meaning to the right hand side for odd particle number in 5d, and the D5 and M5 integrands become the same.
With both the integrands and the supersymmetry factors  agreeing among M5 and D5, we  conclude that both theories give the same amplitudes when reduced to 5d.

The above  reductions imply that the integrands of the $(0,2)$-PT theory reduced to 5d now makes sense for both even and odd numbers of particles, and agrees with the reduction of maximal super Yang-Mills.  Similarly the 5d reductions of $(1,3)$ and $(0,4)$ theories make sense for both odd and even numbers of particles and agree with the 5d maximal supergravity formulae.

\subsection{Dimensional reduction to 4d}\label{sec:4d-massless}
The 6d formalism  similarly  allows for a natural embedding of both 4d massive and massless kinematics. On reduction, the 6d spin spaces each reduce to the sum of the dotted and undotted spin spaces so $\epsilon_A=(\epsilon_\alpha,\epsilon_{\dot \alpha})$. The massive little group in 4d is Spin$(3,\C)=\mathrm{Sl}(2,\C)$ and  we can choose the 6d little group frames so that both SL$(2,\mathbb{C})$-factors align with the massive 4d little group,
\begin{equation}\label{eq:red_massive_4d}
 \kappa_{A}^a=\begin{pmatrix}\kappa_{\alpha}^0&\tilde\kappa^{\dot \alpha\, 0}\\ \kappa^1_{\alpha} & \tilde\kappa^{\dot\alpha\, 1}\end{pmatrix}\,,\qquad \kappa_{\dot a}^A=\begin{pmatrix}\kappa_{0}^\alpha &\kappa_{1}^\alpha\\ \tilde\kappa_{\dot\alpha\,0} & \tilde\kappa_{\dot\alpha\,1}\end{pmatrix}\,.
\end{equation}
Here, $a=0,1$  denote the 4d massive little group indices.
Massive momenta, as well as the mass $m$, are constructed via
\begin{equation}
 k_{\alpha\dot\alpha}=\kappa_{\alpha a}\tilde\kappa_{\dot\alpha b}\epsilon^{ab}\,,\qquad \kappa_{\alpha a}\kappa_{\beta b}\epsilon^{ab} = M\varepsilon_{\alpha\beta}\,,\qquad \tilde\kappa_{\dot\alpha a}\tilde\kappa_{\dot\beta b}\varepsilon^{ab} = \tilde M\epsilon_{\dot\alpha\dot\beta}\,.
\end{equation}
with $M=\tilde M$ and $M^2=m^2$. For more details of the reduction to the Coulomb branch, see  \cite{Geyer:2018xgb,Cachazo:2018hqa}.

From hereon we focus on the reduction to massless kinematics. When  $M=\tilde M=0$, the two spinors  become proportional, and following already from the reduction to 5d, we can  identify the dotted and undotted little groups. We choose  a little-group frame with
$\kappa_{\alpha}^0 = \tilde\kappa_{\dot\alpha}^1=0$ so
\begin{equation}\label{eq:kappa_4d_m=0}
\kappa_\sA^a =  \begin{blockarray}{c@{}cc@{\hspace{0pt}}cl}
        & \mLabel{\alpha} &   \mLabel{\dot\alpha} & & \\
        \begin{block}{(c@{\hspace{0pt}}cc@{\hspace{0pt}}c)l}
        & 0 & \tilde\kappa^{\dot\alpha} &&\mLabel{a=0}  \\
        & \kappa_{\alpha}& 0 &&\mLabel{a=1} \\
    \end{block}
  \end{blockarray}\,,\qquad\qquad
  \kappa_{\dot a}^\sA=\begin{blockarray}{c@{}cc@{\hspace{-4pt}}cl}
        & \mLabel{\dot a =\dot 0} &   \mLabel{\dot a=\dot 1} & & \\
        \begin{block}{(c@{\hspace{0pt}}cc@{\hspace{0pt}}c)l}
        &  \kappa^{\alpha} &0 &&\mLabel{\alpha}  \\
        & 0& -\tilde\kappa_{\dot \alpha} &&\mLabel{\dot\alpha} \\
    \end{block}
  \end{blockarray}\,.
\end{equation}
With this, the polarization data and 2-forms reduce as
\begin{equation}
\epsilon_A=(\epsilon_1\kappa_\alpha, \tilde \epsilon_0 \tilde \kappa_{\dot \alpha})\, , \qquad \epsilon_A\epsilon^B \rightarrow \epsilon_1^2\kappa_\alpha\kappa_\beta\varepsilon_{\dot\alpha\dot\beta}+\epsilon_0^2 \tilde \kappa_{\dot \alpha}\kappa_{\dot\beta}\varepsilon_{\alpha\beta}\,. \label{eq;pol-red4}
\end{equation}
We see that the two components of $\epsilon_{ia}$ are naturally distinguished by helicity. 

\paragraph{Scattering equations.} When reduced to the four-dimensional massless case as in \eqref{eq:kappa_4d_m=0}, the polarized scattering equations  become
\begin{align}\label{eq:full_SE_4d_m=0}
 \mathcal{E}_{i\alpha} &= \sum_j\frac{\la u_i u_j\ra}{\sigma_{ij}}\epsilon_{j\,1} \kappa_{j\alpha} - v_{i\,1}\kappa_{i\alpha} \,,&
 \tilde{\mathcal{E}}_i{}^{\dot\alpha} &= \sum_j\frac{\la u_i u_j\ra}{\sigma_{ij}}\epsilon_{j\,0} \tilde\kappa_{j}{}^{\dot\alpha} - v_{j\,0} \tilde\kappa_{i}{}^{\dot\alpha}\,.
\end{align}
At this stage, the scattering equations have a unified form valid for all MHV sectors simultaneously. They can be reduced to the 4d polarized scattering equations \eqref{polscatt-4d} refined by MHV sector by  dividing the external particles into two sets  with $k$ and $n-k$ particles respectively, corresponding to positive and negative helicities.
This determines  the $\epsilon_{ia}$ up to scale from \eqref{eq;pol-red4}. With this we can embed the massless 4d polarized  scattering equations \eqref{polscatt-4d} into \eqref{eq:full_SE_4d_m=0} with the following consequent choices for the $u_{ia}$ and $v_{ia}$ 
\begin{subequations}\label{eq:sols_red_4d}
\begin{align}
 &\epsilon_{ia} = (0,\epsilon_i) &&  
  \xi_{ia}=v_{ia}=-\frac{1}{\epsilon_i}(1,0) &&u_{ia} = (u_i,0) && i\in -\,, \\
 &  \epsilon_{pa} = (\tilde\epsilon_p,0)  &&\xi_{pa}=v_{pa}=\frac{1}{\tilde \epsilon_p}(0,1) && u_{pa} = (0,u_p)  && p\in +\,.\label{eq:4d-red}
\end{align}
\end{subequations}
This assignment automatically solves the scattering equations $\cE_{i\alpha}=0$ for $i\in -$ and $\tilde\cE_{p\dot\alpha}=0$ for $p\in +$.
Thus  the remaining polarized scattering equations reduce to the refined scattering equations for the N$^{\mathrm{k-2}}$MHV sector
\begin{subequations}
\begin{align}
 \mathcal{E}_{p\alpha} &= \sum_{i\in -}\frac{u_p u_i}{\sigma_{pi}}\, \epsilon_{i\alpha} - \frac{1}{\tilde\epsilon_p}\kappa_{p\alpha}=u_p\lambda_\alpha(\sigma_p)-\frac{1}{\tilde\epsilon_p} \kappa_{p\alpha} \,,\\
 \tilde{\mathcal{E}}_i{}^{\dot\alpha} &= \sum_{p\in +}\frac{u_iu_p}{\sigma_{ip}} \,\tilde\epsilon_{p}{}^{\dot\alpha} - \frac1{\epsilon_i} \tilde\kappa_{i}{}^{\dot\alpha} = u_i\,\tilde\lambda^{\dot\alpha}(\sigma_i) -\frac1{\epsilon_i}  \tilde\kappa_i{}^{\dot\alpha}\,,
\end{align}
\end{subequations}
where we have written \begin{equation}
\epsilon_{i\alpha}=\epsilon_{i1}\kappa_\alpha,\quad \mbox{ for }i\in-, \quad\mbox{ and }\quad\tilde{\epsilon}_{p\dot\alpha}=\tilde\epsilon_{p0}\tilde\kappa_{p\dot\alpha}\, .
\end{equation} 
Thus the 4d refined scattering equations are clearly a subset of the solutions to the dimensionally reduced polarized scattering equations \eqref{eq:full_SE_4d_m=0} for the given choice of polarization data. Conversely,  these are indeed all solutions, since the refined scattering equations have $A(n-3,k-2)$ solutions, where $A$ denotes the Eulerian number. Summing over all sectors, the ansatz \eqref{eq:sols_red_4d} these give the full $(n-3)!$ solutions of the polarized scattering equations.
We will also see below that any division not lining up with the particle helicities has vanishing contribution. 

\paragraph{The reduced determinants.} To study the  reduction of $\det' H$ in terms of the 4d data above, note that $\epsilon_i \sim (0,1)$ for negative helicity particles, and $\epsilon_p \sim (1,0)$ for positive helicities. Thus the entries in the $H$ become $H^k$ with
\begin{equation}
 H^k_{ij}: = \frac{\la \epsilon_i\epsilon_j\ra}{\sigma_{ij}}\,,\qquad H^k_{pq}: = \frac{\left[\tilde\epsilon_p\tilde\epsilon_q\right]}{\sigma_{pq}}\,,\qquad H_{ip}=H_{pi}=0\,.
\end{equation}
for $i,j\in -$ and $p,q\in +$. This agrees with the Hodges matrix \eqref{eq:Hodges} as reviewed in \S\ref{sec:intro-4d}.
In particular, the relations among its entries become the row- and column relations described in \cite{Geyer:2014fka}:
\begin{equation}\label{eq:rels_H_4d}
 \sum_{j\in -} u_jH_{ij}^k =0\,,\qquad \sum_{q\in +}u_qH^k_{pq}=0\,.
\end{equation}

We can now understand how  the polarized scattering equations restrict to the correct MHV sector for a given configuration of particle helicities. To see this, we need to show that if the split in \eqref{eq:sols_red_4d} into $-$ and $+$ does not line up with the helicities of the respective particles, the contribution to the amplitude vanishes. 
But since the integrand is always formulated for the correct MHV sector due to our discussion above, this is just the familiar result of Ref.~\cite{Zhang:2016rzb} that the reduced determinant vanishes when evaluated on scattering equations refined to a different sector.

\paragraph{Measure.} To obtain the correct  measure on reduction to 4d, we have to include the appropriate delta-function restricting the kinematics to 4d. A convenient choice is
\begin{equation}
 \Pi_{4d}:=\int \prod_{i=1}^n \rd k_{i,12}\rd k_{i,34} \!\!\prod_{\substack{j,l=1\\j\neq 1,\,l\neq n}}^{n}\hspace{-7pt}\delta\left(k_{j,12}\right)\,\delta\left(k_{l,34}\right)\,,
\end{equation}
since it reproduces the reduction to $\kappa$  given in \eqref{eq:kappa_4d_m=0}. Note that although we integrate over all $n$ momenta, only $n-1$ delta-functions are included, the remaining constraints follow from momentum conservation. 

It follows from general considerations that  we should have
\begin{lemma}
\begin{align}\label{eq:red_measure_4d}
  d\mu_{n,k}^{4d}
 \;\prod_{i,p}\epsilon_i\tilde  \epsilon_p
=\det{}'H^k\;\Pi_{4d}\;d\mu_n^{\mathrm{pol}} 
\end{align}
so  that $d\mu_n^{4d}$ gives  $\det{}'H^k$ as Jacobian relative to the $d\mu^{\mathrm{pol}}=d\mu^{\mathrm{CHY}}$ on the solutions \eqref{eq:sols_red_4d} refined to the given MHV sector.  In particular $\det{}'H^k$ vanishes on the other MHV sectors.\end{lemma}

The general considerations arise from comparing the CHY gauge and gravity formulae of \eqref{eq:CHY_ampl} to the corresponding 4d ambitwistor string formulae of \eqref{eq:4d_ampl}. 
The first step to notice is that the 
 the gauge theory formulae of \eqref{eq:CHY_ampl} and   \eqref{eq:4d_ampl} are identified if we have 
 \begin{equation}
  d\mu_{n,k}^{4d}=\Pf'(M)d\mu_n^{\mathrm{CHY}}. 
 \end{equation}
Then the fact that the gauge and gravity formulae for CHY are related by exchanging the Parke-Taylor factor for $\Pf'(M)$, whereas for the 4d ambitwistor-string one exchanges the Parke-Taylor for $\det{}'H^k$   suggests that in the $k$th MHV sector
\begin{equation}
\Pf'(M)=\det{}'H^k.
\end{equation}
 This was  shown explicitly in \cite{Zhang:2016rzb,Roehrig:2017wvh}.  
Finally recall that the measure $d\mu^{\mathrm{pol}}_n$ was shown to be equivalent to the CHY measure in \S\ref{sec:reltoCHYmeasure} and putting this together suggests  the lemma. We now prove this explicitly, albeit via 6d.
 

\smallskip

\proof
We have seen above that the 6d polarized scattering equations reduce to the 4d version and so have the correct support restricted to the  given MHV sector.  To calculate the Jacobian, consider a fixed MHV sector, corresponding to the solutions \eqref{eq:sols_red_4d} to the polarized scattering equations. We 
first fix part of the SL$(2,\mathbb{C})_u$ invariance by setting $u_{1\,1}=u_{n\,0}=0$ for $1\in -$ and $n\in +$, giving a contribution to the Jacobian of $u_1u_n$. Similarly, we use the corresponding scattering equations $\mathcal{E}_{1\,1}$ and $\tilde{\mathcal{E}}_{n}{}^{\dot 0}$ to solve for $k_{n,12}$ and $k_{n,34}$, introducing  a Jacobian of $\epsilon_{1\,0}\tilde\epsilon_n^{\dot 1}$.
We used \eqref{eq:4d-red} to solve  the polarized scattering equations that dont survive  in the 4d measure or framework
\begin{subequations}\label{eq:SE_to_solve}
 \begin{align}
  &\mathcal{E}_{i\alpha} := 
 \sum_{j\in -}\frac{\la u_iu_j\ra}{\sigma_{ij}} \epsilon_{j\alpha}  
 - v_{i\,1}\kappa_{i\alpha}=0\,,\\
  &\tilde{\mathcal{E}}_p^{\dot\alpha}\,:=
    \sum_{q\in +}\frac{\la u_pu_q\ra}{\sigma_{pq}} \tilde \epsilon_{q\dot\alpha} \;
  - v_{p\,0}\tilde\kappa_p^{\dot\alpha}=0\,,
 \end{align}
\end{subequations}
for  the variables $u_{i\,1}$, $u_{p\,0}$, $v_{i\,1}$ and $v_{p\,0}$ (using the normalization conditions to fix the other components of $v$).  This gives a further Jacobian that we denote  $J_{\mathrm{pol}}$ so that we have
\begin{align}
\Pi_{4d}\;d\mu_n^{\mathrm{pol}} =
\int d\mu_{n,k}^{4d}\;J_{\mathrm{pol}}\,u_1 u_n\,\epsilon_{1\,0}\,\tilde\epsilon_n^{\dot 1}\prod_{i=1}^n u_i\,,
\end{align}
where the extra factor of $\prod_{i=1}^n u_i$ cancels its inverse explicitly in the definition of the measure $d\mu_n^{4d}$.
The Jacobian matrix whose determinant $J_{\mathrm{pol}}$ arises from solving the polarized scattering equations \eqref{eq:SE_to_solve} has a block-diagonal form due to 
\begin{equation}
 \frac{ \partial \tilde{\mathcal{E}}_p^{\dot\alpha}}{\partial v_{i\,1}}  =  \frac{ \partial \tilde{\mathcal{E}}_p^{\dot\alpha}}{\partial u_{i\,1}} = 0\,,\qquad \frac{ \partial \mathcal{E}_{i\alpha}}{\partial v_{p\,0}}  = \frac{ \partial \mathcal{E}_{i\alpha}}{\partial u_{p\,0}}  = 0\,,
\end{equation}
on the solutions \eqref{eq:4d-red},
so we have $J_{\mathrm{pol}}=J^-J^+$, with $J^-$ and $J^+$ the determinants of the respective block matrices. 
On the solutions \eqref{eq:sols_red_4d}, the entries of the matrix with determinant $J^-$ are  given by
\begin{equation}\label{eq:dE/dv-}
 \frac{ \partial \mathcal{E}_{i\alpha}}{\partial v_{j\,1}} = -\delta_{ij}\kappa_{i\alpha}\,,\qquad  \frac{ \partial \mathcal{E}_{i\alpha}}{\partial u_{j\,1}} = \begin{cases}-u_i\frac{\epsilon_{j\alpha}}{\sigma_{ij}}\,,\qquad i\neq j \\ 
 \sum_{k\in -, k\neq i} u_k   \frac{\epsilon_{k\alpha}}{\sigma_{ik}} \,, \quad i=j.\end{cases}
\end{equation}
The Jacobian $J^-$ is the determinant of this $(2k-1)\times (2k-1)$ matrix (as we have already dealt with $u_{11}$).  To simplify this,  introduce the  index notation $\mathcal{E}^-_{2i-1} \equiv \mathcal{E}_{i,\alpha=0}$ and $\mathcal{E}_{2i}^-\equiv \mathcal{E}_{i,\alpha=1}$  so that the Jacobian $J^-$ is given by
\begin{equation}
 J^- = \varepsilon^{a_1\dots a_{2k-1}}\frac{\partial \mathcal{E}^-_{a_1}}{\partial v_{1\,1}}\dots \frac{\partial \mathcal{E}^-_{a_k}}{\partial v_{k\,1}}\frac{\partial \mathcal{E}^-_{a_{k+1}}}{\partial u_{2\,1}}\dots \frac{\partial \mathcal{E}^-_{a_{2k-1}}}{\partial u_{k\,1}}
\end{equation}
The first equation in \eqref{eq:dE/dv-} gives
$
 \frac{\partial \mathcal{E}^-_{a}}{\partial v_{i\,1}} = \delta^{2i-1}_{a}\kappa_{i\,0} +\delta^{2i}_{a}\kappa_{i\,1}
$
so monomials in the expansion of the determinant with $\p\cE_{2i-1}/\p v_{i1}$ must multiply some $\p\cE_{2i}/\p u_{i 1}$ and similarly $\p\cE_{2i}/\p v_{i1}$ must multiply some $\p\cE_{2i-1}/\p u_{i 1}$ with the opposite sign leading to a contraction on the spinor index.   Thus the sum collapses to one over half the indices, and after some re-ordering  of the terms and relabelling of the indices, we find
\begin{equation}
 J^-=\kappa_{10}\;\varepsilon^{i_2 \dots i_{k-1}}\left(\kappa_{i_2}^{\alpha_2} \frac{\partial\mathcal{E}_{i_2\alpha_2}}{\partial u_{2\,1}} \right)\dots \left(\kappa_{i_k}^{\alpha_k} \frac{\partial\mathcal{E}_{i_k\alpha_k}}{\partial u_{k\,1}} \right) = \kappa_{10}\prod_{\substack{i\in -\\i\neq 1}}\frac{u_i}{\epsilon_i}\;\det\,H_-^{k\, [1]}  =u_1\,\epsilon_{10}\prod_{i\in -} \frac{u_i}{\epsilon_i}
 \;\det{}'H^k_-\,.
\end{equation}
In the second equality, we have used \eqref{eq:dE/dv-} to see that contraction into the respective $\kappa_i$ reproduces the entries of $H_-$, and the last equality holds due to the reduction relations \eqref{eq:rels_H_4d} for the reduced determinant.
Similarly,
\begin{equation}
 J^+= \frac{u_n\,\tilde\epsilon_{n}{}^{\dot 1}\prod_{q\in +} u_q}{\prod_{p\in +} \tilde\epsilon_p}\;\det{}'H_+\,.
\end{equation}
The extra factors $\epsilon_{1\,0}\tilde\epsilon_n^{\dot 1}$  thus cancel against the Jacobian from integrating out $k_{1,12}$ and $k_{n,34}$, the factors of $u$ cancel against the measure and partial gauge fixing,
and we indeed are left with \eqref{eq:red_measure_4d}.$\Box$

\smallskip

As a corollary we briefly mention that for momenta in four dimensions, the $(n-3)!$ solutions to the scattering equations can be refined by MHV degree $k$ with Eulerian number\footnote{The Eulerian number $A(p,q)$ is the number of permutations of $1$ to $p$ where $q$ elements are larger than their
preceding element. They are defined recursively by $A(p,q)=(p-q)A(p-1,q-1)+(q+1)A(p-1,q)$.} 
$A( n-3,k-2)$ in the $k$th sector \cite{Cachazo:2012da, Roiban:2004yf}. The above relation between measures gives 
\begin{corol}
The 4d measure $d\mu^{4d}_{n,k}$ is supported  on the  $A( n-3,k-2)$ solutions to the scattering equations in the $k$th sector.
\end{corol} 
\proof This follows from the fact that $\det{}'H^k$ is supported on the $k$th sector. To see this, define 
the matrices $H^\pm$ by
\begin{equation}
H^+_{ij}=\frac{\la \epsilon_i\epsilon_j\ra}{\sigma_{ij}}\, , \qquad H^-_{ij}=\frac{[ \epsilon_i\epsilon_j]}{\sigma_{ij}} \, , \qquad i\neq j\, , \qquad H_{ii}^\pm=e_i\cdot P(\sigma_i)\, .
\end{equation}
On the one hand, minors of these appear as the blocks in $H^k$.  On the other hand, as explained in \cite{Geyer:2016nsh},
these are gauge fixed versions of the $n\times n$ matrices
appearing in  the Cachazo-Skinner gravity twistor-string-like  formulae \cite{Cachazo:2012pz, Cachazo:2012kg}.  In those papers it is shown that at degree $k-1$ in the twistor-string, appropriate to MHV degree $k-2$,  these matrices have ranks $k-1$ and $n-k-1$ respectively. Thus $\det{}'H^k$ will vanish because one or other block will have insufficient rank when restricted to the inappropriate MHV sector.
$\Box$

\paragraph{Supersymmetry.} The reduction of   the supersymmetry generators $Q_{\sA I}$  and $\tilde Q_{\dot I}^\sA$ on the solutions  \eqref{eq:sols_red_4d}  in  4d give
 \begin{align}
  - \text{ helicity}: && Q_{\alpha }^I = \epsilon_\alpha \,\frac{\partial}{\partial q_I}\,,&& Q^{\dot\alpha}_I = \frac1{\epsilon}\tilde\kappa^{\dot\alpha}q_I\,, \qquad\tilde Q_{\alpha }^{\dot I} = \epsilon_\alpha\,\frac{\partial}{\partial \tilde q_{\dot I}}  \,,&& \tilde Q^{\dot\alpha}_{\dot I} = \frac1{\epsilon}\tilde\kappa^{\dot\alpha}\tilde q_{\dot I}\,,
 \end{align}
 for negative helicity particles. 
where we have raised the Sp$(N)$ R-symmetry indices with the symplectic metric $\Omega$, i.e. $Q_{\alpha I}=\Omega_{IJ}Q_{\alpha }^J$ and $\tilde Q_{\alpha \dot I}=\Omega_{\dot I\dot J}Q_{\alpha }^{\dot J}$. Similarly, for positive helicity,
 \begin{align}
  + \text{ helicity}: && Q_{\alpha}^I =\frac1{\tilde\epsilon} \kappa_\alpha q^I \,,&& Q^{\dot\alpha}_I = \tilde\epsilon^{\dot\alpha}\frac{\partial}{\partial q^I}\,, \qquad\tilde Q_{\alpha}^{\dot I} = \frac1{\tilde\epsilon}\kappa_\alpha \tilde q^{\dot I} \,,&& \tilde Q^{\dot\alpha}_{\dot I} = \tilde\epsilon^{\dot\alpha}\,\frac{\partial}{\partial \tilde q^{\dot I}}\,,
 \end{align}
where $q^I=\Omega^{IJ}q_J$ etc. The index placement is chosen to manifest the embedding of Sp$(N)\times\mathrm{Sp}(\tilde{N})$ into the bigger 4d SU$(\mathcal{N})$ R-symmetry group. We can make this explicit by introducing $q_\scI = (q_I,\tilde q_{\dot I})$ and $\tilde q^\scI = (q^I,\tilde q^{\dot I})$, where $\mathcal{I}=1,\dots,\mathcal{N}=N+\tilde N$ is the SU$(\mathcal{N})$ R-symmetry index in 4d. The supersymmetry generators then become
\begin{subequations}\label{eq:susy-gens-4d}
 \begin{align}
 - \text{ helicity}: &&Q_{\alpha }^\scI = \epsilon_\alpha\frac{\partial}{\partial q_{\scI}} \,,&& Q^{\dot\alpha}_{\scI} = \frac1{\epsilon}\tilde\kappa^{\dot\alpha} q_{\scI} \,, \\
   +\text{ helicity}: &&Q_{\alpha}^\scI = \frac1{\tilde\epsilon}\kappa_\alpha \tilde q^{\scI}\,,&& Q^{\dot\alpha}_{\scI} = \tilde\epsilon^{\dot\alpha}\frac{\partial}{\partial \tilde q^{\scI}}\,,
 \end{align}
\end{subequations}
and the supersymmetry multiplet  takes the familiar form \eqref{eq:4d-multiplet},
\begin{subequations}
 \begin{align}
  &\Phi_- = A^{--} + q_{\scI}\psi^{\scI}_- +\hspace{23pt} q_{\scI}q_{\scJ}\phi^{\scI\scJ}+(q^3)^{\scI}\psi_{\scI}^++q^4 A^{++}\,,\\
  & \Phi_+ = A^{++} + \tilde q^{\scI}\psi_{\scI}^+ + \varepsilon_{\scalebox{0.6}{$\mathcal{IJKL}$}}\tilde q^{\scI}\tilde q^{\scJ}\phi^{\scalebox{0.6}{$\mathcal{KL}$}}+(\tilde q^3)_{\scI}\psi^{\scI}_-+\tilde q^4 A^{--}\,.
 \end{align}
\end{subequations}
Here $q_\scI$ and $\tilde q^\scI$ are conjugate supermomenta, related by a fermionic Fourier transform and $\epsilon\leftrightarrow \tilde\epsilon^{-1}$.

When implementing this reduction in the amplitude, only terms containing one particle of each helicity survive in the  exponential supersymmetry factors due to the form of the solutions \eqref{eq:sols_red_4d} to the 4d scattering equations,
\begin{equation}
 F_N +\tilde F_{\tilde N}\Big|_{4d} = \sum_{\substack{i\in -\\ p\in +}}\frac{ u_i u_p}{\sigma_{ip}}q_{i\scI}\,q_p^\scI=:F_{\scN}^k\,.
\end{equation}
In particular, all local terms of the form $\la \xi_i v_i\ra q_i^2$ vanish due to $\xi_i^a =v_i^a$.
As reviewed in \cref{sec:intro-4d}, this is one of the standard supersymmetry representations in 4d, sometimes referred to  as the link representation \cite{He:2012er}.

Combining the above results, we find that the 6d amplitudes for super Yang-Mills and supergravity reduce correctly to the 4d amplitudes \eqref{eq:4d_ampl}. The reduced determinant in the numerator cancels against the Jacobian from the measure, and we have 
\begin{equation}
 \mathcal{A}_n\Big|_{4d} = \Pi_{4d}\;\int d\mu_n^{\mathrm{pol}} \;\det{}'H \,\mathcal{I}_n^{\mathrm{h}}\;e^{F_\sN+\tilde F_\stN}=\sum_k\int d\mu_{n,k}^{4d} \;\,\prod_{i,p} \epsilon_i\tilde \epsilon_p\;\mathcal{I}_n^{\mathrm{h}}\;e^{F_\scN}=\mathcal{A}_n^{4d}\,,
\end{equation}
with $\mathcal{I}_n^{\mathrm{h}} = \mathrm{PT}(\alpha)$ for super-Yang-Mills, $\mathcal{I}_n^{\mathrm{h}} =\det{}'H$ for supergravity, and  $\mathcal{I}_n^{\mathrm{h}} =\det{}'A$ for Born-Infeld.

\section{Super-BCFW in 6d}\label{sec:BCFW}
In this section, we give a proof of the gravity and Yang-Mills formulae  using BCFW recursion  \cite{Britto:2004ap, Britto:2005fq}, c.f. \cref{thm:BCFW}.  This is  a powerful on-shell tool that has been used to prove a variety of explicit amplitude representations. This technique has two main ingredients. The first is to introduce a deformation of the formula for the amplitude depending on a complex parameter $z$, and to use complex analysis to reconstruct the amplitude in terms of its residues at poles in $z$. The second key ingredient in the argument is the factorization property of amplitudes.   We know from the Feynman diagram representation of amplitudes that they are multilinear  in the polarization vectors and rational in the momenta.  The only poles arise from propagators, so that they can only arise  along \emph{factorization channels}, where partial sums of the momenta go on shell. At tree-level, factorization is the statement that the residues at such poles  are tree amplitudes on each side of the propagator. This then allows us to identify the residues in $z$ in terms of lower point amplitudes, setting up the recursion.  In the following we give more details of the generalities of this argument.  In \S\ref{sec:BCFW-shift} we introduce the complex shift adapted to our formulae. In \S\ref{sec:factorization} we prove that our formulae factorize correctly; this includes also our brane formulae giving a key check on these also.
In \S\ref{sec:bdy} we show that there is no pole as the deformation parameter is taken to infinity in our formulae, completing the BCFW recursion proof of our supersymmetric gauge and gravity formulae  \eqref{eq:integrands}.

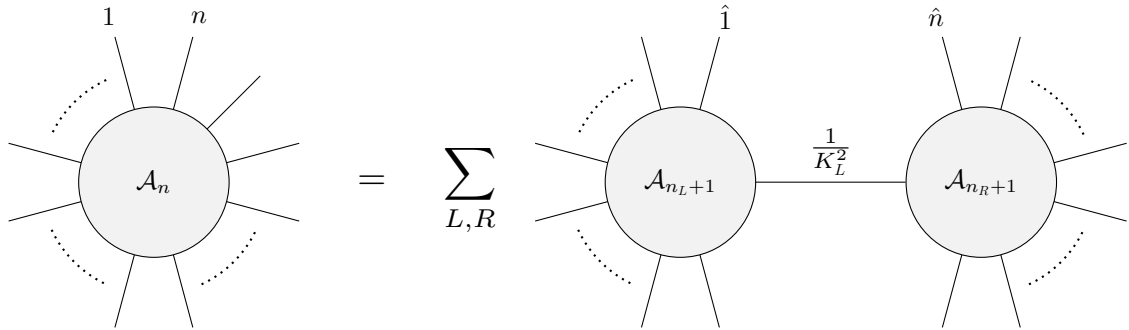
\begin{figure}[ht]
\centering
 \begin{tikzpicture} [scale=1]
  \draw [fill,light-gray] (0,0) circle [radius=1];
  \draw  (0,0) circle [radius=1];
  \node at (0,0) {$\mathcal{A}_n$};
  \draw [dotted,thick,domain=115:155] plot ({1.5*  cos(\x)}, {1.5 * sin(\x)});
  \draw [dotted,thick,domain=205:245] plot ({1.5*  cos(\x)}, {1.5 * sin(\x)});
  \draw [dotted,thick,domain=295:335] plot ({1.5*  cos(\x)}, {1.5 * sin(\x)});
  \draw [domain=0.966:1.932] plot ({\x},{ tan(15)*\x});
  \draw [domain=-0.966:-1.932] plot ({\x},{ tan(15)*\x});
  \draw [domain=0.966:1.932] plot ({\x},{ -tan(15)*\x});
  \draw [domain=-0.966:-1.932] plot ({\x},{ -tan(15)*\x});
  \draw [domain=0.259:0.518] plot ({\x},{ tan(75)*\x});
  \draw [domain=-0.259:-0.518] plot ({\x},{ tan(75)*\x});
  \draw [domain=0.259:0.518] plot ({\x},{ -tan(75)*\x});
  \draw [domain=-0.259:-0.518] plot ({\x},{- tan(75)*\x});
  \draw [domain=0.707:1.414] plot ({\x},{ tan(45)*\x});
 \node at (0.6,2.2) {$n$};
 \node at (-0.6,2.2) {$1$};
 \node at (2.9,0) {\scalebox{1.4}{$=$}};
 \node at (4.2,-0.1) {\scalebox{1.4}{$\displaystyle \sum_{L,R}$}};
 \draw [fill,light-gray] (7,0) circle [radius=1];
 \draw  (7,0) circle [radius=1];
 \node at (7,0) {$\mathcal{A}_{n_\ssL+1}$};
  \draw [domain=0.259:0.518] plot ({\x+7},{ tan(75)*\x});
  \draw [domain=-0.259:-0.518] plot ({\x+7},{ tan(75)*\x});
  \draw [domain=0.259:0.518] plot ({\x+7},{ -tan(75)*\x});
  \draw [domain=-0.259:-0.518] plot ({\x+7},{- tan(75)*\x});
  \draw [domain=-0.966:-1.932] plot ({\x+7},{- tan(15)*\x});
  \draw [domain=-0.966:-1.932] plot ({\x+7},{ tan(15)*\x});
  \draw [dotted,thick,domain=115:155] plot ({7+1.5*  cos(\x)}, {1.5 * sin(\x)});
  \draw [dotted,thick,domain=205:245] plot ({7+1.5*  cos(\x)}, {1.5 * sin(\x)});
  \node at  (7.6,2.2) {$\hat 1$};
 \draw (8,0) -- (10,0);
 \node at (9,0.4) {\scalebox{1.2}{$\frac{1}{K_\ssL^2}$}};
 \draw [fill,light-gray] (11,0) circle [radius=1];
 \draw  (11,0) circle [radius=1];
 \node at (11,0) {$\mathcal{A}_{n_\ssR+1}$};
 \draw [domain=0.259:0.518] plot ({\x+11},{ tan(75)*\x});
  \draw [domain=-0.259:-0.518] plot ({\x+11},{ tan(75)*\x});
  \draw [domain=0.259:0.518] plot ({\x+11},{ -tan(75)*\x});
  \draw [domain=-0.259:-0.518] plot ({\x+11},{- tan(75)*\x});
  \draw [domain=0.966:1.932] plot ({\x+11},{- tan(15)*\x});
  \draw [domain=0.966:1.932] plot ({\x+11},{ tan(15)*\x});
  \draw [dotted,thick,domain=25:65] plot ({11+1.5*  cos(\x)}, {1.5 * sin(\x)});
  \draw [dotted,thick,domain=295:335] plot ({11+1.5*  cos(\x)}, {1.5 * sin(\x)});
  \node at  (10.4,2.2) {$\hat n$};
 \end{tikzpicture}
\caption{A diagrammatic representation of the BCFW relation \eqref{eq:BCFW}.}
\label{fig:BCFW}
\end{figure}

BCFW shifts are generally based on  the following one-parameter deformation of the external momenta,
\begin{equation}\label{eq:BCFW_shift_momenta_vector}
\hat k_{1\mu}=k_{1\mu}+z\,q_{\mu}\,,\qquad
 \hat{k}_{n\mu}=k_{n\mu}-z\,q_{\mu}\,,
\end{equation}
with $q^2=q\cdot k_1=q\cdot k_n=0$. Cauchy's theorem applied to $\cA/z$ then gives an equality between  the original undeformed amplitude at $z=0$ and the sum over all other residues at the possible factorisation channels of the amplitude and at $\infty$. If 
\begin{equation}\label{eq:bdy_BCFW}
  \lim_{z\rightarrow \infty}\mathcal{A}(z)=0\,,
\end{equation}
we say that there are no boundary terms
at $z=\infty$.  The residue theorem then expresses the amplitude at $z=0$ as a sum over products of lower point amplitudes $\mathcal{A}_{n_\sL+1}$ arising at and  $\mathcal{A}_{n_\sR+1}$, with $n_\sL+1$ and $n_\sR+1=(n-n_\sL)+1$ particles respectively, but at shifted values of $z$
\begin{equation}\label{eq:BCFW}
 \mathcal{A}_n = \sum_{L,R}\mathcal{A}_{n_\sL+1}\left(z_\sL\right)\, \frac{1}{k_\sL^2}\,\mathcal{A}_{n_\sR+1}\left(z_\sL\right)\,.
\end{equation}
The sum runs over partitions of the $n$ particles into two sets $L$ and $R$, with one of the deformed momenta  in each subset, $1\in L$ and $n\in R$.
In the propagator, $k_{\sL}=\sum_{i\in L}k_i$ denotes the (undeformed, off-shell)  momentum, whereas the amplitudes are evaluated on the on-shell deformed momentum $\hat k_{\sL}=\sum_{i\in L}k_i+ z_\sL\,q$  with $z_\sL=-k_\sL^2 / 2q\cdot k_\sL$.  See also \cref{fig:BCFW} for a diagrammatic represenation of the recursion.  For particles transforming in non-trivial representations of the little group, the BCFW shift \eqref{eq:BCFW_shift_momenta_vector} has to be extended to the polarization vectors as well \cite{ArkaniHamed:2008yf}, and the boundary terms vanish if the shift vector $q_\mu$ is chosen to align with the polarization vector of one of the shifted particles, $q_\mu = e_{1\,\mu}$. In this case the sum over partitions in the BCFW recursion relation \eqref{eq:BCFW} also includes a sum over a complete set of propagating states, labeled for example by their polarization data for gluons or gravitons.

The recursion \eqref{eq:BCFW}  has been a useful tool to prove novel amplitude representations. In particular, it guarantees that any expression satisfying factorization\footnote{including the correct 3-particle amplitudes} and the boundary condition \eqref{eq:bdy_BCFW}  is a representation of the amplitude.
In \S\ref{sec:BCFW-shift} we adapt the shift to our formulae, in \S\ref{sec:factorization}, we show that our amplitudes factorize correctly, and in \S\ref{sec:bdy} we verify that our boundary terms \eqref{eq:bdy_BCFW} vanish.

\subsection{The BCFW shift for 6d spinors}\label{sec:BCFW-shift}
The higher dimensional BCFW-shifts discussed in the literature (e.g.  \cite{ArkaniHamed:2008yf,Cheung:2009dc,Boels:2012ie}) are ambidextrous, and this makes it difficult to  verify that the boundary terms vanish.
We need to adapt \eqref{eq:BCFW_shift_momenta_vector}
 to the spinor-helicity formalism in 6d.   Such shifts were introduced in  Ref.~\cite{Cheung:2009dc}, but, as discussed in appendix \ref{sec:BCFW_Cliff+Donal}, this does not sit naturally  within the framework of the chiral scattering equations. We therefore introduce a novel  BCFW  shift to start the recursion in  the 6d spinor-helicity formalism. Our shift vector $q_\mu$ does  not coincide with the polarization vector $e_{1\,\mu}$, but is instead related to the chiral polarization data of both shifted particles $1$ and $n$.

\paragraph{Fundamental spinors. } 

 We choose instead the following chiral  BCFW shift, dependent on the (chiral) polarization data of the shifted particles:
\begin{equation}\label{eq:BCFW_shift}
 \hat{\kappa}_{1\sA}^a=\kappa_{1\sA}^a+ z\, \epsilon_1^a\, \epsilon_{n\sA}\,,\qquad  
 \hat{\kappa}_{n\sA}^a=\kappa_{n\sA}^a+ z \,\epsilon_n^a\,\epsilon_{1\sA}\,.
\end{equation}
This shift evidently leaves the polarization spinors $\epsilon_\sA$ invariant, but shifts the spinors $\langle v_1\kappa_{1\,\sA}\rangle$ and  $\langle v_n\kappa_{n\,\sA}\rangle$ featuring in the polarized scattering equations by a term proportional to the polarization spinor of the other particle,
\begin{subequations}
 \begin{align}
   &  \left\langle v_1\hat\kappa_{1\,\sA}\right\rangle= \left\langle v_1\kappa_{1\,\sA}\right\rangle + z \epsilon_{n\sA}\,,      
                   &&\hat{ \epsilon}_{1\sA}=\epsilon_{1\sA}\,, \\
   &  \left\langle v_n\hat\kappa_{n\,\sA}\right\rangle=\left\langle v_n\kappa_{n\,\sA} \right\rangle+ z \epsilon_{1\sA}\,,      
                   &&\hat{ \epsilon}_{n\sA}=\epsilon_{n\sA}\,. 
 \end{align}
\end{subequations}
The invariance of the polarization spinors $\epsilon_{1,n}$ ensures that the shift is well-defined, in the sense that the `shift-spinors' $\delta\kappa_{1\sA}^a\equiv \epsilon_1^a\, \epsilon_{n\sA}$ and $\delta\kappa_{n\sA}^a\equiv \epsilon_n^a\, \epsilon_{1\sA}$ are themselves unaffected. This mirrors the usual BCFW shift, where the  vector $q_\mu$ does \emph{not} transform. It is easily seen that the spinorial deformation \eqref{eq:BCFW_shift} is indeed a valid  vectorial BCFW shift \eqref{eq:BCFW_shift_momenta_vector}. 
However, in contrast to the usual construction the shift vector $q_{\sA\sB}$ is composed of the polarization spinors of \emph{both} particles 1 and $n$,
\begin{equation}\label{eq:BCFW_shift_q}
 q_{\sA\sB}= 2\epsilon_{n\,[\sA}\epsilon_{1\,\sB]}\,.
\end{equation}
It is clear that the shift preserves momentum conservation from the vector representation  \eqref{eq:BCFW_shift_momenta_vector}, and it  preserves Maxwell's equations by construction. Since the shift vector $q_{\sA\sB}$ is constructed from the polarization spinors of both particles, it  is not only orthogonal to the momenta of the shifted particles, $q^2=q\cdot k_1=q\cdot k_n=0$, but also to their polarization vectors $e_{1}$ and $e_n$,  $q\cdot e_1=q\cdot e_n=0$. We will verify in \S\ref{sec:bdy} that this defines a `good' BCFW shift, in the sense that the boundary terms vanish for Yang-Mills theory and gravity.
  We  discuss the comparison with shifts of other authors in  \cref{sec:BCFW_Cliff+Donal}.

\paragraph{Anti-fundamental spinors. } We will see that the chiral BCFW shift \eqref{eq:BCFW_shift} ties in well with the polarized scattering equations. However, for ambidextrous theories such as super Yang-Mills or supergravity however, the shift for spinors in the anti-fundamental representation plays an equally important role. The anti-fundamental shift
\begin{equation}\label{eq:BCFW_shift_momenta_antichiral}
\hat k_{1}^{\sA\sB}=k_{1}^{\sA\sB}+z\,q^{\sA\sB}\,,\qquad
 \hat{k}_{n}^{\sA\sB}=k_{n}^{\sA\sB}-z\,q^{\sA\sB}\,,
\end{equation}
is of course related to the chiral one via  $q^{\sA\sB}= \varepsilon^{\sA\sB\sC\sD} q_{\sC\sD}$, but this does not fully determine the shift  of the anti-chiral spinors $\hat\kappa^\sA_{\dot a}$. We will use this freedom to choose a BCFW shift where both deformations $\delta\kappa^\sA_{1\dot a}$ and $\delta\kappa^\sA_{n\dot a}$ are proportional to the \emph{same} spinor $\tilde\epsilon^A$, \footnote{The choice of $\tilde\epsilon^\sA$ in the anti-fundamental shift will turn out to be crucial in proving that the boundary terms vanish. However, it is also the key distinction from previously defined shifts like the covariant shift of \cite{Cheung:2009dc}. We discuss this in more detail in \cref{sec:BCFW_Cliff+Donal}.}
\begin{subequations}\label{eq:BCFW_antichiral}
\begin{align}
 \hat\kappa_{1}{}^\sA_{\dot a} & = \kappa_{1}{}^\sA_{\dot a}-z\,\tilde \epsilon ^\sA \left(\epsilon_{n\,\sB}\kappa_1{}^\sB_{\dot a}\right)\,,\\
 \hat\kappa_{n}{}^\sA_{\dot a} & = \kappa_{n}{}^\sA_{\dot a}-z\,\tilde \epsilon ^\sA \left(\epsilon_{1\,\sB}\kappa_n{}^\sB_{\dot a}\right)\,.
\end{align}
\end{subequations}
The spinor $\tilde\epsilon^A$ is constructed such that it is a valid choice for $\tilde\epsilon^\sA_1= \tilde\epsilon^\sA$ and $\tilde\epsilon^\sA_n= \tilde\epsilon^\sA$,
\begin{equation}\label{eq:def_tilde_epsilon_anti_v1}
 \tilde\epsilon^\sA=\epsilon_{1\,a} \kappa_n{}^\sA_{\dot a}\,\left(\kappa_n{}^\sB_{\dot a}\,\kappa_1{}_\sB^a\right)^{-1}+
 \epsilon_{n\,a} \kappa_1{}^\sA_{\dot a}\,\left(\kappa_1{}^\sB_{\dot a}\,\kappa_n{}_\sB^a\right)^{-1}\,.
\end{equation}
The first term corresponds to the canonical choice for $\tilde\epsilon^\sA_1$, constructed in complete analogy to \eqref{eq:def_tilde_epsilon}, where we have chosen the reference spinor $\kappa_{*}=\kappa_{n}$. The second term is similarly the canonical choice for $\tilde\epsilon^\sA_n$ with reference spinor $\kappa_{*}=\kappa_{1}$. Due to this choice of reference spinor, the second term is proportional to $\kappa_{1\,\dot a}^\sA$, and is thus pure gauge for particle 1. An analogous argument shows that the first term is pure gauge for particle $n$. Thus we can choose $\tilde\epsilon^\sA_1= \tilde\epsilon^\sA$ and $\tilde\epsilon^\sA_n= \tilde\epsilon^\sA$, and we have the useful relations
\begin{equation}\label{eq:useful_rel_tilde_epsilon}
 \tilde\epsilon^\sA\,\kappa_{1\sA}^a=\epsilon_1^a\,,\qquad \tilde\epsilon^\sA\,\kappa_{n\sA}^a=\epsilon_n^a\,.
\end{equation}
The anti-fundamental BCFW deformation then leads to the standard shift \eqref{eq:BCFW_shift_momenta_antichiral} for the momenta, but where the shift vector $q$ is again determined by the chiral polarization spinors of both shifted particles,
\begin{equation}
 q^{\sA\sB}=2\tilde\epsilon^{[\sA}k_1^{\sB]\sC}\epsilon_{n\,\sC}
                       =-2\tilde\epsilon^{[\sA}k_n^{\sB]\sC}\epsilon_{1\,\sC}\,.
\end{equation}
The latter equality follows from the definition of $\tilde \epsilon^{\sA}$ and the relations \eqref{eq:useful_rel_tilde_epsilon}.\footnote{using $\varepsilon^{\sC\sA_1\sA_2\sA_3} \varepsilon_{\sC \sB_1\sB_2\sB_3} =3!\,\delta^{\sA_1}_{[\sB_1}\delta^{\sA_2}_{\sB_2}\delta^{\sA_3}_{\sB_3]}$.}  Using the same identities, it is also readily verified that $q^{\sA\sB}$ indeed satisfies $q^{\sA\sB}\varepsilon_{\sA\sB\sC\sD}=q_{\sC\sD}$ as claimed above.\\

While not manifest in \eqref{eq:BCFW_antichiral}, the `shift-spinors' (defined by $\hat\kappa_{1,n} {}^\sA_{\dot a}=\kappa_{1,n} {}^\sA_{\dot a}+\delta\kappa_{1,n} {}^\sA_{\dot a}$)
\begin{equation}
 \delta\kappa_{1} {}^\sA_{\dot a}=-\tilde \epsilon ^\sA \left(\epsilon_{n\,\sB}\kappa_1{}^\sB_{\dot a}\right)\qquad \text{and}\qquad  \delta\kappa_{n} {}^\sA_{\dot a}=-\tilde \epsilon ^\sA \left(\epsilon_{1\,\sB}\kappa_n{}^\sB_{\dot a}\right)\,,
\end{equation}
are themselves invariant under the BCFW deformation. To see this, let us focus on $\delta\kappa_{1} {}^\sA_{\dot a}$, and recall that $\epsilon_{n\sB}$ is unaffected by the shift. Then $\left(\epsilon_{n\,\sB}\kappa_1{}^\sB_{\dot a}\right)$ does not transform because $\tilde \epsilon^\sA$ is orthogonal to $\epsilon_{n\sA}$ as we have seen in \eqref{eq:useful_rel_tilde_epsilon}, so the only deformation can come from $\tilde \epsilon^\sA$ itself. To see how $\tilde \epsilon^\sA$ behaves under BCFW, it is useful to rewrite its definition \eqref{eq:def_tilde_epsilon_anti_v1} as
\begin{equation}\label{eq:def_tilde_epsilon_v2}
 \tilde\epsilon^\sA=-\frac{1}{k_1\cdot k_n}\left(k_n^{\sA\sB}\epsilon_{1\sB}+k_1^{\sA\sB}\epsilon_{n\sB}\right)\,.
\end{equation}
In this form, the relations \eqref{eq:useful_rel_tilde_epsilon} are manifest, and it is clear that it transforms at most linearly in $z$ because the denominator is invariant due to $q\cdot k_1=q\cdot k_n=0$. However, neither of the polarization spinors $\epsilon_{1\sB}$ and $\epsilon_{n\sB}$ transform, and 
\begin{equation}
 q^{\sA\sB}\epsilon_{1\sB} = \tilde\epsilon^{[\sA}k_1^{\sB]\sC}\epsilon_{n\sC}\epsilon_{1\sB}=0\,.
\end{equation}
Therefore $ \tilde\epsilon^\sA$ as well as the shift-spinors $\delta\kappa_{1} {}^\sA_{\dot a}$ and $\delta\kappa_{n} {}^\sA_{\dot a}$ are invariant under the BCFW deformation, and the shift \eqref{eq:BCFW_antichiral} is well-defined.\\

\paragraph{Shifting the supermomenta.}  In the R-symmetry preserving supersymmetry representation, the supershift is \emph{not} implemented via a linear shift in the fermionic variables, but rather by a multiplicative exponential factor
\begin{equation}\label{eq:shift-ferm-fact}
 \mathcal{I}_n\rightarrow\hat{\mathcal{I}}_n\,\exp\left(-z\,q_{1\sI}q_{n\sJ}\Omega^{\sI\sJ}\right)\,.
\end{equation}
This is clearly the fermionic Fourier transform of the standard linear super-BCFW shift in the little-group preserving representation, see e.g. Ref.~\cite{Boels:2012ie}. As expected, the Fourier Transform interchanges linear shifts of the variables in $z$ with a multiplication by an exponential factor.

To see this explicitly, consider the amplitude in the little-group preserving representation of \cref{eq:ampl_susy}, obtained from the R-symmetry representation via a fermionic half-Fourier transform as discussed in  \cref{sec:susy-factor},
\begin{equation}
   \int \prod_{i=1}^n d^Nq_i^l  \; \prod_je^{-q_j^l\eta_{\xi l}}\;e^{F_N}\Bigg|_{q_l=\eta_{\epsilon l}}=\prod_{i}\delta^{0|N}\left( \sum_{j}\frac{\la u_i u_j\ra}{\sigma_{ij}}\la \epsilon_j \eta_{j}^l\ra -\la v_i\eta_{i}^l\ra\right)\,.
\end{equation}
On the right, we have grouped the fermionic variables into a little-group spinor $\eta_a^l$, with $\eta_{i \epsilon }^l = \la \epsilon_i \eta_i^l\ra$ and $\eta_{\xi i}^l = \la \xi_i \eta_i^l\ra$. In this representation, the fermionic BCFW-shift mirrors the shift in the chiral spinors,
\begin{align}\label{eq:shift_eta}
 \hat{\eta}_1^{la}&= \eta_1^{la} +z\,\epsilon_1^a  \la\epsilon_n \eta_n^l\ra&
 \hat{\eta}_n^{la}&= \eta_n^{la} +z\,\epsilon_n^a  \la\epsilon_1 \eta_1^l\ra 
 %
\end{align}
Our discussion from the polarized scattering equations is then directly applicable to the fermionic case: only $\la v_{1,n}\hat\eta_{1,n}\ra$ are shifted, while   $\la \epsilon_{1,n}\hat\eta_{1,n}\ra$ remain invariant. In particular, all $z$-dependence resides in the delta-functions 
\begin{equation}
 \delta^{0|N}\left( \sum_{j}\frac{\la u_1u_j\ra}{\sigma_{1j}}\la \epsilon_j \eta_{j}^l\ra -\la v_1\eta_{1}^l\ra-z \la \epsilon_n\eta_{n}^l\ra\right)\;\delta^{0|N}\left( \sum_{j}\frac{\la u_n u_j\ra}{\sigma_{nj}}\la \epsilon_j \eta_{j}^l\ra -\la v_n\eta_{n}^l\ra-z \la \epsilon_1\eta_{1}^l\ra\right)
\end{equation}
We can then transform back to the R-symmetry preserving representation, where the $z$-dependent terms combine to give the exponential of \eqref{eq:shift-ferm-fact}, while the other terms give back the usual supersymmetry factor $e^F$. \footnote{It is of course sufficient to only transform the fermionic variables in $1$ and $n$ to see this. Alternatively, we can also choose to perform a \emph{full} fermionic Fourier transform on only one of the particles,  e.g. $n$,
\begin{equation}
 \int d^{2N}\!q_n \;e^{-q_{nI}\eta_n^{I}}e^F=\delta^{0|2N}\left(\sum_i\frac{\la u_i u_n\ra}{\sigma_{in}}\Omega^{IJ}q_{iJ}\right)\;\exp\left(\frac1{2}\sum_{i,j\neq n}\frac{\la u_i u_j\ra}{\sigma_{ij}}\Omega^{IJ}q_{iI}q_{jJ}-\eta_n^I\right)\,.
\end{equation}
This clearly comes at the expense of having to treat the two shifted particles differently. In this case, we choose the following BCFW shift for the new fermionic variables $\eta_n$:
\begin{equation}
 \hat \eta^I_n = \eta_n^I +z\, \Omega^{IJ}q_{1J}\,.
\end{equation}
After transforming back to the R-symmetry breaking representation, this leads to the same exponential factor. } We thus conclude that the BCFW shift  amounts to the insertion of an exponential factor $\exp\left(-z\,q_{1\sI}q_{n\sJ}\Omega^{\sI\sJ}\right)$ in the integrand of the exponential supersymmetry representation. Due to the chiral nature of the spinorial  shift, it is only necessary to shift the chiral supermomenta, so \emph{no} corresponding factor  $\exp\big(-z\,\tilde q_{1\dot \sI}\tilde q_{n\dot \sJ}\tilde \Omega^{\dot \sI\dot \sJ}\big)$ appears in the integrand.

\paragraph{Reduction to 4d.} Under dimensional reduction, the 6d  shift  \eqref{eq:BCFW_shift} reduces to the well-known BCFW shift in four dimensions. To see this, consider the case where the particles $1$ and $n$ have negative and positive helicity respectively. In the conventions of \cref{sec:4d-massless}, this can be embedded into 6d via
\begin{align}
 & \epsilon_{1a}=(0,\epsilon_1)\,,  &&\epsilon_{na}=(\tilde{\epsilon}_n,0)\,.
\end{align}
The six-dimensional shift \eqref{eq:BCFW_shift} for fundamental spinors then reduces straightforwardly to the usual BCFW shift in four dimensions,
 \begin{align}
 &\hat\kappa_{1\,A}^a=
   \begin{pmatrix}0 & \tilde \kappa_{1}^{\dot \alpha} \\ \kappa_{1\alpha} & 0\end{pmatrix}+ z \epsilon_1\tilde{\epsilon}_n   \begin{pmatrix}0 & \tilde{\kappa}_{n}^{\dot\alpha} \\ 0& 0\end{pmatrix}\,,  
& \hat\kappa_{n\,A}^a=
  \begin{pmatrix}0 & \tilde \kappa_{n}^{\dot \alpha} \\ \kappa_{n\alpha} & 0\end{pmatrix}-z\epsilon_1\tilde{\epsilon}_n\begin{pmatrix}0 & 0 \\ \kappa_{1\alpha} & 0\end{pmatrix}\,,
\end{align}
up to the manifest scale $\epsilon_1\tilde\epsilon_n$ in the polarization data, which could be absorbed into $z$. The shift vector $q_{\alpha\dot\alpha}=\epsilon_{1\alpha}\tilde\epsilon_{n\dot \alpha}$ again agrees with the usual choice up to the polarization-dependent scale $\epsilon_1\tilde\epsilon_n$. Proving that the shift of the anti-fundamental spinors gives the same results is a little more involved due to $\delta\kappa_{1,n}^\sA \sim\tilde \epsilon^\sA$ (rather than $\delta\kappa_1^\sA \sim\tilde\epsilon_1^\sA$ and $\delta\kappa_n^\sA\sim\tilde\epsilon_n^\sA$ respectively). Using the definition \eqref{eq:def_tilde_epsilon_v2} for $\tilde\epsilon^\sA$, we find
\begin{equation}
 \tilde\epsilon^\sA = \left(+\frac{\tilde \epsilon_n\,\kappa_1^\alpha}{\la 1n \ra},\;\;-\frac{\epsilon_1\,\tilde \kappa_{n\dot\alpha}}{\left[ 1n \right]}\right)^T\,,\qquad 
 \epsilon_{n\sB}\kappa_{1\dot a}^\sB = \left(0,\;\,-\tilde\epsilon_n\,\left[1n\right]\right)\,, \qquad 
 \epsilon_{1\sB}\kappa_{n\dot a}^\sB = \left(\epsilon_a\,\la 1n\ra,\;\,-0\right)\,.
\end{equation}
Inserting this into \eqref{eq:BCFW_antichiral} then  leads to the following shift for the  anti-fundamental spinors;
 \begin{align}\label{eq:4d-BCFW-shift}
  &\hat\kappa_{1\,\dot a}^\sA = \begin{pmatrix} \kappa_1^\alpha & 0\\ 0 & -\tilde\kappa_{1\dot\alpha}\end{pmatrix}
  +z\epsilon_1\tilde\epsilon_n \begin{pmatrix}0 & \frac{\tilde\epsilon_n[1n]}{\epsilon_1\la 1n\ra}\kappa_1^\alpha \\ 0 & -\tilde \kappa_{n\dot\alpha}\end{pmatrix} \,,&
   &\hat \kappa_{1n\,\dot a}^\sA = \begin{pmatrix} \kappa_n^\alpha & 0\\ 0 & -\tilde\kappa_{n\dot\alpha}\end{pmatrix}
  -z\epsilon_1\tilde\epsilon_n \begin{pmatrix}\kappa_1^\alpha & 0 \\ -\frac{\epsilon_1\la 1n\ra}{\tilde\epsilon_n[1n]}\tilde \kappa_{n\dot\alpha} &0\end{pmatrix}\,.
 \end{align}
To see that this gives the same four-dimensional shift, note that the off-diagonal entries are proportional to $\kappa_1^\alpha$ and $\tilde\kappa_{n\dot \alpha}$ respectively, so they can be absorbed into a ($z$-dependent) 6d little group transformation of $\kappa_{1\,\dot a}^\sA$ and $\kappa_{n\,\dot a}^\sA$. Moreover, since one of the off-diagonal terms always vanishes, this little group transformation leaves the diagonal entries unaffected, and we get\footnote{To be explicit, the relevant little group transformations are  
\begin{equation}
 U_{\dot a}^{\dot b}=\begin{pmatrix}1 & b \\ c & 1\end{pmatrix}\,,\qquad \text{ with }\;\;b=-z\,\tilde\epsilon_n^2\frac{[1n]}{\la 1n\ra}\,,\;\;c=0\;\;\text{ for particle 1}\,,\qquad b=0\,,\;\;c=z\,\epsilon_1^2\frac{\la 1n\ra}{[1n]}\;\;\text{ for particle n}\,.
\end{equation}}
\begin{subequations}
\begin{align}
  &\hat\kappa_{1\,\dot a}^\sA \simeq U_{\dot a}^{\dot b}\hat\kappa_{1\,\dot b}^\sA= \begin{pmatrix} \kappa_1^\alpha & 0\\ 0 & -\tilde\kappa_{1\dot\alpha}\end{pmatrix}
  +z\epsilon_1\tilde\epsilon_n \begin{pmatrix}0 & 0 \\ 0 & -\tilde \kappa_{n\dot\alpha}\end{pmatrix} \,,\\
  &\hat\kappa_{n\,\dot a}^\sA \simeq U_{\dot a}^{\dot b}\hat\kappa_{n\,\dot b}^\sA= \begin{pmatrix} \kappa_n^\alpha & 0\\ 0 & -\tilde\kappa_{n\dot\alpha}\end{pmatrix}
  -z\epsilon_1\tilde\epsilon_n \begin{pmatrix}\kappa_1^\alpha & 0 \\ 0 &0\end{pmatrix}\,,
 \end{align}
 \end{subequations}
 in agreement with our result from the chiral spinors \eqref{eq:4d-BCFW-shift}.
Above, we have used $\simeq$ to indicate that the relations  hold up to a 6d little-group rotation. We emphasize that the need for this additional little-group rotation to bring $\hat\kappa_{1,n}^\sA$ into diagonal form was expected from the embedding of 4d kinematics into 6d, see \S\ref{sec:4d-massless}: even after restricting to 4d massless kinematics, $\kappa^\alpha_{\dot a}$ are only required to be proportional, in general an additional little-group rotation is  needed to bring it into the diagonalized form of \eqref{eq:kappa_4d_m=0}.

\subsection{Factorization}\label{sec:factorization}
For scattering-equations-based amplitude representations, it is well-known that factorization of the momenta arises from factorization of the of the moduli-space  $\mathfrak{M}_{0,n}$ of $n$-points on the Riemann-sphere modulo Mobius transformations \cite{Dolan:2013isa}.  The boundary  $\p\mathfrak{M}_{0,n}$ of $\mathfrak{M}_{0,n}$ consists of loci where a collection of points $\sigma_i$ for $i\in L$ come together at a point.  This is understood geometrically as a limit where the Riemann surface $\Sigma=\CP^1$ decomposes into two $\CP^1$s, $\Sigma_L$ and $\Sigma_R$ joined at a node, with the $\sigma_i$, $i\in L$ on $\Sigma_L$ and the rest on $\Sigma_R$.  We denote by $ \widehat{\mathfrak{M}}_{0,n}$  the Deligne-Mumford compactification of the moduli space of marked Riemann surfaces \cite{Deligne:1969}, obtained by including  such nodal surfaces of genus zero, with arbitrarily many components and nodes, but with at least 3 marked points/nodes on each component.

Singularities in the integrand $\mathcal{I}_n$ for any theory only depend on the kinematic data via polynomials. All poles in the formula  stem from those in the $\sigma_{ij}$ and $\la u_i u_j\ra$ which come from the boundary of the moduli space  $\partial \mathfrak{M}_{0,n}^{\mathrm{pol}}$.  Here the moduli space $ \mathfrak{M}_{0,n}^{\mathrm{pol}}$ encodes the locations of the punctures $\sigma_i$ as well as the values for $u_i,\,v_i$, modulo the symmetry group $\mathrm{SL}(2,\mathbb{C})_\sigma\times\mathrm{SL}(2,\mathbb{C})_u$. However, the additional boundary components in $ \mathfrak{M}_{0,n}^{\mathrm{pol}}$ correspond to spurious singularities involving the polarization data as  seen for example in \eqref{u34} and other formulae in  \S\ref{sec:low-point}.  But, for super Yang-Mills and supergravity theories, we have proven linearity of $\det{}'H$ in the polarization data in \S\ref{sec:poldata}. Thus, all poles of the integrand originate from boundaries of the moduli space of the Riemann sphere  $\partial \widehat{\mathfrak{M}}_{0,n}\subset \partial \mathfrak{M}_{0,n}^{\mathrm{pol}}$. 

At tree level, $\partial \widehat{\mathfrak{M}}_{0,n}$ is the union of components $\partial_{\sL,\sR} \widehat{\mathfrak{M}}_{0,n}$ that correspond to separating degenerations  that split the sphere $\Sigma$ into two components, $\Sigma_\sL$ and $\Sigma_\sR$ partitioning the punctures into $L\cup R$, with $R$ the complement of $L$ so $n=n_\sL+n_\sR$,
\begin{equation}\label{eq:dM}
 \partial_{\sL,\sR} \widehat{\mathfrak{M}}_{0,n}\simeq \widehat{\mathfrak{M}}_{0,n_L+1}\times\widehat{\mathfrak{M}}_{0,n_R+1}\,.
\end{equation}
The  component $\partial_{\sL,\sR} \widehat{\mathfrak{M}}_{0,n}$ can be parametrized by gluing two  Riemann spheres $\Sigma_\sL$ and $\Sigma_\sR$ as follows. Choose a marked point on each sphere,  $\sigma_\sR\in \Sigma_\sR$ and $x_\sL\in\Sigma_\sL$, and remove the disks $|\sigma-\sigma_\sR|<\varepsilon^{1/2}$ and $|x-x_\sL|<\varepsilon^{1/2}$, where $\varepsilon$ is the parameter governing the degeneration. Then we can form a single Riemann surface by identifying,
\begin{equation}\label{eq:param_dM}
 \left(x-x_{\sL}\right)\left(\sigma-\sigma_{\sR}\right)=\varepsilon \,.
\end{equation}
The component  $\partial_{L,R} \widehat{\mathfrak{M}}_{0,n}$ corresponds to the limiting case $\varepsilon\rightarrow 0$.
Often we simplify this degeneration by choosing  $x_\sL=\infty$, where \eqref{eq:param_dM} becomes
\begin{equation}
 \sigma=\sigma_\sR+\varepsilon \tilde x,,\qquad\quad\text{with } \tilde x=x^{-1}\,.
\end{equation}

Let us briefly review how factorization works in the CHY formalism. 
\begin{lemma}
Suppose that the marked points $\sigma_i$ satisfy the scattering equations
\begin{equation}\label{eq:SE_CHY_v2}
 \mathcal{E}_i:=\sum_{j\neq i} \frac{k_i\cdot k_j}{\sigma_{ij}}=0\,,
\end{equation}
then $\{\sigma_i\}\in \partial_{\sL,\sR} \widehat{\mathfrak{M}}_{0,n}$ implies   $k_L^2=0$ where $k_\sL=-\sum_{i\in L}k_i$. 
\end{lemma}\label{lem:CHY-fact}
\proof This follows by inserting \eqref{eq:param_dM} into the following combination of the scattering equations 
\begin{equation}\label{eq:SE_null_mom}
0=\sum_{i\in L}\sigma_{i\sR}\mathcal{E}_i=\sum_{i,j\in L}\tilde x_{i\sL}\frac{k_i\cdot k_j}{ \tilde x_{ij}}=\frac12\sum_{i,j\in L} k_i\cdot k_j=\frac{1}{4}k_\sL^2\,,
\end{equation}
where the second equality holds to order $O(\varepsilon)$ as the denominator is $O(1)$ for $j\in R$. 
$\Box$
\smallskip 

Thus, in the degeneration limit  $k_\sL$ is null, and the propagator goes on-shell. The scattering equations further ensure that the CHY measure $d\mu_n^{\CHY}$ mirrors the behaviour of the moduli space at the boundary \cite{Dolan:2013isa, Geyer:2015jch},
\begin{equation}
 d\mu_n^{\CHY}=\frac{\varepsilon^{2\left(n_\ssL-1\right)}}{\prod_{i\in L}x_{i\sL}^4}\frac{d\varepsilon}{\varepsilon}\,\delta\left(k_\sL^2 - \varepsilon\mathcal{F}\right)\;d\mu_{n_\ssL+1}^{\CHY}\;d\mu_{n_\ssR+1}^{\CHY}\,.
\end{equation}
Each `half integrand' $\mathcal{I}_n$ for Yang-Mills theory and gravity -- either a Pfaffian or a Parke-Taylor factor  -- also factorizes into two subamplitudes, linked by a sum over states  in the internal propagator,
\begin{equation}\label{eq:fact_int}
 \mathcal{I}_n^{\mathrm{h}} = \varepsilon^{-\left(n_\ssL-1\right)}\prod_{i\in L}x_{i\sL}^2\,\;\sum_{\mathrm{states}}\mathcal{I}_{n_\ssL+1}^{\mathrm{h}} \,\mathcal{I}_{n_\ssR+1}^{\mathrm{h}} \,.
\end{equation}
Combining the measure and the integrand, we see that gravity and Yang-Mills amplitudes in the CHY-representation factorize correctly, in accordance with \eqref{eq:BCFW}.

In the rest of this section, we will follow a similar strategy to the one outlined above for the CHY formalism, and first establish the map between the polarized scattering equations and factorization channels. Based on this, we determine how the measure $d\mu_n^{\mathrm{pol}}$ behaves on the boundary of the moduli space. In line with the equivalence between the polarized measure and the CHY measure established in \cref{sec:reltoCHYmeasure}, we find 
\begin{equation}\label{eq:fact_pol-measure}
 d\mu_n^{\mathrm{pol}}= \frac{\varepsilon^{2\left(n_L-1\right)}}{\prod_{i\in L}x_{i\sL}^4}\frac{d\varepsilon}{\varepsilon}\,\frac{d^8\kappa_\sA^a}{\mathrm{vol}\,\mathrm{SL}(2,\mathbb{C})}\;d\mu_{n_L+1}^{\mathrm{pol}}\;d\mu_{n_R+1}^{\mathrm{pol}}\,.
\end{equation}
The delta-functions $\delta\left(k_\sL^2 - \varepsilon\mathcal{F}\right)$ enforcing that $\varepsilon\sim k_\sL^2\sim k_\sR^2$ are part of the momentum conservation contained in the polarized measure.
Finally, we show that the integrands obey \eqref{eq:fact_int}, and that the sum over states is encoded in a suitable superspace integral,
\begin{equation}
  \sum_{\mathrm{states}}\mathcal{I}_{n_\ssL+1}^{\mathrm{h}} \,\mathcal{I}_{n_\ssR+1}^{\mathrm{h}} =\int d^{2N} \!q_\sL d^{2N} \!q_\sR\;  \mathcal{I}_{n_\ssL+1}^{\mathrm{h}} \,\mathcal{I}_{n_\ssR+1}^{\mathrm{h}}\;\left(\la \epsilon_\sL \epsilon_\sR\ra^{N}\,e^{i\la \epsilon_\ssL \epsilon_\ssR\ra^{-1} \,q_{\sL\sI}q_{\sR\sJ}\Omega^{\sI\sJ}}\right)\,.
\end{equation}
The formulae based on the polarized scattering equations thus factorize as expected for super Yang-Mills and supergravity amplitudes.

\subsubsection{Polarized scattering equations and measure}\label{sec:fact_SE}
\paragraph{Factorisation of the polarized scattering equations.}  We wish to find an analogue of lemma \ref{lem:CHY-fact} for the factorization properties of the polarized scattering equations. We have
\begin{lemma}
Define  $\epsilon_{\sR\sA}^a:=\sum_{i\in L}u_i^a\epsilon_{i\sA}$. Factorization  $\{\sigma_i\}\in \partial_{\sL,\sR} \widehat{\mathfrak{M}}_{0,n}$ and the polarized scattering equations then implies  the factorization
\begin{equation}\label{eq:def_epsilon_R_final}
 \epsilon_{\sR\sA}^a:=\sum_{i\in L}u_i^a\epsilon_{i\sA}=u_\sR^a\epsilon_{\sR\sA}\,.
\end{equation}
\end{lemma}

\proof We consider the form  \eqref{polscatt-k}
\begin{equation}
\epsilon_{i[\sA}\cE_{i\sB]}:=\sum_j \frac{\langle u_i u_j\rangle \epsilon_{j[\sB}\epsilon_{i\sA]}}{\sigma_{ij}}-k_{i\sA\sB}=0\, ,\label{polscatt-k1}
\end{equation}
and by analogy with \eqref{eq:SE_null_mom}
consider the sum 
\begin{equation}\label{eq:def_epsilon_L_deriv}
0=\sum_{i\in L}\sigma_{i\sR}\epsilon_{i[\sA}\cE_{i\sB]}=\sum_{i,j\in L} \tilde x_{j\sL}\frac{\langle u_i u_j\rangle \epsilon_{j[\sB}\epsilon_{i\sA]}}{\tilde x_{ij}} =\frac{1}{2}\sum_{i,j\in L}\langle u_i u_j\rangle \epsilon_{j[\sB}\epsilon_{i\sA]}\equiv \left\langle\epsilon_{\sR[\sA}\epsilon_{\sR\sB]}\right\rangle\, .
\end{equation}
Again, the second equality holds to order $O(\epsilon)$ in the degeneration parameter, and in the last equality, we have introduced the spinor $\epsilon_{\sR\sA}^a:=\sum_{i\in L}u_i^a\epsilon_{i\sA}$.
The relation \eqref{eq:def_epsilon_L_deriv}  tells us that $\langle\epsilon_{\sR[\sA}\epsilon_{\sR\sB]} \rangle=O(\varepsilon)$, so to leading order in $\varepsilon$, $\epsilon_{\sR\sA}^a$ factorizes into an SL$(4)$ spinor and a little group spinor, 
$ \epsilon_{\sR\sA}^a=u_\sR^a\epsilon_{\sR\sA}$
for some $u_\sR^a, \epsilon_{\sR\sA}$ as desired. $\Box$

\smallskip  

\begin{corol}
In the degeneration limit, the original worldsheet spinor $\lambda(\sigma)$ 
thus induces a spinor $\lambda^{(\sR)}(\sigma)$ on the  sphere $\Sigma_\sR$, with
\begin{equation}\label{eq:def_lambda_R}
\lambda^{(\sR)}_{\sA}{}^a(\sigma)=\sum_{p\in R} \frac{u_p^a\epsilon_{p\sA}}{\sigma-\sigma_p} + \frac{u_\sR^a\epsilon_{\sR\sA}}{\sigma-\sigma_\sR} \,,\qquad\text{where }u_\sR^a\epsilon_{\sR\sA}=\sum_{i\in L}u_i^a\epsilon_{i\sA}\, .
\end{equation}
\end{corol}
By an extension of the same argument, $\lambda(\sigma)$  also  induces a spinor $\lambda^{(\sL)}(\sigma)$ on the  sphere $\Sigma_\sL$, which can be seen as follows. Since $\lambda(\sigma)$ is a worldsheet spinor, the combination $\lambda(\sigma)\sqrt{d\sigma}$ is invariant under the inversion  $(\sigma-\sigma_\sR)(x-x_\sL)=\varepsilon$, 
\begin{equation}
\lambda_A^a(\sigma)\sqrt{d\sigma}=\lambda_A^a(x)\sqrt{dx}  \, ,\qquad\text{with }\,\, \lambda_A^a(x)=\sum_{i=1}^n \frac{w_i^a\,\epsilon_{i\sA}}{x-x_i}\,,
\end{equation}
where $w_i^a$ denote the little group spinors in the coordinates $x$. The invariance of $\lambda_A^a(\sigma)\sqrt{d\sigma}$ then implies that the $u_i$  transform as worldsheet spinors of the local bundle at the marked point $\sigma_i$,
\begin{equation}\label{eq:uw}
 \frac{u_i^a\,\sqrt{d \sigma}}{\sigma-\sigma_i}=\frac{w_i^a\,\sqrt{d x}}{x-x_i}\qquad\text{and thus }\,\, u_i^a=\frac{i\varepsilon^{1/2} }{x_{i\sL}}w_i^a\,  .
\end{equation}
At this stage, the same reasoning as above ensures that $\lambda(x)$ descends to $\lambda^{(\sL)}(x)$ on $\Sigma_\sL$  with
\begin{equation}\label{eq:def_lambda_L}
\lambda^{(\sL)}_\sA{}^a(x)= \sum_{i\in L} \frac{w_i^a\epsilon_{i\sA}}{x-x_i} + \frac{w_\sL^a \epsilon_{\sL\sA}}{x-x_\sL} \, , \qquad \text{where }\,\, w^a_\sL\epsilon_{\sL\sA}=\sum_{p\in R} w_p^a \epsilon_{p\sA}\, .
\end{equation}

\smallskip

In the CHY amplitude representation, the relation \eqref{eq:SE_null_mom} makes it clear that the scattering equations map the boundary of the moduli space to a factorization channel of the amplitude.  
To see this from \eqref{eq:def_epsilon_L_deriv}, note that  momentum conservation on each subsphere (encoded in the polarized scattering equations) gives
\begin{equation}\label{eq:k_L_in_spinors}
 k_{\sR\,\sA\sB}=-\sum_{p\in R} k_p{}_{\,\sA\sB}=\epsilon_{\sR[\sA}\sum_{p\in R}\frac{\left\langle u_pu_\sR\right\rangle}{\sigma_{p\sR}}\epsilon_{p\sB]}\,,
\end{equation}
where we have used the form \eqref{polscatt-k1} of the polarized scattering equations on  $\Sigma_\sR$, 
\begin{equation}
 \epsilon_{p[\sA}\mathcal{E}_{p\sB]}^{(\sR)}=\sum_{q\in R}\frac{\left\langle u_pu_q\right\rangle}{\sigma_{pq}}\epsilon_{p[\sA}\epsilon_{q\sB]}+\frac{\left\langle u_pu_\sR\right\rangle}{\sigma_{p\sR}}\epsilon_{\sR[\sA}\epsilon_{p\sB]}-k_p{}_{\,\sA\sB}=0\,,
\end{equation}
and the first term does not contribute due to the antisymmetry in the SL$(4)$ spinor index. The relations \eqref{eq:k_L_in_spinors} guarantees that to leading order in the degeneration parameter $\varepsilon$, the internal momentum $k_\sR$ is on-shell, $k_\sR^2=O(\varepsilon)$, and the boundary of the moduli space indeed corresponds to a factorization channel of the amplitude.
The same reasoning can also be applied to the momentum $k_\sL$ on the sphere $\Sigma_\sL$,
\begin{equation}
 k_{\sL\,\sA\sB}=-\sum_{i\in L} k_i{}_{\,\sA\sB}=\sum_{\substack{i\in L\\p\in R}}\frac{\left\langle w_i w_p\right\rangle}{x_{i\sR}}\epsilon_{p[\sA}\epsilon_{i\sB]}=\sum_{\substack{ i\in L\\p\in R}}  \frac{\left\langle u_i u_p\right\rangle}{\sigma_{\sR p}}\epsilon_{p[\sA}\epsilon_{i\sB]}=-k_{\sR\,\sA\sB}\,,
\end{equation}
so $k_\sL$ goes on-shell as $\varepsilon\rightarrow 0$ and  $k_\sL=-k_\sR$, as expected for a factorization channel.
Here, the second identity follows again from the polarized scattering equations on $\Sigma_\sL$, the third from the  degeneration relations \eqref{eq:uw} for $u_p$ and $w_i$, and the last from the definition of $\epsilon_\sR$ and the relation \eqref{eq:k_L_in_spinors} for $k_\sR$. \\

\paragraph{The scaling weights in $\varepsilon$.} Before proceeding further, it is helpful to take a closer look at the scaling in the  parameter $\varepsilon$ in the degeneration limit $\varepsilon\ll 1$. Near the boundary of the moduli space, a marked point $i$ lies on the sphere $\Sigma_\sL$ if $x_{i\sL}\sim 1$ is of order one, and similarly a point $p$ lies on $\Sigma_\sR$ if $\sigma_{p\sR}\sim 1$. Using the parametrization \eqref{eq:param_dM} of the degeneration, this gives immediately  
\begin{subequations}\label{eq:x_sigma_OoM}
 \begin{align}
  &  i\in L: && x_{i\sL}\sim 1\,, \qquad \sigma_{i\sR}\sim\varepsilon\,,\\
  &  p\in R: && x_{p\sL}\sim \varepsilon\,, \qquad \sigma_{p\sR}\sim 1\,.
 \end{align}
\end{subequations}
As a direct consequence, the separation $x_{ij}\sim 1$ of two marked points $i,j$ that lie on $\Sigma_\sL$ is of order one in the degeneration limit (and $\sigma_{pq}\sim 1$ for $p,q$ on $\Sigma_\sL$). Using \cref{unique} on the spheres $\Sigma_\sL$ and $\Sigma_\sR$, we can also infer the scaling of little-group invariants constructed from $u$'s and $w$'s. \Cref{unique} implies that
 there only exist solutions to the polarized scattering equations if all terms in $\la w_i\lambda^{(\sL)}_\sA(x_i)\ra$ and $\la u_p\lambda^{(\sR)}_\sA(\sigma_p)\ra$ remain of order one.
For  points $i,j\in L$ and $p,q\in R$, this means
\begin{subequations}
 \begin{align}
  &  i, j\in L: && \la w_i w_\sL\ra \sim 1\,, \qquad \la w_i w_j \ra \sim 1\,,\\
  &  p,q\in R: && \la u_p u_\sR \ra \sim 1 \,, \qquad \,\,\la u_p u_q \ra \sim 1 \,,
 \end{align}
\end{subequations}
and the order of all other contractions follows from the relation \eqref{eq:uw} between $u$ and $w$ and \eqref{eq:x_sigma_OoM}.\footnote{So, for example, $\la u_i w_\sL\ra \sim\varepsilon^{1/2}$ and $\la u_i w_j \ra \sim\varepsilon^{1/2}$.}
We can further use  the definitions of $u_\sR$ and $w_\sL$ to derive the order of the remaining spinor brackets: from the dominant balance in $\la u_i u_\sR\ra \epsilon_{\sR\sA}$, $\la w_p w_\sL\ra\epsilon_{\sL\sA}$, $\la u_p u_\sR\ra\epsilon_{\sR\sA}$ and $\la w_\sL u_\sR\ra \epsilon_{\sR\sA}$, we find respectively
 \begin{align}
  i\in L\,,\;p\in R\,:\qquad\qquad  \lla u_i u_\sR\rra \sim \varepsilon\,, \qquad
   \lla w_p w_\sL \rra \sim \varepsilon\,, \qquad \lla u_i u_p \rra \sim 1\,,\qquad \lla u_\sR w_\sL \rra \sim \varepsilon^{1/2} \,.
 \end{align}
 Summarizing the above discussion, we have seen that both the worldsheet spinor $\lambda(\sigma)\sqrt{d\sigma}$ and the polarized scattering equations descend to the subspheres, with leading terms of order one throughout the degeneration,
\begin{equation}
 \prod_{i=1}^n \delta^4\left( \mathcal{E}_i\right) = \prod_{i\in L} \delta^4\left( \mathcal{E}_i^{(\sL)}\right)\prod_{p\in R} \delta^4\left( \mathcal{E}_p^{(\sR)}\right)
\end{equation}
where the scattering equations on the subspheres $\Sigma_\sL$ and $\Sigma_\sR$ are given by the usual construction,
\begin{subequations}
\begin{align}
& \mathcal{E}_i^{(\sL)}=\left\langle w_{i}\lambda^{(\sL)}_\sA(x_i) \right\rangle-\left\langle v_i \kappa_{i\sA}\right\rangle && i\in L \,,\\
& \mathcal{E}_p^{(\sR)}=\left\langle u_{p}\lambda^{(\sR)}_\sA(\sigma_p) \right\rangle-\left\langle v_p \kappa_{p\sA}\right\rangle && p\in R\,.
\end{align}
\end{subequations}
We stress that in contrast to the CHY formalism, the polarized scattering equations  do \emph{not} contribute powers of the degeneration parameter $\varepsilon$ to the measure. As we will see below, the factor of $\epsilon^{2\left(n_\ssL-1\right)-1}$ instead comes entirely from the integration over the variables $(\sigma_i,u_i)$.

\paragraph{Factorization of the measure.}
Armed with the insights on how the polarized scattering equations behave on the boundary of the moduli space, let us now take a closer look at the measure. The degeneration of the measure $d^{n-3}\sigma$ on the sphere is entirely analogous to the CHY case, but it provides a good introduction and we will review it here for completeness. 

For any values of the degeneration parameter, M\"obius invariance on the sphere allows us to fix three marked points, two of which we choose to lie on one subsphere in the limit $\varepsilon\ll1$, $\sigma_{p_1},\sigma_{p_2}\in \Sigma_\sR$, and one on the other, $x_{i_1}\in \Sigma_\sL$.\footnote{In the ambitwistor string, this has a particularly elegant interpretation in terms of picture changing operators. We start out on the Riemann sphere with $n$ vertex operators and $n-3$ picture changing operators. In the degeneration limit, the only non-trivial assignment of these onto the two subspheres correlates the number of picture changing operators with the number of vertex operators as described above. All other possibilities give zero after integration over the ghost zero modes.} At the boundary of the moduli space, we have the further freedom to fix the junction points $\sigma_\sR$, $x_\sL$ of the two spheres, as well as one additional point $\sigma_{i_2}$ on  $\Sigma_\sL$.  To leading order in $\varepsilon$, the Jacobian $J^{\text{m\"ob}}$ for this gauge fixing becomes the Jacobian $J^{\text{m\"ob}}_\sR$ for gauging $\{\sigma_{p_1},\sigma_{p_2},\sigma_\sR\}\subset\Sigma_\sR$, \footnote{While the degeneration appears to treat $\Sigma_\sL$ and $\Sigma_\sR$ differently, their roles can be interchanged by starting from a parametrization of the sphere in $x$-coordinates instead of $\sigma$.}
\begin{equation}
 J^{\text{m\"ob}}=\sigma_{i_1 p_1} \sigma_{p_1 p_2}\sigma_{p_2 i_1}=\sigma_{\sR p_1} \sigma_{ p_1 p_2}\sigma_{ p_2\sR}=J^{\text{m\"ob}}_\sR\,.
\end{equation}
Together with  the differentials $\prod_{p\in R}d\sigma_p$, which  descend directly to $\Sigma_\sR$, this Jacobian gives the usual M\"obius invariant measure on $\Sigma_\sR$. For the punctures $\sigma_i$ with $i\in L$ on the other hand, we find from \eqref{eq:param_dM}
\begin{equation}
 d\sigma_i=-\frac{\varepsilon}{x_{i\sL}^2}\,dx_i\,,\qquad\qquad d\sigma_{i_2}=\frac{x_{i_1 i_2}}{x_{i_1\sL}x_{ i_2\sL}}\, d\varepsilon\,.
\end{equation}
Combining these factors gives both the correct differentials and the Jacobian $J^{\text{m\"ob}}_\sL$ for the measure on $\Sigma_\sL$ after gauge-fixing $\{x_{i_1},x_{i_2},x_\sL\}$. Putting this all together, the measure on the moduli space of marked Riemann spheres factorizes as
\begin{align}
 \frac{\prod_{i=1}^n d\sigma_i}{\mathrm{vol} \,\mathrm{SL}(2,\mathbb{C})}
& = \frac{\varepsilon^{n_\ssL-2}d\varepsilon}{\prod_{i\in L}x_{i\sL}^2} \,\, \Bigg(\left(x_{ i_1 i_2}x_{ i_2\sL}x_{\sL i_1}\right)\hspace{-8pt}\prod_{\substack{i\in L\\ i\neq  i_1, i_2}} dx_i \Bigg)\,\,\Bigg(\left(\sigma_{\sR p_1} \sigma_{p_1  p_2}\sigma_{p_2\sR}\right)\hspace{-8pt}\prod_{\substack{p\in R\\ p\neq  p_1,  p_2}} d\sigma_p\Bigg)\\
 &= \frac{\varepsilon^{n_\ssL-2}d\varepsilon}{\prod_{i\in L}x_{i\sL}^2}\,\, \frac{dx_\sL\prod_{i\in L} dx_i}{\mathrm{vol} \,\mathrm{SL}(2,\mathbb{C})_\sL} \,\,\frac{d\sigma_\sR\prod_{p\in R} d\sigma_p}{\mathrm{vol} \,\mathrm{SL}(2,\mathbb{C})_\sR}
\end{align}

Consider next the part of the measure dependent on $v$. By the same argument as above, \cref{unique} ensures that all $v_i$ for $i=1,\dots,n$ remain of order one throughout the degeneration. The part of the measure involving $v$'s, including the delta-functions encoding the normalization, thus factorize directly into the contributions on each subsphere,
\begin{equation}\label{eq:fact_v}
 \prod_{i=1}^n d^2 v_i\,\delta\Big(\la v_i\epsilon_i\ra-1\Big)= \prod_{i\in L} d^2 v_i\,\delta\Big(\la v_i\epsilon_i\ra-1\Big)\; \prod_{p\in R} d^2v_p\,\delta\Big(\la v_p\epsilon_p\ra-1\Big)\,.
\end{equation}
In contrast to the measure $d^{n-3}\sigma$ for the punctures however, the right hand side of \eqref{eq:fact_v} does not yet give the full $v$-dependence of the measure on $\Sigma_\sL$ and $\Sigma_\sR$, because we are missing the contributions $v_\sL$ and $v_\sR$ from the junction points. We will see later how these extra variables are defined and in what form they appear in the amplitude.\\ 

For the $u$-dependent part of the measure, it will again be convenient to first work with a gauge-fixed measure, and restore gauge invariance on each sphere $\Sigma_{\sL,\sR}$ after factorization. In the same manner as for the punctures $\sigma_{i}$, we gauge the SL$(2,\mathbb{C})_u$ by  fixing two moduli on $\Sigma_\sR$ and one on $\Sigma_\sL$ (c.f. \eqref{eq:u_gauge})
\begin{equation}
 u_{p_1 a} = \left(1\,,\;0\right)\,,\qquad \la u_{i_1} u_*\ra=0\,,\qquad\qquad \text{for }p_1\in R\,,\;i_1\in L\,,
\end{equation}
where $u_*$ is an arbitrarily chosen reference spinor.   For convenience, let us also introduce $u_*^\perp$, normalized such that $\la u_* u_*^\perp\ra=1$. The usual Faddeev-Popov procedure gives the Jacobian $
 J^u=\lla u_{ i_1} u_{p_1} \rra \,\lla u_{p_1}u_*\rra $, and thus the $u$-part of the measure  becomes
\begin{equation}\label{eq:measure_u}
 \frac{\prod_{i=1}^n d^2u_i}{\mathrm{vol} \,\mathrm{SL}(2,\mathbb{C})_u}=
 \frac{\varepsilon^{n_\ssL}}{\prod_{i\in L}x_{iR}^2} \lla w_{ i_1} u_{p_1} \rra \,\lla u_{p_1}u_*\rra
\Biggl(d\la w_{i_1} u_*^\perp\ra \prod_{\substack{i\in L\\ i\neq i_1}}d^2 w_i \Biggr)
\Biggl( \prod_{p\in R}d^2 u_p \Biggr)\,,
\end{equation}
where we used that the $u_i$ transform as worldsheet spinors of the local bundles at $\sigma_i$, see \eqref{eq:uw}. As was the case for the marked points $\sigma_i$, this does not fully fix the SL$(2,\mathbb{C})$ gauge on each component sphere at the boundary of the moduli space, and we have the further  freedom to  fix both of the `junction moduli' $w_\sL$ on $\Sigma_\sL$, as well as one component of $u_\sR\in \Sigma_\sR$. As above, the right side is not yet in a recognizably factorized form, but misses components of the Jacobians for gauge-fixing on the subspheres, as well as the measure for one of the junction moduli $d\la u_{\sR}\, u_*^\perp\ra$. 

For a full factorization of the measure, we are also still missing the delta-functions enforcing the polarized scattering equations on the junction points, as well as an integral over the internal momentum in the propagator, $d^6k_\sL=d^8 \kappa_{\sL\sA}^a / \mathrm{vol}\,\mathrm{SL}(2,\mathbb{C})$. We introduce these, as well as all missing factors discussed above, by  inserting a conveniently chosen factor of 1, \footnote{This is quickly checked: First note that a quick weight count in the spinors $\kappa_\sL$ shows that the right hand side is weightless in $\kappa_\sL$, and indeed the Faddeev-Popov Jacobian from fixing the SL$(2,\mathbb{C})$ freedom cancels against (part of) the Jacobian from solving the scattering equations. We can make this explicit e.g. by fixing $\kappa_{\sL 0}^a$, as well as $\epsilon_{\sL 1}$. Then
\begin{equation}
 J_{\mathrm{SL}(2)_\kappa} = k_{\sL\,01} \epsilon_{\sL 0}\,,\qquad J_{\mathrm{pol}}^{-1}=\varepsilon^{-1}\, \la u_\sL w_\sR\ra \la w_\sR u_*\ra\,\, k_{\sL\,01} \epsilon_{\sL 0}\,,
\end{equation}
and thus the integral indeed gives one.}
\begin{align}\label{eq:measure_1=}
 1=\varepsilon^{-1} \int \frac{d^8\kappa_\sL}{\mathrm{vol}\, \mathrm{SL}(2,\mathbb{C})}\,  &d\la u_{\sR}\, u_*^\perp\ra \,d^2 v_\sL \,d^2v_\sR\,\,\, \la u_\sR w_\sL\ra \la w_\sL\, u_*\ra\,\,\delta\left(\lla v_\sL\epsilon_\sL\rra -1\right)\,\delta\left(\lla v_\sR\epsilon_\sR\rra-1 \right)  \\
& \delta^4\left( \left\langle w_\sL \lambda^{(\sL)}_\sA(x_\sL)\right\rangle-\left\langle v_\sL\kappa_{\sL\sA}\right\rangle\right)\,\delta^4\left( \left\langle u_\sR \lambda^{(\sR)}_\sA(\sigma_\sR)\right\rangle-i\left\langle v_\sR\kappa_{\sL\sA}\right\rangle\right)\,.\nonumber 
\end{align}
The spinors $\kappa_\sL$ encode the intermediate momentum $k_\sL=-k_\sR$, with \footnote{We have chosen a little-group frame where $\kappa_\sR=i\kappa_\sL$ to simplify the expression.}
\begin{equation}
 k_{\sL\,\sA\sB}=\lla \kappa_{\sL\sA}\kappa_{\sL\sB} \rra\,, \qquad k_{\sR\,\sA\sB}= -k_{\sL\,\sA\sB}=\la (i\kappa_{\sL\sA})(i\kappa_{\sL\sB}) \ra=: \la \kappa_{\sR\sA}\kappa_{\sR\sB} \ra\,,
\end{equation}
 and the integral fully localizes on the normalization conditions for $v_\sL$ and $v_\sR$, as well as the delta-functions  enforcing the  scattering equations at the node
\begin{subequations}\label{eq:SE_at_node}
\begin{align}
&\left\langle w_{\sL}\lambda^{(\sL)}_\sA(x_\sL) \right\rangle= \sum_{i\in L} \frac{\left\langle w_i w_\sL\right\rangle}{x_{i\sL}}\epsilon_{i\sA}=\lla v_\sL \kappa_{\sL\sA}\rra\,,\\
& \left\langle u_{\sR}\lambda^{(\sR)}_\sA(\sigma_\sR) \right\rangle = \sum_{p\in R} \frac{\left\langle u_p u_\sR\right\rangle}{\sigma_{p\sR}}\epsilon_{p\sA}=\left\langle v_\sR \kappa_{\sR\sA}\right\rangle=i\lla v_\sR \kappa_{\sL\sA}\rra  \,.
\end{align}
\end{subequations}
The little group-spinors $\epsilon_{\sL,\sR}^a$ relate $\kappa_{\sL\sA}^a$ to the previously defined  are defined objects $\epsilon_{\sL\sA}$ and $\epsilon_{\sR\sA}$ via $\epsilon_{\sL a}\kappa_{\sL\sA}^a=\epsilon_{\sL\sA}$ and $\epsilon_{\sR a}\kappa_{\sR\sA}^a=\epsilon_{\sR\sA}$. By directly comparing \eqref{eq:SE_at_node} to the definitions \eqref{eq:def_epsilon_R_final} and \eqref{eq:def_lambda_L}  of $\epsilon_{\sL\sA}$ and $\epsilon_{\sR\sA}$, we find that
\begin{align}
   \lla u_\sR \lambda^{(\sR)}_\sA(\sigma_\sR)\rra=-i\varepsilon^{-1/2} \,\lla u_\sR w_\sL \rra \epsilon{\sL\sA}\,, &&
  \lla w_\sL \lambda^{(\sL)}_\sA(x_\sL)\rra=-i\varepsilon^{-1/2} \,\lla u_\sR w_\sL \rra \epsilon{\sR\sA}\,.
\end{align}
so that the nodal scattering equations are indeed consistent with our previous definitions. Note that despite the factors of $\varepsilon^{-1/2}$, the right side is of order one due to $\la u_\sL w_\sR\ra \sim \varepsilon^{1/2}$. The nodal scattering equations thus imply that the variables $(\epsilon_\sL,v_\sL)$ and $(\epsilon_\sR,v_\sR)$ are related by
\begin{equation}\label{eq:v_L=e_R}
 v_\sL^a =\varepsilon^{-1/2}\la u_\sR w_\sL\ra \epsilon_\sR^a\,,\qquad v_\sR^a =-\varepsilon^{-1/2}\la u_\sR w_\sL\ra \epsilon_\sL^a\,,
\end{equation}
and so the integration over $v_\sL$ and $v_\sR$ should be understood as an integration over the polarization choices of the particle running through the cut propagator.

We can now combine the elaborate factor of 1 in  \eqref{eq:measure_1=} with the remaining part of the measure as follows. It evidently provides the missing factors for the $v$-dependent part of the measure and the polarized scattering equations to factorize correctly, as well as the missing measure $d\la u_\sL u_*^\perp\ra$ for the $u$-dependent part. Using a Schouten identity and dropping terms of subleading order in $\varepsilon$, we can further combine the factors $\la w_{i_1} u_{p_1} \ra  \la u_{p_1} u_*\ra$ from the measure and  $\la u_\sR w_\sL\ra \la w_\sL u_*\ra $ from \eqref{eq:measure_1=} to give the missing Jacobians for gauge-fixing the $u$'s and $w$'s on $\Sigma_{\sL,\sR}$,
\begin{equation}
 \la w_{i_1} w_\sL\ra  \la u_{p_1}  u_\sR \ra \la u_{p_1} u_*\ra \la w_\sL u_*\ra = J^u_{\sR} J^w_{\sL} \,.
\end{equation}
Combining everything, the $u$-part of the measure factorizes with the expected degeneration factor
\begin{equation}
\varepsilon^{-1} \la u_\sL w_\sR\ra \la w_\sR u_*\ra\, d\la u_{\sL} u_*^\perp\ra \,
\frac{\prod_{i=1}^n d^2u_i}{\mathrm{vol} \,\mathrm{SL}(2,\mathbb{C})_u} 
=
\frac{\varepsilon^{n_\ssL-1}}{\prod_{i\in L}x_{i\sL}^2}\,\,\frac{d^2 w_\sL \prod_{i\in \sL}d^2w_i}{\mathrm{vol} \,\mathrm{SL}(2,\mathbb{C})_w^\ssL} \,\frac{d^2 u_\sR \prod_{p\in R}^n d^2u_p}{\mathrm{vol} \,\mathrm{SL}(2,\mathbb{C})_u^\ssR} \,,
\end{equation}
and so the polarized measure $d\mu_n^{\mathrm{pol}}$  indeed factorizes as \eqref{eq:fact_pol-measure},
\begin{equation}
 d\mu_n^{\mathrm{pol}}= \frac{\varepsilon^{2\left(n_L-1\right)}}{\prod_{i\in L}x_{i\sL}^4}\frac{d\varepsilon}{\varepsilon}\,\frac{d^8\kappa_\sA^a}{\mathrm{vol}\,\mathrm{SL}(2,\mathbb{C})}\;d\mu_{n_L+1}^{\mathrm{pol}}\;d\mu_{n_R+1}^{\mathrm{pol}}\,.
\end{equation}

\subsubsection{Factorization of the integrands}\label{sec:fact-integrands}
\paragraph{Parke-Taylor factors and the reduced determinants.}
The Parke-Taylor factors factorize as usual; when all punctures $i\in L$ are consecutive in the colour-ordering $\alpha$, then
\begin{equation}
 \mathrm{PT}(\alpha) = \varepsilon^{-\left(n_\ssL-1\right)}\prod_{i\in L} x_{i\sL}^2\;\, \mathrm{PT}(\alpha_\sL) \mathrm{PT}(\alpha_\sR)\, ,
\end{equation}
where $\mathrm{PT}(\alpha_\sL)$ denotes the Parke-Taylor factor on the $\Sigma_\sL$, with the ordering $\alpha_\sL=\alpha\big|_\sL\cup {x_\sL}$. 
If the marked points $i\in L$ do not appear in a consecutive order in $\alpha$, the pole is of lower order of $\varepsilon$, and there is no factorization in this channel.\\

The factorization of the reduced determinant is similarly straightforward. On the boundary of the moduli space, its components  are given by
\begin{align}
&H_{ij}=\frac{x_{i\sL}x_{j\sL}}{\varepsilon}\frac{\epsilon_{i\sA}\epsilon_j^\sA}{x_{ij}} \,,& &H_{ip}=\frac{\epsilon_{i\sA}\epsilon_p^\sA}{\sigma_{\sR p}}\,,
&  H_{pi}=\frac{\epsilon_{p\sA}\epsilon_i^\sA}{\sigma_{p\sR}}\,,& &H_{pq}=\frac{\epsilon_{p\sA}\epsilon_q^\sA}{\sigma_{pq}} \,.
\end{align}
Due to the permutation invariance of the reduced determinant, we can make a convenient choice and  remove one row and column from each side, $i_1, i_2\in L$ and $p_1, p_2\in R$, 
\begin{equation}
 \det{}' H = \frac{\det H^{[i_1 p_1]}_{[i_2 p_2]}}{\la u_{i_1} u_{p_1}\ra [\tilde u_{i_2}\tilde u_{p_2}]}=\varepsilon^{-n_\ssL}\prod_{i\in L}x_{i\sL}^2\,\,\frac{\det H_\sL{}^{[i_1 \sL]}_{[i_2 \sL]}\,\det H_\sR{}^{[p_1 \sR]}_{[p_2 \sR]}}{\la w_{i_1} u_{p_1}\ra [\tilde w_{i_2}\tilde u_{p_2}]}\,.
\end{equation}
In the last step, we have identified the leading term in $\varepsilon$ as determinants of $H_\sL$ and $H_\sR$ respectively, with the rows and columns associated to $x_\sL$ and $\sigma_\sR$ removed. Using the Schouten identity $\la w_{i_1} u_{p_1} \ra  \la u_\sR w_\sL\ra  = \la w_{i_1} w_\sL\ra  \la u_{p_1}  u_\sR \ra$ (to leading order in $\varepsilon$), as well as the relations \eqref{eq:v_L=e_R}, the reduced determinant becomes
\begin{equation}\label{eq:fact_detH}
 \det{}' H =\varepsilon^{-\left(n_\sL-1\right)}\frac{1}{\la \epsilon_\sL\epsilon_\sR\ra[ \tilde \epsilon_\sL\tilde \epsilon_\sR]} \, \prod_{i\in L} x_{i\sL}^2 \,\, \det{}'H_\sL\,\det{}'H_\sR\,.
\end{equation}
To see that this is  the correct factorization behaviour for the bosonic case, let us compare \eqref{eq:fact_detH} to the sum over states. To implement this sum in our framework, we  introduce again a global basis for the little group space of the internal particle. With $\epsilon_\sL^a$ and $\epsilon_\sR^a$ as defined above, it is natural to choose the other basis elements (on each component sphere)
\begin{align}\label{eq:def_xi_L}
 \xi_\sL^a=\frac{\epsilon_\sR^a}{\la \epsilon_\sR\epsilon_\sL\ra}\,,\qquad\qquad \xi_\sR^a=\frac{\epsilon_\sL^a}{\la \epsilon_\sL\epsilon_\sR\ra}\,,
\end{align}
i.e. we choose the same basis $(\epsilon_\sL,\xi_\sL)$ (up to normalization constants) for both the left and the right component sphere. 
Consider now the amplitude $A_n:=A_n^{\epsilon_1\tilde\epsilon_1\dots}$ with all external particles in states at the top of the multiplet.\footnote{For readability, we suppress the $\epsilon$- indices for external particles below.} Then the sum over states reads
\begin{align}\label{eq:sum_states_bos_prelim}
 \sum_{\mathrm{states}}A_{n_\ssL+1}A_{n_\ssR+1} 
 &=\varepsilon_{ab} \varepsilon_{\dot a \dot b}\,A^{a\dot a}_{n_\ssL+1} A^{b\dot b}_{n_\ssR+1} 
 =\xi_{\sL[a}\epsilon_{\sL | b]} \, \tilde \xi_{\sL[\dot a}\tilde \epsilon_{\sL | \dot b]} \,A^{a\dot a}_{n_\ssL+1} A^{b\dot b}_{n_\ssR+1} \\
 & = \frac{A_{n_\ssL+1}^{\epsilon_\ssL \tilde \epsilon_\ssL} A_{n_\ssR+1}^{\epsilon_\ssR \tilde \epsilon_\ssR} }{\la \epsilon_\sL\epsilon_\sR\ra[ \tilde \epsilon_\sL\tilde \epsilon_\sR]}
 +\frac{\la \epsilon_\sL\epsilon_\sR\ra}{[ \tilde \epsilon_\sL\tilde \epsilon_\sR]} A_{n_\ssL+1}^{\xi_\ssL \tilde \epsilon_\ssL}  A_{n_\ssR+1}^{\xi_\ssR \tilde \epsilon_\ssR}
 +\frac{[ \tilde \epsilon_\sL\tilde \epsilon_\sR]}{\la \epsilon_\sL\epsilon_\sR\ra} A_{n_\ssL+1}^{\epsilon_\ssL \tilde \xi_\ssL}  A_{n_\ssR+1}^{\epsilon_\ssR \tilde \xi_\ssR}
 +\frac{\la \epsilon_\sL\epsilon_\sR\ra}{[ \tilde \epsilon_\sL\tilde \epsilon_\sR]} A_{n_\ssL+1}^{\xi_\ssL \tilde \xi_\ssL}  A_{n_\ssR+1}^{\xi_\ssR \tilde \xi_\ssR}\,.\nonumber
\end{align}
In the second equality, we have used the definition \eqref{eq:def_xi_L}  of the little group basis choice for the internal particle, and contracted the polarization data back into the amplitudes. While this does not yet look reminiscent of the factorization property \eqref{eq:fact_detH}, let us take a closer look at the amplitudes $A_{n_\ssL+1}^{\xi_\ssL\tilde\epsilon_\ssL}$ etc., arising from contracting $\xi_\sL$ or $\xi_\sR$ in the respective subamplitudes. Using either the supersymmetry representation or the results of \S\ref{sec:multiplet}, the (half-) integrand of these amplitudes is given by\footnote{The other integrands are $\mathcal{I}^{\mathrm{h}}=\det{}'H\,[ \tilde\xi_\sL \tilde v_\sL]$ for  $A_{n_\ssL+1}^{\epsilon_\ssL\tilde\xi_\ssL}$ and $\mathcal{I}^{\mathrm{h}}=\det{}'H\,\la \xi_\sL v_\sL\ra\,[ \tilde\xi_\sL \tilde v_\sL]$ for  $A_{n_\ssL+1}^{\xi_\ssL\tilde\xi_\ssL}$ respectively. } $\det{}'H\,\la \xi_\sL v_\sL\ra$. However, due to \eqref{eq:v_L=e_R}, $v_\sL^a = \xi_\sL^a$, and so all of these amplitudes vanish,
\begin{equation}
 A_{n_\ssL+1}^{\xi_\ssL\tilde\epsilon_\ssL}=A_{n_\ssL+1}^{\epsilon_\ssL\tilde\xi_\ssL}=A_{n_\ssL+1}^{\xi_\ssL\tilde\xi_\ssL} = 0\,,
\end{equation}
and similarly for $A_{n_\ssR+1}$. The sum over states thus simplifies drastically, and only the first term contributes,
\begin{align}\label{eq:sum_states_bos}
 \sum_{\mathrm{states}}A_{n_\ssL+1}A_{n_\ssR+1} 
 &=\frac1{\la \epsilon_\sL\epsilon_\sR\ra[ \tilde \epsilon_\sL\tilde \epsilon_\sR]}A_{n_\ssL+1}^{\epsilon_\ssL \tilde \epsilon_\ssL} A_{n_\ssR+1}^{\epsilon_\ssR \tilde \epsilon_\ssR} \,.
\end{align}
Thus the reduced determinant indeed factorizes as expected   for gluon amplitudes, c.f. \eqref{eq:fact_detH}.

\paragraph{The sum over states in the supersymmetry representation.} 
Before discussing factorization of the full supersymmetric amplitudes, let us first derive an expression for the sum over states as an integral over the fermionic variables of propagating particle.  For readability, we focus on the chiral case below, all statements extend straightforwardly to $\mathcal{N}=(N,\tilde N)$ supersymmetry.
In general, these fermionic integrals  take the form
\begin{equation}\label{eq:susy-fact_general}
 \mathcal{A}_n = \frac1{k_\sL^2}\int d^{2N} \!q_\sL d^{2N} \!q_\sR\;  \mathcal{A}_{n_\ssL+1}  \mathcal{A}_{n_\ssR+1}\;G(q_\sL,q_\sR)\,
\end{equation}
where  $G(q_\sL,q_\sR)$ is a `gluing factor' for the internal propagator that depends on the choice of supersymmetry representation, and 
is determined -- up to an overall normalization-- by supersymmetric invariance.
This can be seen as follows. The left hand side of \eqref{eq:susy-fact_general} vanishes under the  full supersymmetry generator $Q_{\sA\sI}$. 
Using further that
\begin{equation}
 Q_{\sA\sI} \mathcal{A}_{n_\ssL+1}  = -Q_{\sL\, \sA\sI} \mathcal{A}_{n_\ssL+1}\,,
 \qquad 
 Q_{\sA\sI} \mathcal{A}_{n_\ssR+1}  = -Q_{\sR\,\sA\sI} \mathcal{A}_{n_\ssR+1}\,,
 \qquad
 Q_{\sA\sI} \,G(q_\sL,q_\sR) = 0\,,
\end{equation}
due to the supersymmetric invariance of the amplitudes on the right, we find that $G$ has to satisfy
\begin{equation}\label{eq:cond_G}
 0=\int d^{4N} \!q_\sL d^{4N} \!q_\sR\;  \Big(\big(Q_{\sL\,\sA\sI}\mathcal{A}_{n_\ssL+1} \big) \mathcal{A}_{n_\ssR+1}+\mathcal{A}_{n_\ssL+1}  \big(Q_{\sR\,\sA\sI}\mathcal{A}_{n_\ssR+1}\big)\Big) \;G(q_\sL,q_\sR)\,.
\end{equation}
Using the explicit form of the supersymmetry representation \eqref{SUSY-factor}, we can easily verify that this is solved by\footnote{We can see this as follows. Using the explicit form of the supersymmetry representation,  the condition \eqref{eq:cond_G} contains two terms proportional to $\epsilon_\sL$ and $\epsilon_\sR$ respectively, 
$ C_\sL+C_\sR=0$, with
 \begin{align*}
 C_\sL&= \int d^{2N} \!q_\sL d^{2N} \!q_\sR\; e^{F_\ssL+F_\ssR}   \;G(q_\sL,q_\sR)\;  \epsilon_{\sR\sA}\left(-\la v_\sL v_\sR\ra q_{\sL \sI}+ i\sum_{p\in R}\frac{\la u_{\sR}u_{p}\ra}{\sigma_{\sR p}}q_{p\sI}\right)\\
 C_\sR&= \int d^{2N} \!q_\sL d^{2N} \!q_\sR\; e^{F_\ssL+F_\ssR} \;G(q_\sL,q_\sR)\; \epsilon_{\sL\sA}\left(i\la  v_\sL v_\sR\ra q_{\sR \sI}+ \sum_{i\in L}\frac{\la w_{\sL}w_{i}\ra}{x_{\sL i}} q_{i\sI}\right)\,.
 \end{align*}
Then we can straightforwardly integrate out $q_\sR$ in $C_\sL$ (and $q_\sL$ in $C_\sR$) using  the ansatz \eqref{eq:gluing-factor-susy} for $G$ and the vanishing of the local terms in the supersymmetry factors at the node $\la \xi_\sL v_\sL\ra=\la \xi_\sR v_\sR\ra =0$, 
and confirm that indeed $C_\sL =C_\sR=0$. }
\begin{equation}\label{eq:gluing-factor-susy}
 G(q_\sL,q_\sR) = \big|G(0,0)\big|\exp\left(\frac{ i\, q_{\sL\sI} q_{\sR\sJ}\Omega^{\sI\sJ}}{\la \epsilon_\sL  \epsilon_\sR\ra}\right)\,.
\end{equation}
To further fix the normalization $\big|G(0,0)\big|$, we compare the factorization for external gluons from \eqref{eq:susy-fact_general}  to the sum over states \eqref{eq:sum_states_bos}. In the notation $A_n:=A_n^{\epsilon_1\tilde\epsilon_1\dots}$, the fermionic integrals give
\begin{equation}\label{eq:sum-states-from-int}
 A_n=\,\frac1{k_\sL^2}\big|G(0,0)\big|\left( \frac1{ \la \epsilon_\sL\epsilon_\sR\ra^{2N}}\, A_{n_\ssL+1}^{\epsilon_\ssL}  A_{n_\ssR+1}^{\epsilon_\sR} +\dots+A_{n_\ssL+1}^{\xi_\sL}  A_{n_\ssR+1}^{\xi_\sR}\right)\,,
\end{equation}
where we used $A_{n_\ssL+1}^{\epsilon_\ssL}$ to indicate that the particle flowing through the on-shell propagator is in the top state of the chiral supersymmetry multiplet, parametrized by $\epsilon_\sL$. For the terms $A_{n_\ssL+1}^{\xi_\ssL} $ with the propagating particle at the bottom of the multiplet, we have used the consistency of the integrands with the supersymmetry representation, see \S\ref{sec:multiplet}.\footnote{As discussed above, these terms vanish if all external particles are in the top state of the multiplet.} By matching \eqref{eq:sum-states-from-int} to the sum over states \eqref{eq:sum_states_bos}, the normalization  is given by
\begin{equation}
 \big|G(0,0)\big| = \la \epsilon_\sL\epsilon_\sR\ra^N\,,
\end{equation}
and the fermionic integral representing the sum over states in the R-symmetry preserving supersymmetry representation takes the form
\begin{equation}\label{eq:susy-ampl-fact}
 \mathcal{A}_n = \frac{1}{k_\sL^2}\int d^{2N} \!q_\sL d^{2N} \!q_\sR\;  \mathcal{A}_{n_\ssL+1}  \mathcal{A}_{n_\ssR+1}\;\left(\la \epsilon_\sL \epsilon_\sR\ra^{N}\,e^{i\la \epsilon_\ssL \epsilon_\ssR\ra^{-1} \,q_{\sL\sI}q_{\sR\sJ}\Omega^{\sI\sJ}}\right)\,.
\end{equation}

\paragraph{Factorization of the supersymmetry factors.}
Given that the measure and the integrands factorize correctly, we can isolate the supersymmetry factors in the relation \eqref{eq:susy-ampl-fact}. To prove that the superamplitudes factorize correctly, we thus need to show that at the boundary of the moduli space
\begin{equation}\label{eq:susy-fact}
 e^{F}\Big|_{\partial\mathfrak{M}} \stackrel{!}{=}\la \epsilon_\sL \epsilon_\sR\ra^{2N} \int d^{2N} \!q_\sL d^{2N} \!q_\sR\;  e^{F_\ssL+F_\ssR}\;e^{i\la \epsilon_\ssL \epsilon_\ssR\ra^{-1} \,q_{\sL\sI}q_{\sR\sJ}\Omega^{\sI\sJ}}\,,
\end{equation}
Our strategy will be to first calculate the left side of this equation, and then simplify the right to see that they match. On the left, the parametrization of $(\sigma_i,u_i)$ on the boundary gives
\begin{equation}\label{eq:F_fact-form}
 F\Big|_{\partial\mathfrak{M}}=\underbrace{\frac{1}{2}\sum_{i,j\in L} \frac{\left\langle w_i w_j\right\rangle}{x_{ij}}q_{i\sI}q_{j\sJ}\Omega^{\sI\sJ}}_{:=\hat F_\sL}+\underbrace{\frac{1}{2}\sum_{p,q\in R} \frac{\left\langle u_p u_q\right\rangle}{\sigma_{pq}}q_{p\sI}q_{q\sJ}\Omega^{\sI\sJ}}_{:=\hat F_\sR}+\sum_{\substack{i\in L\\ p\in R}} \frac{\left\langle u_p u_i\right\rangle}{\sigma_{p\sR}}q_{i\sI}q_{p\sJ}\Omega^{\sI\sJ}\,.
\end{equation}
Here, we have introduced the factors $\hat F_\sL$ and $\hat F_\sR$ for later convenience.\footnote{The `hat'-notation is intended as a reminder that these are not yet the factors $F_\sL$ and $F_\sR$ for the subamplitudes since they do not include the contributions from the junction point.} On the right hand side, we can integrate  out $q_\sL$ and $q_\sR$, 
\begin{align}\label{eq:fact_susy_intermediate}
 \la \epsilon_\sL \epsilon_\sR\ra^{2N}& e^{\hat F_\sL+\hat F_\sR}\int d^{2N} \!q_\sL \;  \prod_I \delta\!\left( i \la v_\sL  v_\sR\ra\, q_{\sL\sI}+  \sum_{p\in R}\frac{\la u_{\sR}u_{p}\ra}{\sigma_{\sR p}}q_{p\sI}\right)\;\exp\left(\sum_{i\in L}\frac{\la w_i w_\sR\ra}{x_{i\sL}}q_{i\sI}q_{\sL\sJ}\Omega^{\sI\sJ}\right)\nonumber\\
 &=e^{\hat F_\sL+\hat F_\sR}\exp\left(-\varepsilon^{-1/2}\la \epsilon_\sL\epsilon_\sR\ra\sum_{\substack{i\in L\\p 
 \in R}}\frac{\la u_i w_\sL\ra\la u_{\sR}u_{p}\ra}{\sigma_{\sR p}}q_{i\sI}q_{p\sJ}\Omega^{\sI\sJ}\right)\,,
\end{align}
where,  we have used that $\exp(\la \xi_\sR v_\sR \ra q_\sR^2)=1$ due to $v_\sR=\xi_\sR$.
To simplify the exponent in the last line, we use a Schouten identity and the relations \eqref{eq:v_L=e_R} for the polarization spinors of the propagating particle  to obtain to leading order
\begin{equation}
 \la u_i w_\sL\ra\la u_{\sR}u_{p}\ra=-\la u_i u_p\ra\la w_\sL u_{\sR}\ra+\la u_i u_\sR\ra\la u_{p} w_\sL\ra = -\varepsilon^{1/2}\frac{\la u_i u_p\ra}{\la \epsilon_\sL\epsilon_\sR\ra}+O\left(\varepsilon^{3/2}\right)\,.
\end{equation}
The exponent thus agrees with \eqref{eq:F_fact-form}, and so our formulae \eqref{eq:ampl_susy} factorize as expected of amplitudes in super Yang-Mills theory and supergravity.\\

As an aside, we give an alternative way of deriving the factorization of the supersymmetry factors that mirrors the bosonic discussion of the polarized scattering equations more closely. First, note that the delta-functions in the first line of \eqref{eq:fact_susy_intermediate} can be solved  in  analogy to the bosonic case \eqref{eq:def_epsilon_L_deriv} by
\begin{subequations}
\begin{align}
 u_\sR^a\, q_{\sR\sI}&=\sum_{i\in L}u_i^a\, q_{i\sI}+\theta_{\sR\sI}^a\,,&&\la w_\sL \theta_{\sR\sI}\ra=0\,,\\
 w_\sL^a\, q_{\sL\sI}&=\sum_{p\in R}w_p^a\, q_{p\sI}+\theta_{\sL\sI}^a\,,&&\la u_\sR \theta_{\sL\sI}\ra=0\,.
\end{align}
\end{subequations}
Here the $2N$ conditions imposed by the delta-functions  have been replaced by  $4N$ constraints, but supplemented by $2N$  degrees of freedom encoded in  $\theta_\sL$ and $\theta_\sR$. We can now solve the constraints $\la w_\sL \theta_{\sR\sI}\ra=0$ by taking $\theta_{\sR\sI}^a = \alpha_\sR\, w_\sL^a\theta_{\sR\sI}$, and similarly for $\theta_\sL$. For convenience, we have defined $\theta_\sR$ to be of order one, and kept a normalization factor $\alpha_\sR$ explicit. Contracting the resulting relations into $u_i$ (or $w_p$ respectively) gives the dominant balance $\alpha_\sR = \varepsilon^{1/2}$, and so we are  left with
\begin{align}
 u_\sR^a\, q_{\sR\sI}&=\sum_{i\in L}u_i^a\, q_{i\sI}+\varepsilon^{1/2}\,w_\sL^a\theta_{\sR\sI}\,,\qquad
 w_\sL^a\, q_{\sL\sI}=\sum_{p\in R}w_p^a\, q_{p\sI}+\varepsilon^{1/2}\,u_\sR^a\theta_{\sL\sI}\,,
\end{align}
on support of the delta-functions. The exponent then directly gives the correct factorization \eqref{eq:susy-fact}.

\paragraph{Factorization of $\pf{}'A$ and the M5 half-integrand $\cI_{\mathrm{M5}}^{\mathrm{h}}$.}  While the brane theories are not known to satisfy a BCFW recursion, the above treatment of the intergands can be extended easily to prove that the M5 and D5 amplitudes factorize correctly. It would be interesting to extend this to a full soft recursion as introduced in \cite{Cheung:2015ota}, but this is beyond the scope of this paper. \\

Let us first consider the Pfaffian $\pf{}'A$. On a boundary $\partial_{\sL,\sR}\mathfrak{M}_{0,n}$, the matrix entries become
\begin{equation}
 A_{ij} = \frac{x_{i\sL}x_{j\sL}}{\varepsilon}\,\frac{k_i\cdot k_j}{x_{ij}}\,,\qquad A_{ip} = \frac{k_i\cdot k_p}{\sigma_{Rp}}\,,\qquad A_{ip} = \frac{k_p\cdot k_q}{\sigma_{pq}}\,.
\end{equation}
If $n_\sL$, $n_\sR$ are odd (so the subamplitudes $\cA_{n_\ssL+1}$ and $\cA_{n_\ssR+1}$ have an even number of particles), it is convenient to define $\pf{}'A$ by reducing on $i\in L$, $p\in R$. Since the block-matrix proportional to $\varepsilon^{-1}$ is of even rank $n_\sL -1$, the reduced Pfaffian then factorizes as
\begin{equation}
 \pf{}'A = \frac{(-1)^{i+p}}{\sigma_{ip}}\pf A^{[ip]}=\varepsilon^{-\frac{1}{2}\left(n_\ssL-1\right)}\prod_{j\in L}x_{j\sL}\;\pf{}' A_\sL \;\pf{}' A_\sR\,.
\end{equation}
Here, the powercounting  of $\varepsilon$ is due to the removed row and column $i\in L$. 

On the other hand, if $n_\sL$, $n_\sR$ are even, i.e. we are studying factorization channels into subamplitudes with an odd number of particles, it is still convenient to reduce on $i\in L$, $p\in R$ to avoid leading-order cancellations. In contrast to the odd case however, the factorization now involves a sum over states as shown in \cite{Dolan:2013isa}, and the leading order term gives  $\pf{}'A \sim \varepsilon^{-\left(\frac{n_\ssL}{2}-1\right)}$. For amplitudes with half-integrand  $\mathcal{I}_n^{\mathrm{h}}=\det{}'A$, there are thus no factorization channels with odd-point subamplitudes, and for $n_\sL$ even, we indeed find
\begin{equation}
 \det{}'A = \varepsilon^{-\left(n_\ssL-1\right)}\prod_{j\in L}x_{j\sL}^2\;\det{}' A_\sL \;\det{}' A_\sR \,,
\end{equation}
as expected for half-integrands.\\

The calculation of the factorization of $\pf U^{(2,0)}$ featuring in the M5 half-integrand is  more involved due to the structure of its entries, and we have delegated the discussion to  appendix \cref{sec:fact_U}. The final property for odd $n_\sL$ however is very compact,
\begin{equation}
 \pf U^{(2,0)} =\varepsilon^{\frac{n_\ssL - 1}{2}}\, \frac{\la \epsilon_\sL \epsilon_\sR\ra^2}{\prod_{j\in L}x_{j\sL}} \, \pf U^{(2,0)}_\sL\,\pf U^{(2,0)}_\sR\,,
\end{equation}
and gives the following factorization of the M5 half-integrand,
\begin{equation}
 \mathcal{I}_{\mathrm{M5}}^{\mathrm{h}} = \varepsilon^{-\left(n_\ssL - 1\right)}\frac{\prod_{j\in L}x_{j\sL}^2}{\la \epsilon_\sL \epsilon_\sR\ra^2}\;\mathcal{I}_{\mathrm{M5},\sL}^{\mathrm{h}} \,\mathcal{I}_{\mathrm{M5},\sR}^{\mathrm{h}} \,.\label{eq:fact_M5}
\end{equation}
Repeating the arguments used in the factorization of the reduced determinant $\det{}'H$, the only non-vanishing contribution to sum over states comes from the top of the multiplet, in agreement with \eqref{eq:fact_M5}.  We thus conclude that the brane amplitudes also factorize correctly.\\

As discussed above, for the brane theories factorization into odd-point subamplitudes is ruled out by the presence of $\det{}'A$ in the integrand. On the other hand, the novel  formulae in the web of theories in \cref{tabletheories} are composed of  $\mathcal{I}_{\mathrm{M5}}^{\mathrm{h}}$ with another half-integrand that supports factorization channels with odd-particle subamplitudes (such as the Parke-Taylor factor for the $(2,0)-$PT formulae). From this perspective, we would also like to study the factorization of $\mathcal{I}_{\mathrm{M5}}^{\mathrm{h}}$ for even $n_\sL $. A straightforward counting shows that in that case $\pf U^{(2,0)}\sim \varepsilon^{\frac{n_\ssL}{2}}$, so $\mathcal{I}_{\mathrm{M5}}^{\mathrm{h}}$ does give a non-zero contribution to factorization channels with even $n_\sL$. While it would be interesting to pursue this further to gain some insights into the $(2,0)-$PT formulae, or construct odd-particle versions, this is beyond the scope of this paper.

\subsection{Boundary terms}\label{sec:bdy}
As we have seen in \S\ref{sec:factorization} and 
\S\ref{sec:3pt}, the  formulae \eqref{eq:ampl_susy} based on the polarized scattering equations factorize correctly, and reproduce the correct three-particle Yang-Mills and gravity amplitudes. To demonstrate that they satisfy the BCFW recursion relation -- and are thus representations of the tree-level amplitude -- we still need to show that  the boundary terms in the BCFW recursion relation vanish,
\begin{equation}
 \lim_{z\rightarrow \infty}\mathcal{A}(z)=0\,.
\end{equation}
We will follow a similar strategy to the one employed in the discussion of factorization, and discuss first how the polarized scattering equations and the measure behave under the BCFW deformation \eqref{eq:BCFW_shift} and \eqref{eq:BCFW_antichiral},
\begin{subequations}
\begin{align}
\hat{\kappa}_{1\sA}^a&=\kappa_{1\sA}^a+ z\, \epsilon_1^a\, \epsilon_{n\sA}\,, 
& \hat\kappa_{1}{}^\sA_{\dot a} & = \kappa_{1}{}^\sA_{\dot a}-z\,\tilde \epsilon ^\sA \left(\epsilon_{n\,\sB}\kappa_1{}^\sB_{\dot a}\right)\,,\\
  \hat{\kappa}_{n\sA}^a&=\kappa_{n\sA}^a+ z \,\epsilon_n^a\,\epsilon_{1\sA}\,, & 
  \hat\kappa_{n}{}^\sA_{\dot a} & = \kappa_{n}{}^\sA_{\dot a}-z\,\tilde \epsilon ^\sA \left(\epsilon_{1\,\sB}\kappa_n{}^\sB_{\dot a}\right)\,.
\end{align}
\end{subequations}
As expected from the equivalence of the polarized measure $d\mu^{\mathrm{pol}}_n$ and the CHY-measure $d\mu^{\CHY}_n$, we find that the measure scales as $z^{-2}$,
\begin{equation}
 \lim_{z\rightarrow \infty}d\mu^{\mathrm{pol}}_n=z^{-2}d\tilde\mu^{\mathrm{pol}}_n\,,
\end{equation}
and thus only integrands scaling at most as $\mathcal{I}_n\sim z$ as $z\rightarrow\infty$ can give vanishing boundary terms. In the case of super Yang-Mills theory and supergravity, we find that $e^{F+\tilde F}\sim z^0$, and $\det'{}H\sim z^0$ while $\mathrm{PT}(\alpha)\sim z$ for colour-ordered partial amplitudes where the shifted particles 1 and $n$ are adjacent, and $\mathrm{PT}(\alpha)\sim z^0$ otherwise.  Putting this together, the supergravity and super Yang-Mills expressions scale as 
\begin{equation}
 \mathcal{M}(z)\sim z^{-2}\,,\qquad\qquad \mathcal{A}(z)\sim z^{-1}\,,
\end{equation}
in the large-$z$ limit, so the boundary terms vanish in both cases. We conclude that the formulae based on the polarized scattering equations  satisfy the BCFW recursion relation, and thus give representations of the respective tree-level amplitudes.

\subsubsection{The polarized scattering equations}
\paragraph{Polarized scattering equations and measure.} A crucial feature of the BCFW deformation of the fundamental spinors is that it leaves the polarization spinors $\epsilon_{1,n}$ of the shifted particles invariant. The polarized scattering equations  are thus unaffected for all particles $i\neq 1,n$, and become
\begin{subequations}
 \begin{align}
  \hat{\mathcal{E}}_i& = \sum_j \frac{\left\langle u_i u_j\right \rangle}{\sigma_{ij}}\epsilon_{j\sA}  -\left\langle v_i\kappa_{i\sA}\right\rangle
  \,, \\
  \hat{\mathcal{E}}_1& = \sum_{j\neq n} \frac{\left\langle u_1 u_j\right \rangle}{\sigma_{1j}}\epsilon_{j\sA}  -\left\langle v_1\kappa_{1\sA}\right\rangle + \left(\frac{\left\langle u_1 u_n\right\rangle}{\sigma_{1n}} - z \right)\epsilon_{n\sA}\,, \\
  \hat{\mathcal{E}}_n& = \sum_{j\neq 1} \frac{\left\langle u_n u_j\right \rangle}{\sigma_{nj}}\epsilon_{j\sA}  -\left\langle v_n\kappa_{n\sA}\right\rangle + \left(\frac{\left\langle u_1 u_n\right\rangle}{\sigma_{1n}} - z \right)\epsilon_{1\sA}\,.
 \end{align}
\end{subequations}
In the large-$z$ limit, the scattering equations $\mathcal{E}_1$ and $\mathcal{E}_n$  require that $\sigma_{1n}\sim z^{-1}$ while $\left\langle u_1 u_n\right\rangle\sim 1$ remains of order one.  We can refine this dominant balance by explicitly solving for the difference $\sigma_{n1}=z^{-1}\langle u_n u_1\rangle$ to leading order, which suggests the following change of variables:
\begin{equation}\label{eq:dom_bal_sigma}
  \sigma_{n}= \sigma_1+z^{-1} \langle u_n u_1\rangle + z^{-2}y_n\,,
\end{equation}
The shifted polarized scattering equations are indeed manifestly independent of $z$ when expressed in terms of the variables $\sigma_1$ and $y_n$,\footnote{We have omitted higher order terms in $z^{-1}$ in $\hat{\mathcal{E}}_i$ and $\hat{\mathcal{E}}_n$.}
\begin{subequations}\label{eq:shifted_pol_SE}
 \begin{align}
  \hat{\mathcal{E}}_i& = \sum_{j\neq 1,n} \frac{\left\langle u_i u_j\right \rangle}{\sigma_{ij}}\epsilon_{j\sA}  +\frac{1}{\sigma_{i1}}\big(\left\langle u_i u_1\right \rangle\epsilon_{1\sA}+\left\langle u_i u_n\right \rangle\epsilon_{n\sA}\big) -\left\langle v_i\kappa_{i\sA}\right\rangle
  \,, \\
  \hat{\mathcal{E}}_1& = \sum_{j\neq n} \frac{\left\langle u_1 u_j\right \rangle}{\sigma_{1j}}\epsilon_{j\sA}  -\left\langle v_1\kappa_{1\sA}\right\rangle + \frac{y_n}{\left\langle u_1 u_n\right \rangle^2}\,\epsilon_{n\sA}\,, \\
  \hat{\mathcal{E}}_n& = \sum_{j\neq 1} \frac{\left\langle u_n u_j\right \rangle}{\sigma_{1j}}\epsilon_{j\sA}  -\left\langle v_n\kappa_{n\sA}\right\rangle + \frac{y_n}{\left\langle u_1 u_n\right \rangle^2}\,\epsilon_{1\sA}\,.
 \end{align}
\end{subequations}
Let us define a new polarized measure $d\tilde  \mu^{\mathrm{pol}}_n$ in analogy to \eqref{measure6d}, but now using the $z$-independent scattering equations \eqref{eq:shifted_pol_SE} as well as the new variable $y_n$ specifying the marked point $\sigma_n$. Then the shifted measure $d\hat{\mu}^{\mathrm{pol}}_n$ obeys
\begin{equation}
  \lim_{z\rightarrow \infty}d\hat{\mu}^{\mathrm{pol}}_n=z^{-2}d\tilde\mu^{\mathrm{pol}}_n\,,
\end{equation}
due to $d\sigma_n=z^{-2}\,dy_n$. This makes is clear that only theories with integrands scaling at most as $\mathcal{I}_n\sim z$ for large $z$ will have vanishing boundary terms in the BCFW recursion relation.

\paragraph{Anti-chiral scattering equations.} While the anti-chiral equivalent to the polarized scattering equations does not play a prominent role in the amplitude expressions, we will need the behaviour of the variables $\tilde u_i^{\dot a}$ to determine the scaling behaviour of the integrands. On support of the chiral polarized scattering equations, the marked points $\sigma_1$ and $\sigma_n$ factorize in the large-$z$ limit,
\begin{equation}
\sigma_{n}= \sigma_1+z^{-1} \langle u_n u_1\rangle + z^{-2}y_n\,.
\end{equation}
Using this, the anti-chiral scattering equations are given to order $\mathcal{O}(z)$ by
\begin{align*}
\hat{\mathcal{E}}_i\Big|_{\mathcal{O}(z)} &= -z\left( \frac{[\tilde u_i\tilde u_1]}{\sigma_{i1}}\left(\epsilon_n{}_\sC\, \epsilon_1{}^\sC\right) 
    + \frac{[\tilde u_i\tilde u_n]}{\sigma_{i1}}\left(\epsilon_1{}_\sC\, \epsilon_n{}^\sC\right) \right)\tilde\epsilon ^\sA\\
  \hat{\mathcal{E}}_1\Big|_{\mathcal{O}(z)} &=z \frac{[\tilde u_1\tilde u_n]}{\langle u_1 u_n\rangle}\epsilon_n^A +z\,y_n \frac{[\tilde u_1\tilde u_n]}{\langle u_1 u_n\rangle^2} \left(\epsilon_1{}_\sC\, \epsilon_n{}^\sC\right) \tilde\epsilon ^\sA-z^2 \frac{[\tilde u_1\tilde u_n]}{\langle u_1 u_n\rangle} \left(\epsilon_1{}_\sC\, \epsilon_n{}^\sC\right) \tilde\epsilon ^\sA +z \left(\epsilon_n{}_\sC\, [v_1\kappa_1{}^\sC]\right) \tilde\epsilon ^\sA  \\
  \hat{ \mathcal{E}}_n\Big|_{\mathcal{O}(z)} &=z \frac{[\tilde u_n\tilde u_1]}{\langle u_n u_1\rangle}\epsilon_1^A+z\,y_n \frac{[\tilde u_n\tilde u_1]}{\langle u_n u_1\rangle^2} \left(\epsilon_n{}_\sC\, \epsilon_1{}^\sC\right) \tilde\epsilon ^\sA  -z^2 \frac{[\tilde u_n\tilde u_1]}{\langle u_n u_1\rangle} \left(\epsilon_n{}_\sC\, \epsilon_1{}^\sC\right) \tilde\epsilon ^\sA +z \left(\epsilon_1{}_\sC\, [v_n\kappa_n{}^\sC]\right) \tilde\epsilon ^\sA \,.
\end{align*}
Due to the terms proportional to $z^2$ as well as the different spinors in $\hat{ \mathcal{E}}_1$ and $\hat{ \mathcal{E}}_n$, the only dominant balance for this set of equations is  $[\tilde u_1\tilde u_n]\sim z^{-1}$. 
We will parametrize this balance by 
\begin{equation}
 \tilde u_n^{\dot a} = \frac{[\tilde w_n \tilde u_n]}{[\tilde w_n \tilde u_1]}\tilde u_1^{\dot a} +z^{-1} \tilde w_n^{\dot a}\,.
\end{equation}
Using this, the anti-chiral polarized  scattering equations simplify to
\begin{subequations}
\begin{align}
      \hat{\mathcal{E}}_i\Big|_{\mathcal{O}(z)} &= -z\frac{[\tilde u_i\tilde u_1]}{\sigma_{i1}}\left( \left(\epsilon_n{}_\sC\, \epsilon_1{}^\sC\right) 
    +\frac{[\tilde w_n \tilde u_n]}{[\tilde w_n \tilde u_1]}\left(\epsilon_1{}_\sC\, \epsilon_n{}^\sC\right) \right)\tilde\epsilon ^\sA\,,\\
  \hat{\mathcal{E}}_1\Big|_{\mathcal{O}(z)} &=z\left(  -\frac{[\tilde u_1\tilde w_n]}{\langle u_1 u_n\rangle} \left(\epsilon_1{}_\sC\, \epsilon_n{}^\sC\right) + \left(\epsilon_n{}_\sC\, [v_1\kappa_1{}^\sC]\right) \right)\tilde\epsilon ^\sA \label{eq:anti-chiral-BCFW_E1} \\
   \hat{\mathcal{E}}_n\Big|_{\mathcal{O}(z)} &=z \left(-\frac{[\tilde u_1\tilde w_n]}{\langle u_1 u_n\rangle} \left(\epsilon_n{}_\sC\, \epsilon_1{}^\sC\right) + \left(\epsilon_1{}_\sC\, [v_n\kappa_n{}^\sC]\right)\right) \tilde\epsilon ^\sA \label{eq:anti-chiral-BCFW_En}
\end{align}
\end{subequations}
Together with the normalization condition $[v_1 \epsilon_1]=1$, the leading order of $\hat{\mathcal{E}}_1$  determines $v_{1}$ to order one; in other words we can set $v_{1}=v^*_{1}+z^{-1}\tilde v_{1}$ where $\mathcal{E}_1\big|_{\mathcal{O}(z)}(v_1^*)=0$, and similarly for $v_n$.
All remaining scattering equations $\hat{\mathcal{E}}_i$ are solved to leading order by
\begin{equation}\label{eq:tilde_u_n_large-z}
 \tilde u_n^{\dot a} =- \frac{\left(\epsilon_n{}_\sB\, \epsilon_1{}^\sB\right) }{\left(\epsilon_1{}_\sC\, \epsilon_n{}^\sC\right) }\tilde u_1^{\dot a} +z^{-1} \tilde w_n^{\dot a}\,.
\end{equation}
Changing variables to $\{\sigma_i,\tilde u_i^{\dot a},v_i^{\dot a}\}$ for $i\neq 1,n$ and  $\{\sigma_1, \tilde u_1^{\dot a},\tilde v_1^{\dot a}\}$ and $\{ y_n, \tilde w_n^{\dot a},\tilde v_n^{\dot a}\}$ thus renders the anti-chiral scattering equations manifestly independent of $z$ as $z\gg 1$.

\subsubsection{Supersymmetry}
As discussed in \cref{sec:BCFW-shift}, in the R-symmetry preserving supersymmetry representation the supershift is implemented via  multiplication by an exponential factor
\begin{equation}\label{eq:shift-ferm-fact_v2}
 e^{F+\tilde F}\rightarrow e^{\hat F+\tilde F}=e^{F-z\,q_{1\sI}q_{n\sJ}\Omega^{\sI\sJ}}\,e^{\tilde F}\,,
\end{equation}
rather than a linear shift in the fermionic variables. From the solutions to the antifundamental polarized scattering equations \eqref{eq:tilde_u_n_large-z}, it is easily checked that $\tilde F$ remains of order one in the limit  $z\rightarrow \infty$, so the only $z$-dependent term is proportional to  $\la u_1 u_n\ra /\sigma_{1n} -z$. On the support of the polarized scattering equations \eqref{eq:dom_bal_sigma}, this combination remains of order one, and as a consequence, so does $\hat F$;
\begin{equation}
 \hat F =\frac1{2}\sum_{\substack{i,j\\i,j\neq n}}\frac{\la u_i u_j\ra}{\sigma_{ij}}q_{i\sI}q_{j\sJ}\Omega^{\sI\sJ}+\sum_{i\neq 1}\frac{\la u_i u_n\ra}{\sigma_{i1}}q_{i\sI}q_{n\sJ}\Omega^{\sI\sJ}+\frac{y_n}{\la u_1 u_n\ra^2}q_{1\sI}q_{n\sJ}\Omega^{\sI\sJ}\,.
\end{equation}
The supersymmetry factors are thus of order one in the large-$z$-limit, $e^{F+\tilde F}\sim z^0$. Alternatively, this can be seen from the little-group preserving representation, where the fermionic-delta functions \eqref{eq:delta-rep} and the shift \eqref{eq:shift_eta}  manifestly mirror the polarized scattering equations. As $z\rightarrow\infty$, the same argument as for the polarized scattering equations thus guarantees that the delta-functions remain of order one.

\subsubsection{The integrand}
\paragraph{The Parke-Taylor factor. } The large-$z$ limit for the colour half-integrand $\mathrm{PT}(\alpha)$ is familiar from the original $d$-dimensional CHY amplitude representation. Since the Parke-Taylor factor only depend on the moduli of the marked Riemann sphere, its behaviour as $z\rightarrow\infty$ is determined by \eqref{eq:dom_bal_sigma}.
\begin{equation}
  \sigma_{n}= \sigma_1+z^{-1} \langle u_n u_1\rangle + z^{-2}y_n\,.
\end{equation}
For colour-ordered Parke-Taylor factors, we thus find
\begin{equation}
 \widehat{\mathrm{PT}}(\alpha)\equiv\prod_{i=1}^n\frac{1}{\sigma_{\alpha(i)\alpha(i+1)}} \sim\begin{cases}z & \alpha^{-1}(1)=\alpha^{-1}(n)\pm1\,,
                                                    \\ 1 &\text{otherwise}\,,\end{cases}
\end{equation}
so the colour half-integrands are of order $z$ if the legs $1$ and $n$ are adjacent in the colour-ordering $\alpha$ and of order $z^0$ otherwise.

\paragraph{The reduced determinant.} In contrast to the Parke-Taylor factor, the reduced determinant  $\det{}'\hat H$ depends on $z$ not only via the marked points  $\sigma_{1n}\sim z^{-1}$, but also via the anti-chiral spinors $\hat \epsilon_1^\sA$ and $\hat \epsilon_n^\sA$. There is however no chiral  contribution of order $z$ since $\hat \epsilon_{1\,\sA}= \epsilon_{1\,\sA}$ and $\hat \epsilon_{n\,\sA}= \epsilon_{n\,\sA}$, and so all $z$-dependence stems from the columns 1 and $n$,
\begin{subequations}
\begin{align}
 \hat H_{i1} &= -z \frac{\epsilon_{i\,\sA}\tilde\epsilon^\sA}{\sigma_{i1}}\left(\epsilon_{n\,\sB}\epsilon_1^\sB\right) + \frac{\epsilon_{i\,\sA}\epsilon_1^\sA}{\sigma_{i1}}\,, \\
  \hat H_{in} &= -z \frac{\epsilon_{i\,\sA}\tilde\epsilon^\sA}{\sigma_{i1}}\left(\epsilon_{1\,\sB}\epsilon_n^\sB\right) + \frac{\epsilon_{i\sA}\epsilon_n^\sA}{\sigma_{i1}}-\frac{\epsilon_{i\,\sA}\tilde\epsilon^\sA}{\sigma_{i1}^2}\left(\epsilon_{1\,\sB}\epsilon_n^\sB\right) \langle u_nu_1\rangle\,.
\end{align}
\end{subequations}
The entries $\hat H_{1n}$, $\hat H_{n1}$ as well as the diagonal entries $\hat{H}_{11}$ and $\hat{H}_{nn}$ depend quadratically on $z$, and we find to subleading order
\begin{subequations}
\begin{align}
 \hat H_{1n}& =- z^2\,\frac{\epsilon_{1\,\sA}\tilde\epsilon^\sA}{\langle u_nu_1\rangle}\left(\epsilon_{1\,\sB}\epsilon_n^\sB\right)+z\,y_n\,\frac{\epsilon_{1\,\sA}\tilde\epsilon^\sA}{\langle u_nu_1\rangle^2}\left(\epsilon_{1\,\sB}\epsilon_n^\sB\right)+z\,\frac{\epsilon_{1\,\sA}\epsilon_n^\sA}{\langle u_nu_1\rangle}\,,\\
 \hat H_{n1}& = +z^2\,\frac{\epsilon_{n\,\sA}\tilde\epsilon^\sA}{\langle u_nu_1\rangle}\left(\epsilon_{n\,\sB}\epsilon_1^\sB\right)-z\,y_n\,\frac{\epsilon_{n\,\sA}\tilde\epsilon^\sA}{\langle u_nu_1\rangle^2}\left(\epsilon_{n\,\sB}\epsilon_1^\sB\right)-z\,\frac{\epsilon_{n\,\sA}\epsilon_1^\sA}{\langle u_nu_1\rangle}\,,
\end{align}
\end{subequations}
which uniquely determines $\hat H_{11}$ and $\hat H_{nn}$ from linearity relations \eqref{eq:co-rank} among the columns of $\hat H$. We remark that all remaining diagonal entries $\hat H_{ii}$ are independent\footnote{Here and below, independence of $z$ refers to the large-$z$ limit, and thus only entails independence to order $z^0$, with possible contributions of order $z^{-1}$ that vanish as $z\rightarrow\infty$.} of $z$, as can be seen from the (row) linearity relation
\begin{equation}\label{eq:BCFW_def_H_ii}
 \langle u_i u_n\rangle \hat H_{ii} = -\sum_{j\neq n,i} \langle u_j u_n\rangle \hat H_{ji}= -\sum_{j\neq n,i} \langle u_j u_n\rangle  H_{ji}\,,
\end{equation}
which is manifestly of order one. All $z$-dependence of $\hat H$ is thus confined to the columns $1$ and $n$, suggesting that we define the reduced determinant by removing these columns. Naively this would imply $\det{}'\hat H\sim z$ because of the denominator factor  $[\tilde u_1 \tilde u_n]=z^{-1}[\tilde u_1 \tilde w_n]$, but its coefficient  vanishes, as can be seen from a judicious choice of row and column operations on $\hat H$.\footnote{Recall from \cref{lemma:red-det} that the reduced determinant is invariant  under row and column operations.} In practice, however, it is easier to extract the large-$z$ behaviour by using row- and column operations to remove the $z$-dependence from one of the two columns, say column 1, and reduce on a different column.

To make this explicit, let us construct a new matrix $\hat H'$ whose column 1 is independent of $z$ (apart from $\hat H'_{11}$ and $\hat H'_{n1}$, which will still be removed),
\begin{equation}
 \hat H'_{i1} = \hat H_{i1}-\frac{\epsilon_{n\,\sB}\epsilon_1^\sB}{\epsilon_{1\,\sB}\epsilon_n^\sB}\hat H_{in}\,,\qquad\qquad  \tilde w'{}^{\dot a}_n = \tilde u^{\dot a}_n +  \frac{\epsilon_{n\,\sB}\epsilon_1^\sB}{\epsilon_{1\,\sB}\epsilon_n^\sB} \tilde u^{\dot a}_1\,.
\end{equation}
Due to \cref{lemma:red-det}, the reduced determinants agree,  $\det{}' H'=\det{}' H$, and in particular so do their large-$z$-limits.
But by construction, $H'$ only depends on $z$ via the $n$-th column and the entries $H_{1n}$ and $H_{nn}$, so we can manifestly remove all dependence on $z$ by reducing on the rows 1 and $n$ and the columns $i\neq 1$ and $n$,
\begin{equation}
 \det{}'\hat H=  \det{}'\hat H'=\frac{1}{\left\langle u_1 u_n\right \rangle \left(\left[ \tilde u_{ i}\tilde u_n\right] +  \frac{\epsilon_{n\,\sB}\epsilon_1^\sB}{\epsilon_{1\,\sB}\epsilon_n^\sB}\left[ \tilde u_{ i}\tilde u_1\right]  \right)}\det \hat H{}^{[1n]}_{[i n]}\,.
\end{equation}
The expression on the right hand side is now manifestly of order $\mathcal{O}(z^0)$.

\paragraph{Yang-Mills theory and gravity.} Over the last section, we derived that
\begin{equation}
 e^{F+\tilde F}\sim z^0\,,\qquad \sum_{\alpha\in S_n/\mathbb{Z}_n}\mathrm{PT}(\alpha)\sim z\,,\qquad\det{}'H\sim z^0\,,
\end{equation}
in the large-$z$ limit. Combining this with the behaviour of the measure, we find that the boundary terms in  supergravity and super Yang-Mills both vanish as expected,
\begin{equation}
 \mathcal{M}(z)\sim z^{-2}\,,\qquad\qquad \mathcal{A}(z)\sim z^{-1}\,.
\end{equation}
This completes the BCFW-recursion proof of our formulae.\\

As a brief aside, we mention here the curious observation that our brane formulae  also do not receive boundary contributions in the BCFW recursion, despite their poor behaviour for large momenta. Though we are not aware of a discussion of this in the literature, this is also true for the D-brane amplitudes in the usual CHY--framework, and just relies on the additional observation that $\pf{}'A\sim z^0$ in the large-$z$ limit, which in turn follows from similar row- and column operations on $A$ as are used on $M$ to show that $\pf{}'M\sim z^0$.  It would be interesting to investigate this cancellation  from the field theory perspective.

\section{Discussion}\label{sec:discussion} 

In this article we have argued that the polarized scattering equations provide a natural generalization of the twistor and ambitwistor supersymmetric formulae from four dimensions.  
They lead to  formulae for a full spectrum of supersymmetric gauge, gravity and brane theories in six-dimensions.  These formulae are furthermore shown to factorize properly as a consequence of properties of the polarized scattering equations themselves, as described in \S\ref{sec:fact_SE}.  This led to a proof of the main formulae by BCFW recursion. 

There remain issues that are not optimally resolved in our framework.  Because the solutions to the polarized scattering equations themselves depend on the polarization data, it is no longer obvious that the formulae we obtain are linear in each polarization vector as they need to be, although the proof is relatively straightforward.  As shown in \S\ref{sec:PSE}, there is an $n+2$ dimensional vector space of potential solutions to the polarized scattering equations whose dimensionality is then then reduced by choice of polarization spinors.   It should be possible to develop this further to produce 
formulae that are manifestly linear in the polarization data, or alternatively with free little-group indices as is more usually in higher-dimensional  spinor-helicity frameworks.  

There remain many avenues for further development and investigation.  One is the treatment of massive amplitudes in four and perhaps five dimensions.  Here there is ongoing work both by the authors of this paper and \cite{Wen:2020qrj}, who further apply these to construct formulae for loop amplitudes for brane and other theories in four dimensions.    Further avenues are as follows.

\paragraph{Grassmannians, polyhedra, and equivalence with other formulations.}  In four dimensions, twistor-string formulae for amplitudes, and indeed general BCFW terms,  can be embedded  as $2n-4$-dimensional cycles in the Grassmannian $G(k,n)$ for amplitudes with $k$ negative helicity particles, \cite{Bullimore:2009cb,ArkaniHamed:2009sx}. 

In \cite{Cachazo:2018hqa} it was similarly shown that their 6d formulae could be embedded into a Lagrangian Grassmannian, i.e., the Grassmannian $LG(n,2n)$ of Lagrangian $n$-spaces in a symplectic $2n$-dimensional vector space.  Ref.~\cite{Schwarz:2019aat} further discussed how the polarized scattering equation formulation  of \cite{Geyer:2018xgb} and this paper can also be embedded in the same Grassmannian, allowing one to see that the two formulations are essentially gauge equivalent representations.  
In the formulation in this paper, an element of the Grassmannian can be represented as an $n\times 2n$ matrix $C^{ia}_l$ with $a$ being the little group index for $k_i$ and $l$ being also a particle index.\footnote{For \cite{Cachazo:2018hqa,Schwarz:2019aat} this $l$-index is replaced by $ak$ where $a$ is the global little group index, and $k=0,\ldots,(n-2)/2$ indexes a basis in the space of polynomials on $\C$ of degree $(n-2)/2  $.} The symplectic form is given by $\Omega_{ia jb}=\varepsilon_{ab}\delta_{ij}$ and the condition that $C^{ia}_l$ defines an element of the Lagrangian Grassmannian is that
\begin{equation}
C^{ia}_lC^{jb}_m\Omega_{iajb}=0\, .
\end{equation}
This skew form is natural in the sense that it arises from momentum conservation in the form
\begin{equation}
\kappa_{iA}^{a}\kappa_{jB}^{b}\Omega_{iajb}=0\, .
\end{equation}
The Grassmannian integral formula then takes the form
\begin{equation}
\int_\Gamma d\mu\;\cI\int \prod_j\delta^4(C_j^{ia}\kappa_{iaA})\, .
\end{equation} 
Here $\cI$ is a theory dependent integrand, $\Gamma$ a cycle in the Grassmannian of dimension $4n-6$, and $d\mu$ a measure on $\Gamma$.  
Our data embeds into the Grassmannian by
\begin{equation}
C^{ai}_j= \frac{\la u_iu_j\rangle}{\sigma_{ij}}\epsilon_i^a-\delta^i_jv_i^a\, ,
\end{equation}
with $\Gamma$ parametrized by $(\sigma_i,u_i,v_i)$ subject to the constraints $\la v_i\epsilon_i\ra=1$ and modulo the M\"obius transformations on the $\sigma_i$, and SL$(2)$ on the $u_i$.
A different parametrization\footnote{In the notation of those references, the $4n-6$-cycles are parametrized by $(\sigma_i, w_{ia}^b)$ subject to a normalization of the determinants of the $W_{ia}^b$ in terms of the $\sigma_i$. } for $\Gamma$ is given in \cite{Cachazo:2018hqa}, and in \cite{Schwarz:2019aat} it was argued that the two representations are gauge equivalent in  $LG(n,2n)$.

In this paper in \S\ref{sec:poldata}, the argument for linearity of the reduced determinants in the polarization data relies on a map between solutions to the polarized scattering equations that have different polarization data.  This map should therefore similarly arise from an analogous gauge transformation in the Grassmannian $LG(n,2n)$.

Polyhedra such as the amplituhedron \cite{Arkani-Hamed:2013jha} emerge when BCFW cycles in a Grassmannian are united into one geometric object whose combinatorics are determined by a certain positive geometry.   The original  amplituhedron was adapted to momentum twistor or Wilson-loop descriptions of $N=4$ super Yang-Mills amplitudes  \cite{Hodges:2009hk, Mason:2009sa,Mason:2010yk}, but there is, at least as yet, no analogue of this in six dimensions. The version of the 4d amplituhedron  ideas that are most natural in the context of the Grassmannian descriptions here is that described in \cite{Damgaard:2019ztj}, a $2n-4$-dimensional space.  It follows from the above that the analogue in 6d should therefore be a $4n-6$ dimensional space.  In our context this space will then be naturally embedded in $\R^{4n}$ (perhaps projected onto some quotient) as the image of the positive Lagrangian Grassmannian $LG_+(n,2n)$ under the map
\begin{equation}
Y_{lA}=C^{ia}_l\kappa_{iaA}\, .
\end{equation}
There is of course an anti-chiral version also.  It remains to explore these frameworks.

\paragraph{Worldsheet models in 6d.} Another gap in our description is to identify ambitwistor string models that underly the formulae.  Ambitwistor-string models that admit vertex operators that yield the polarized scattering equations and supersymmetry factors were introduced in \cite{Geyer:2018xgb}, together with worldsheet matter that provides the reduced determinants.  However, these were chiral, and combining both chiralities to produce the gauge and gravity formulae has so far proved problematic: there are  constraints needed to identify the two otherwise independent chiral halves.  However, as seen here such constraints dont seem to matter too much at the level of the formulae.  The chiral models would seem to be a better bet for the various $(N,0)$ theories, but for these the worldsheet matter required to provide the integrands has yet to be identified.   The issues facing the 6d worldsheet models are resolved on reduction and we plan to write about this elsewhere.

\paragraph{Higher dimensions.} Representations of ambitwistor space, in terms of twistor coordinates with little-group indices exist in higher dimensions also.  Furthermore, naive ambitwistor models in those coordinates lead to higher-dimensional analogues of the polarized scattering equations.   A  discussion of such models was given in \cite{Geyer:2019ayz}. Again one can obtain supersymmetric ampltude formulae without worrying too much about the detailed implementation of the models.  In particular, there are many more constraints required to restrict the representation to ambitwistor space as in the space of null geodesics, and again these were not implemented in any systematic way.   Indeed closely related models were proposed over the years by Bandos  and coworkers \cite{Bandos:1999pq, Bandos:2006nr, Bandos:2017kdq, Bandos:2017eof, Bandos:2017zap, Bandos:2014lja}.  Bandos takes the attitude that the additional constraints should not be imposed, and instead that it should be possible to find  genuine M-theory physics in these extra degrees of freedom \cite{Bandos:1999pq, Bandos:2019zqp,Bandos:2019qlg}.

\paragraph{Gerbe amplitudes.} In addition to the well understood gauge, gravity and brane formulae, we also obtain more controversial formulae with $(2,0)$, $(3,1)$ and  $(4,0)$ supersymmetry.  The linear super-multiplets are Gerbe-like analogues of YM and gravity theories in the sense that Gerbes, self-dual closed 3-forms, appear in the multiplets.  In particular in the $(2,0)$ case with the Parke-Taylor factor in the integrand,  there is an important and much studied theory  with $(2,0)$ supersymmetry that one might hope to say something about.  This theory is expected to reduce to super-Yang-Mills in five dimensions as indeed our $(2,0)$ formulae with a Parke-Taylor does for even numbers of particles.  In six dimensions however, this is thought to be a strongly coupled theory and so shouldnt give rise to meaningful amplitudes.  It has  furthermore been argued that there are no invariant three point amplitudes for such models in 6d \cite{Huang:2010rn}. On the other hand, the four point formulae has  $s$ and $t$ singularities \eqref{4pt-02-PT}, so that soft limits should give a nontrivial limit involving the 3-point amplitude.  Thus such soft limits are likely to be  ambiguous and not make sense.  Similar issues arise for the other Gerbe-like theories with $(3,1)$ and $(4,0)$ supersymmetry. See \S\ref{Sec:Other} for more discussion and  \cite{Cachazo:2018hqa} where for  more detail in the context of the little-group preserving representation.

The amplitude formulae we obtain are problematic for odd particle number. Being ratios of Pfaffians of matrices whose size depends on the particle number $n$, one obtains zero divided by zero for odd $n$ and like the 3-particle case, might not  have a sensible meaning.  For the $(N,N)$-theories, analogous formulae can also be obtained, but identities such as \eqref{eq:detH=pfA} allow us to  obtain a well-defined non-zero formula when $n$ is odd.  Such relations also hold for the Gerbe theories reduced to 5d because they coincide with the reductions of $(N,N)$ theories. However,  we have not been able to find such relations in 6d.  Thus the prognosis for some physical interpretation of these formulae is not clear. Some reasonable definition must be found for odd $n$ that is compatible with factorization, see the discussion after \eqref{eq:fact_M5} for additional details.  
If so, a further test will be to investigate massive modes on reduction to 5d as the R-symmetry of reduced $(0,2)$ massive modes is distinct from that of $(1,1)$ massive modes. For massive modes the little group in 5d is still SO$(4)$ with spin group  $SL(2)\times SL(2)$. Thus the dotted and undotted scattering equations remain distinct and there is no longer an identification between the $U^{(a,b)}$ for fixed $a+b$.  There is therefore no  clear analogue of \eqref{Pf-det} so analogues of the odd-point formulae for 5d massive modes reduced from 6d massless modes remain problematic.  

 There  are speculations that such  theories might play an important role in M-theory  \cite{Hull:2000zn, Borsten:2017jpt,Henneaux:2017xsb,Henneaux:2018rub} so 
despite all these issues,  these formulae perhaps deserve further study as one of the few handles we have on the possible interactions in such theories.

\section*{Acknowledgements}
GA is supported by the EPSRC under grant EP/R513295/1. YG gratefully acknowledges support from the CUniverse research promotion project ``Toward World-class Fundamental Physics'' of Chulalongkorn University (grant reference CUAASC), as well as support from the National Science Foundation Grant PHY-1606531 and the Association of Members of the Institute for Advanced Study (AMIAS).  LJM  is grateful to the EPSRC for support under grant EP/M018911/1.

We are grateful to Leron Borsten, Hadleigh Frost, Alfredo Guevara, Matthew Heydeman, John Schwarz and  Congkao Wen for discussions.

\appendix
\section{Appendices}
\subsection{Direct proof of permutation invariance of \texorpdfstring{$H$}{H}}\label{sec:perm}
As an alternative to the abstract proof in \cref{lemma:perm-inv}, we can show directly that the  reduced determinant $\det{}'H$ is permutation invariant by using row and column operations, as well as the constraints
\begin{equation}\label{eq:constr_6d}
 \sum_i u_i^a H_{ij} = 0\,,\qquad \sum_i \tilde u_i^{\dot a} H_{ji} = 0\,.
\end{equation}
Recall  the definition \eqref{gendet2} of the reduced determinant;
\begin{equation}\label{eq:red_det_v2}
 \det{}'(H): =(-1)^{i_1+i_2+j_1+j_2}\frac{\det\left(H^{[i_1i_2]}_{[j_1j_2]}\right)}{\langle u_{i_1} u_{i_2}\rangle [u_{j_1} u_{j_2}]}\,,
\end{equation}
where $H^{[i_1i_2]}_{[j_1j_2]}$ denotes the matrix $H$ with rows $i_1$ and $i_2$ and columns $j_1$ and $j_2$ removed,
\begin{equation}
 \det\left(H^{[i_1i_2]}_{[j_1j_2]}\right) = \frac{\partial^2}{\partial H_{i_1 j_1}\partial H_{i_2 j_2}}\det(H)\,.
\end{equation}
By definition, $\det{}'(H)$ is s invariant under exchanging two particle labels $i,j\neq i_{1,2},j_{1,2}$, since the determinant picks up a sign under each exchange of  rows or  columns. To prove permutation invariance, we thus only need to show that the reduced determinants obtained from removing different rows or columns are identical.
Moreover, it is clearly sufficient to consider the case of different choices for the row $i_2$, all other cases are straightforward extensions.
To be specific, consider $\det(H^{[1\,2]}_{[n-1\,n]})$ and $\det(H^{[1\,3]}_{[n-1\,n]})$, and let us suppress the subscript ${}_{[n-1\,n]}$ for the removed columns to keep the expressions readable. 
Then the reduced determinant  \eqref{eq:red_det_v2} is permutation invariant if 
\begin{equation}
 \langle u_1 u_3\rangle\det\left(H^{[1\,2]}\right) =-\langle u_1 u_3\rangle \det\left(H^{[1\,2]}\right)\,.
\end{equation}
First, multiply the row in $H^{[1\,2]}$ associated to particle 3 by $\langle u_1 u_3\rangle $ (and similarly for $H^{[1\,3]}$),
\begin{equation}
 \widehat{H}^{[1\,2]}_{3i}=\langle u_1 u_3\rangle H_{3i}\,,\qquad \widehat{H}^{[1\,3]}_{2i}=\langle u_1 u_2\rangle H_{2i}\,.
\end{equation}
The determinants of the hatted matrices are then  related to the original determinants via
\begin{equation}\label{eq:hat_H_6d}
 \det\left(\widehat{H}^{[1\,2]}\right) =\langle u_1 u_3\rangle \det\left(H^{[1\,2]}\right)\,,\qquad \det\left(\widehat{H}^{[1\,3]}\right) =\langle u_1 u_2\rangle \det\left(H^{[1\,3]}\right)\,.
\end{equation}
To compare the two determinants $ \det\widehat{H}^{[1\,2]}$ and $ \det\widehat{H}^{[1\,3]}$, proceed as follows: Multiply each row $\widehat{H}_{ji}^{[1\,2]}$ associated to particle $j\neq 3$ by $\langle u_1 u_j\rangle$, and add it to the   row $\widehat{H}^{[1\,2]}_{3i}$,
\begin{equation}\label{eq:H12i=H13i}
 \widehat{H}^{[1\,2]}_{3i}= \sum_{j\neq 1,2}\langle u_1 u_j\rangle H_{ji} = -\langle u_1 u_2\rangle H_{2i} = -\widehat{H}^{[1\,3]}_{2i}\,.
\end{equation}
In the second equality, we have used the constraint \eqref{eq:constr_6d}, and the last identity follows from our definitions above. In particular, note that \eqref{eq:H12i=H13i} holds for $i=2$ as well, so there is no subtlety associated to the diagonal entries.
Since row and column operations leave the determinant invariant, we can thus conclude that 
\begin{equation}
  \det\left(\widehat{H}^{[1\,2]}\right)= -\det\left(\widehat{H}^{[1\,3]}\right)\,,
\end{equation}
and permutation invariance follows by using \eqref{eq:hat_H_6d}.

\medskip

Note that we can easily use the same idea to show that $\det(H)=0$. In this case, we follow the same steps as above, but now for the unreduced matrix $H$. Again, we define 
\begin{equation}
 \widehat{H}_{2i}=\langle u_* u_2\rangle H_{2i}\,,
\end{equation}
for any reference spinor $u_*$ in the little group. The determinants are again related by $ \det\widehat{H} =\langle u_* u_2\rangle  \det\left(H\right)$. As before, we can use the constraint equations, together with convenient row operations on the matrix (adding $\langle u_* u_j\rangle H_{ji}$ to $\widehat{H}_{2i}$).  However, since no rows have been removed from the matrix, this time we find
\begin{equation}
 \widehat{H}_{2i}= \sum_{j}\langle u_* u_j\rangle H_{ji} =0\,,
\end{equation}
and so the determinant vanishes.

\medskip

We can also extend this proof to the determinant with only one row and column removed, $H^{[1]}_{[n]}=\frac{\partial}{\partial H_{1n}}\det(H)=0$: Proceed as above, but choose $u_*=u_1$ to coincide  with the removed row. Then again
\begin{equation}
 \widehat{H}^{[1]}_{2i}= \sum_{j\neq 1} \langle u_1 u_j\rangle  H_{ji} =0\,,
\end{equation}
since the term from the omitted row does not contribute to the constraint when $u_*=u_1$, and we conclude $H^{[1]}_{[n]}=\frac{\partial}{\partial H_{1n}}\det(H)=0$.

\subsection{Comparison to other BCFW shifts in higher dimensions} \label{sec:BCFW_Cliff+Donal}
For generic polarization data of the particles $1$ and $n$, the BCFW shift \eqref{eq:BCFW_shift_q} differs from the BCFW shift for Yang-Mills theory and gravity of \cite{ArkaniHamed:2008yf}, as well as the 6d spinorial shift of \cite{Cheung:2009dc}. In these,  for gluons and gravitons, the shift vector is chosen to align with the polarization of one of the shifted particles, $q_\mu=e_{1\,\mu}$, to ensure that the boundary terms vanish.\footnote{In addition, we also have to work in a gauge where $q_\mu=e_{1\,\mu}$ does not transform under the shift.} In the 6d spinor-helicity formalism, the polarization vector $e_1$ is given by (c.f. \S\ref{YM-int})
\begin{equation}\label{eq:pol_vec}
 e_{1\,\sA\sB}=\epsilon_{1\,[\sA}\tilde\epsilon_{1\,\sB]}\,,\qquad \text{with } \epsilon_{1\,\sA}=\epsilon_{1\,a}\kappa_{1\,\sA}^a \text{ and } \epsilon_1^\sA=\tilde\epsilon_1^{\dot a}\kappa^\sA_{1\,\dot a}=\tilde \epsilon_{1\,\sB}\, k_1^{\sA\sB}\,.
\end{equation}
Due to the gauge freedom $e_\mu\sim e_\mu+k_\mu$,  the spinor $\tilde \epsilon_{1\,\sA}$ is only defined up to terms  proportional to $\kappa_{1\,\sA}^a$. Up to this freedom, a canonical choice \cite{Cheung:2009dc} is given by
\begin{equation}\label{eq:def_tilde_epsilon}
 \tilde \epsilon_{1\,\sA} = \tilde \epsilon_{1\, \dot a}\kappa_{*}{}_\sA^b\,\left(\kappa_{*}{}_\sB^b\,\kappa_1{}^\sB_{\dot a}\right)^{-1}\,,
\end{equation}
where $\kappa_{*}{}_\sA^b$ is a reference spinor satisfying $\kappa_{*}{}_\sB^b\,\kappa_1{}^\sB_{\dot a}\neq 0$, and the inverse is defined as the matrix inverse in the little group spaces of the particles $1$ and $n$. This choice for $\tilde \epsilon_{1\,\sA} $ clearly satisfies $\tilde \epsilon_1^{\dot a} =\tilde \epsilon_{1\,\sA} \, \kappa_{1\,\dot a}^{\sA}$, and thus reproduces  $F_\sA^\sB=\big(\gamma_{\mu\nu}\big)\!{}_\sA^\sB\, F^{\mu\nu} =\epsilon_\sA \epsilon^\sB$. 

The spinorial BCFW shift $q_{\sA\sB}= \epsilon_{n\,[\sA}\epsilon_{1\,\sB]}$ is thus only equivalent to the standard BCFW shift $q_{\sA\sB}=e_{1\,\sA\sB}$ if we can choose a little group spinor $v^*_{1a}$ such that 
\begin{equation}
 \epsilon_{n\,\sA} = -\tilde \epsilon_{1\,\sA}+v^*_{1a}\, \kappa_{1\,\sA}^a\,.
\end{equation}
However, for generic momenta and polarization, no such $v^*_{1a}$ exists: upon choosing the reference spinor $\kappa_{*}{}_\sA^b=\kappa_{n}{}_\sA^b$ in \eqref{eq:def_tilde_epsilon}, we see that $q=e_1$ only if the polarization spinors for particles $1$ and $n$ satisfy  $\tilde\epsilon_1^{\dot a} =- \epsilon_{n\,\sA}\, \kappa_{1}^{\sA\,\dot a}$. Thus, the BCFW shift $q_{\sA\sB}= \epsilon_{n\,[\sA}\epsilon_{1\,\sB]}$ generically  differs from those discussed previously in the literature \cite{ArkaniHamed:2008yf,Cheung:2009dc}). Note however that since $q$ is constructed from the chiral polarization spinors of both shifted particles, it  \emph{does} lie in the space of possible polarization vector for both particles. 

\paragraph{Comparison to the 6d BCFW shift of  Cheung \& O'Connell}
In the bosonic case, the super-BCFW shift discussed in \cref{sec:BCFW} is strongly remininscent of the shift used in the work \cite{Cheung:2009dc} of Cheung and O'Connell on the 6d spinor-helicity formalism to derive higher point gluon amplitudes. Here  we compare our shift to that of \cite{Cheung:2009dc}, and comment on the similarities and differences in the resulting recursion relations.

Let us briefly review the work of \cite{Cheung:2009dc}.\footnote{See also \cite{Boels:2012ie} for related work in higher dimensions.} For bosonic Yang-Mills theory, it is advantageous to keep the little-group symmetry manifest, see also \cref{sec:susy} for a discussion on the trade-off between the little-group and R-symmetry for super Yang-Mills. Amplitudes are thus of the form $A_{\boldsymbol{ a\dot a}}^{\LG}:=A^\LG_{a_1\dot a_1\dots a_n\dot a_n}$, which relates to our representation (due to the linearity in the polarization spinors proven in \cref{sec:poldata}) via
\begin{equation}
  A_{\boldsymbol{\epsilon\tilde\epsilon}}= \epsilon_1^{a_1}\tilde\epsilon_1^{\dot a_1}\dots \epsilon_n^{a_n}\tilde\epsilon_n^{\dot a_n}A^\LG_{a_1\dot a_1\dots a_n\dot a_n}\,,\qquad \quad A_{\boldsymbol{\epsilon\tilde\epsilon}}:=A_{\epsilon_1\tilde\epsilon_1\dots \epsilon_n\tilde\epsilon_n} \,.
\end{equation}
The BCFW-shift of Cheung and O'Connell is then designed to keep this little-group symmetry of the amplitude representations manifest. Note that the standard $d$-dimension BCFW recursion relation does not interact well with the little-group preserving amplitude representation, because the shift vector has to be chosen to align  with the polarization of one of the particles,  $q^\mu=e_1^\mu$, see \cite{ArkaniHamed:2008yf}. In the spinor-helicity formalism however, there is no natural candidate for $q=e_1$, essentially by construction. Cheung and O'Connell avoid this complication by studying partially contracted amplitudes of the form
\begin{equation}
 X^{a_1\dot a_1}\,A_{a_1\dot a_1\dots a_n\dot a_n}^\LG\,,
\end{equation}
where $X$ is a little group vector for particle 1. For these amplitudes, they can use the standard BCFW construction and choose the deformation vector to be  $q^\mu=X^{a\dot a}\,e_{1a\dot a}^\mu$, where $e_{1a\dot a}$ is a basis of polarization vectors for particle 1. Requiring that the shift leaves the external momenta on-shell is equivalent to $q^2=\det X=0$, and thus $X^{a\dot a} = x^a \tilde x^{\dot a}$ factorizes, where we can identify
\begin{equation}\label{eq:x=epsilon}
 x^a=\epsilon_1^a\,,\qquad \tilde x^{\dot a}=\tilde \epsilon_1^{\dot a}\,.
\end{equation}
This construction leaves the direction of the deformation free (parametrized by $X$), but still aligns it with the polarization vector of particle 1, since for any $X$ we have $q^\mu=\epsilon_{1}^a\tilde\epsilon_{1}^{\dot a} \,e_{1a\dot a}^\mu=e_1^\mu$. Since linearity in $X^{a\dot a}=\epsilon_{1}^a\tilde\epsilon_{1}^{\dot a}$ is guaranteed, the full little-group-preserving $A_{\boldsymbol{ a\dot a}}^{\LG}$ can still be extracted this way. Having defined this covariantized, but vectorial BCFW shift\footnote{We use the notation $\check k_{1,n}$ here to facilitate the comparison to the chiral shift denoted by $\hat k_{1,n}$.}
\begin{equation}
 \check k_1 = k_1+zq\,,\quad \check k_n = k_n-zq\,,\quad\text{where }q^\mu =e_1^\mu=\epsilon_{1}^a\tilde\epsilon_{1}^{\dot a}\, e_{1a\dot a}^\mu\,\text{ and }\, e_{1\sA\sB}^{a\dot a} = \kappa_{1\sA}^a\kappa_{*\sB}^b\left(\kappa_{*\sC}^b\kappa_{1\dot a}^\sC\right)^{-1}\,,
\end{equation}
Cheung and O'Connell then implement it at the spinorial level as follows:
\begin{subequations}\label{eq:BCFW-shift_COC}
 \begin{align}
   &\check \kappa_{1\sA}^a = \kappa_{1\sA}^a +z \,\epsilon_1^a \,\tilde\epsilon_{1\sA}&& \check \kappa_{1\dot a}^\sA = \kappa_{1\dot a}^\sA -z\, \tilde \epsilon_{1\dot a} \,\tilde\epsilon_1^\sA\,, \\
   &\check \kappa_{n\sA}^a = \kappa_{n\sA}^a +z\, y^a \,\epsilon_{1\sA} && \check \kappa_{n\dot a}^\sA = \kappa_{n\dot a}^\sA -z \,\tilde y_{\dot a} \,\epsilon_1^{\sA}\,.
 \end{align}
\end{subequations}
Here, $\tilde \epsilon_{1\sA}$ and $\tilde\epsilon_1^\sA$ are defined as in \eqref{eq:def_tilde_epsilon}, such that $e_{1\sA\sB}=\epsilon_{1[\sA}\tilde\epsilon_{1\sB]}$, and similarly for the antichiral case. Moreover, 
$y$ and $\tilde y$ are little group spinors of particle $n$, and are determined by the spinors $\kappa_1$, $\kappa_n$, as well as $\epsilon_1^a$ and $\tilde \epsilon_1^{\dot a}$, via 
\begin{equation}
 y_a = \tilde \epsilon_1^{\dot a}\left(\kappa_{n\sA}^a\kappa_1^{\sA\dot a}\right)^{-1}\,,\qquad \tilde y_{\dot a} = \epsilon_1^{a}\left(\kappa_{1\sA}^a\kappa_n^{\sA\dot a}\right)^{-1}\,.
\end{equation}
 Using this shift, the BCFW recursion relation for the little-group preserving representation becomes
\begin{equation}
 \epsilon_1^{a_1}\tilde \epsilon_1^{\dot a_1}A_{a_1\dot a_1\dots a_n\dot a_n}^\LG = \sum_{L}\frac{\varepsilon^{b_\ssL b_\ssR}\varepsilon^{\dot b_\ssL\dot  b_\ssR}}{k_\sL^2}\,\epsilon_1^{a_1}\tilde \epsilon_1^{\dot a_1}A^{\LG\,(\sL)}_{a_1\dot a_1\dots b_\ssL\dot b_\ssL}\left(\hat k_1\dots k_\sL\right)A^{\LG\,(\sR)}_{ b_\ssR\dot b_\ssR\dots a_n\dot a_n}\left(- k_\sL\dots \hat k_n\right)\,.
\end{equation}

\paragraph{The shift of  Cheung \& O'Connell  and the polarized scattering equations.}
Naively, this recursion relation seems quite suitable to the framework based on the polarized scattering equations -- contracting both sides into the remaining $\epsilon$'s and $\tilde \epsilon$'s leads directly to the recursion relation of \cref{sec:BCFW}. This however is not true for  the BCFW shift \eqref{eq:BCFW-shift_COC}, which is inherently ambidextrous, and does not seem natural from the point of view of the (chiral) polarized scattering equations. 
It is difficult to verify that the boundary terms are absent,\footnote{To illustrate this difficulty, note that the scattering equations for $i\neq 1,n$ contain a single term of order $z$,
\begin{equation}
 \mathcal{E}_i\supset \left( \frac{\la u_i u_1\ra}{\sigma_{i1}}+z\, \la \epsilon_n y\ra \frac{\la u_i u_n\ra}{\sigma_{in}} \right)\epsilon_{1\sA}\,.
\end{equation}} 
and it thus doesn't seem feasible to apply the original recursion in our framework. Note that the ambidextrous nature of the shift can be traced back to  the choice of the deformation vector $q=e_1$, and thus seems to be an intrinsic feature of any BCFW-relation closely related to the general-$d$ recursion of \cite{ArkaniHamed:2008yf}.

\paragraph{Comparison.} To illustrate how the our chiral BCFW shift relates to the ambidextrous Cheung and O'Connell shift, it is helpful to recast  \eqref{eq:BCFW-shift_COC} in terms of some still-to-be-specified variables $x$ and $y$, related as before via
\begin{equation}\label{eq:xy-rels}
 y_a = \tilde x^{\dot a}\left(\kappa_{n\sA}^a\kappa_1^{\sA\dot a}\right)^{-1}\,,\qquad \tilde y_{\dot a} = x^{a}\left(\kappa_{1\sA}^a\kappa_n^{\sA\dot a}\right)^{-1}\,.
\end{equation}
We stress that at this point these are the only constraints on the variables $\{x,\tilde x,y,\tilde y\}$, and that $x$ and $\tilde x$ may not align with the polarization of particle 1. The shift \eqref{eq:BCFW-shift_COC} is then given by \footnote{This is in fact the original notation for the BCFW shift given in \cite{Cheung:2009dc}, though with the interpretation of $x=\epsilon$ and $\tilde x=\tilde\epsilon$ as in \eqref{eq:x=epsilon} and \eqref{eq:BCFW-shift_COC}.}
\begin{subequations}\label{eq:BCFW-shift_COC_v2}
 \begin{align}
   &\check{ \kappa}_{1\sA}^a = \kappa_{1\sA}^a +z \,x^a \la y\, \kappa_{n\sA}\ra && \check\kappa_{1\dot a}^\sA = \kappa_{1\dot a}^\sA -z\, \tilde x_{\dot a} \left[ \tilde y\,\kappa_n^\sA\right]\,, \\
   &\check\kappa_{n\sA}^a = \kappa_{n\sA}^a +z\, y^a \la x\,\kappa_{1\sA}\ra && \check \kappa_{n\dot a}^\sA = \kappa_{n\dot a}^\sA -z \,\tilde y_{\dot a} \left[\tilde x\,\kappa_1^\sA\right]\,.
 \end{align}
\end{subequations}
We note that this is the 6d-version of the super BCFW-shift of \cite{Boels:2012ie}, using a slightly modified notation to keep it more in line with \cite{Cheung:2009dc}. 
As above, we use the notation $\check\kappa_{1,n}$ for the shifted variables to make it easier to compare this ambidextrous shift to the chiral one of \cref{sec:BCFW}. 
The shift \eqref{eq:BCFW-shift_COC_v2} can then be chosen to partially agree with the chiral BCFW shift \eqref{eq:BCFW_shift} and \eqref{eq:BCFW_antichiral} by setting 
\begin{equation}\label{eq:xy=epsilon}
 x^a=\epsilon_1^a\,,\qquad y^a = \epsilon_n^a\,,
\end{equation}
which leads to the same shift for fundamental spinors,  $\hat\kappa_{1,n}= \check\kappa_{1,n}$. To see what happens to the  antifundamental spinors, we first observe that the relations \eqref{eq:xy-rels} become
\begin{equation}
 \tilde x_{\dot a} = \epsilon_{n\sA}\kappa_{1\dot a}^\sA\,,\qquad \tilde y_{\dot a}  = \epsilon_1^a\left(\kappa_{1\sA}\kappa_{n}^\sA\right)^{-1}_{a\dot a} = \frac{ \epsilon_{1\sA}\kappa_{n\dot a}^\sA}{k_1\cdot k_n}\,.
\end{equation}
In comparison to \eqref{eq:BCFW_antichiral}, this shift is missing the  `pure gauge' terms of $\tilde \epsilon^\sA$, and so the two shifts do not agree for the antifundamental spinors. While the shift \eqref{eq:xy=epsilon} may be interesting in its own right, the proportionality of the antifundamental shift to $\tilde\epsilon^\sA$ was crucial in proving that the boundary terms vanish.

More generally, we can show that  the antifundamental shift $\check\kappa_{1\dot a}^\sA$ never agrees with $\hat\kappa_{1\dot a}^\sA$  for any choice of $\{x,\tilde x,y,\tilde y\}$. To see this, contract both shifted spinors $\hat\kappa_{1\dot a}^\sA$ and $\check\kappa_{1\dot a}^\sA$ into $\kappa_{1\sA}^a$ (and equivalently for $n$). This vanishes for the chiral shift, $\hat\kappa_{1\dot a}^\sA\,\kappa_{1\sA}^a=0$, but is generically non-zero for the little-group preserving shift, $\check\kappa_{1\dot a}^\sA\kappa_{1\sA}^a\neq0$, and we conclude that $\hat\kappa_{1\dot a}^\sA\neq\check\kappa_{1\dot a}^\sA$.

\paragraph{A more general shift.}  \eqref{eq:BCFW-shift_COC_v2} is not the most general spinor deformation giving rise to the vecorial shift $\check k_1=k_1+z q$, $\check k_n=k_n-z q$. In fact, it is easily checked that we have the freedom to add terms proportional to $x^a \la x\,\kappa_{1\sA}\ra$ to $\check\kappa_{1\sA}^a$ etc,
\begin{subequations}\label{eq:BCFW-shift_COC_v3}
 \begin{align}
   &\check{ \kappa}_{1\sA}^a = \kappa_{1\sA}^a +z \,x^a \Big(\la y\, \kappa_{n\sA}\ra+\alpha_1\,\la x\,\kappa_{1\sA}\ra\Big) && 
   \check\kappa_{1\dot a}^\sA = \kappa_{1\dot a}^\sA -z\, \tilde x_{\dot a} \Big(\left[ \tilde y\,\kappa_n^\sA\right]+\tilde\alpha_1 \left[\tilde x\,\kappa_1^\sA\right]\Big)\,, \\
   &\check\kappa_{n\sA}^a = \kappa_{n\sA}^a +z\, y^a\Big( \la x\,\kappa_{1\sA}\ra +\alpha_n\,\la y\,\kappa_{n\sA}\ra\Big) && 
   \check \kappa_{n\dot a}^\sA = \kappa_{n\dot a}^\sA -z \,\tilde y_{\dot a} \Big( \left[\tilde x\,\kappa_1^\sA\right]+\tilde \alpha_n\left[\tilde y\,\kappa_n^\sA\right]\big)\,.
 \end{align}
\end{subequations}
From the point of view of this more general shift, we can finally understand both the shift of Cheung and O'Connell \eqref{eq:BCFW-shift_COC} and our chiral shift \eqref{eq:BCFW_shift}, \eqref{eq:BCFW_antichiral} as special choices of the free variables. As discussed above, Cheung and O'Connell pick
\begin{equation}
 x^a=\epsilon_1^a\,,\qquad \tilde x^{\dot a}=\tilde \epsilon_1^{\dot a}\,,\qquad \alpha_1=\alpha_n=\tilde \alpha_1=\tilde \alpha_n=0\,,
\end{equation}
whereas our chiral shift corresponds to
\begin{equation}
 x^a=\epsilon_1^a\,,\qquad y^a=\epsilon_n^a\,,\qquad \tilde \alpha_1^{-1}=\tilde\alpha_n = k_1\cdot k_n\,,\qquad \alpha_1=\alpha_n=0\,.
\end{equation}
Note that despite the six degrees of freedom in resolving the vectorial shift, most of the choices for $\{x,\tilde x,y,\tilde y\}$ will not give rise to a `good' BCFW shift for any $\alpha_{1,n}$, $\tilde\alpha_{1,n}$. To our knowledge, the only two options to be found in the literature are the two discussed above: $q=e_1$ (the ambidextrous shift of \cite{ArkaniHamed:2008yf} and  \cite{Cheung:2009dc}), or $q\cdot e_1=q\cdot e_n=0$ (the chiral shift of this paper).\footnote{The latter is of course only possible in $d\geq 6$.}

\subsection{Factorization of \texorpdfstring{$\pf U^{(2,0)}$}{Pf U(2,0)}}\label{sec:fact_U}
In this appendix, we provide details on the following factorization properties of the Pfaffian $\pf U^{(2,0)}$.  
\begin{lemma}\label{lemma:fact_pfU}
 On boundary divisors $\partial_{\sL,\sR}\mathfrak{M}_{0,n}\simeq \mathfrak{M}_{0,n_\ssL+1}\times \mathfrak{M}_{0,n_\ssR+1}$ with odd $n_\sL$ and $n_\sR$,
\begin{equation}
 \pf U^{(2,0)} =\varepsilon^{\frac{n_\ssL - 1}{2}}\, \frac{\la \epsilon_\sL \epsilon_\sR\ra^2}{\prod_{j\in L}x_{j\sL}} \, \pf U^{(2,0)}_\sL\,\pf U^{(2,0)}_\sR\,.
\end{equation}
\end{lemma}

\paragraph{Proof:} Despite the availability of permutation symmetric formulae, it will actually be easier to use the representation \eqref{eq:lemma_Kai}
\begin{equation}
 \pf U^{(2,0)} = \frac{\det U_Y^2}{\det X_Y}
\end{equation}
in terms of $\det X_Y$ and $\det U_Y$, since these readily factorize. Restricting again to odd $n_\sL$ and $n_\sR$ odd, i.e. even subamplitudes, we can choose a partition $Y$ with $\frac{1}{2}\left(n_\sL -1\right)$ particles in $L$, and $\frac{1}{2}\left(n_\sR +1\right)$ particles in $R$, or in other words $|Y\cap L |= \frac{1}{2}\left(n_\sL -1\right)$ and $|Y\cap R |= \frac{1}{2}\left(n_\sR +1\right)$. 

Consider first the factorization of $\det X_Y$. Using the above partition, $X$ decomposes into a block-diagonal form, with
\begin{equation}
 X_Y =   \begin{blockarray}{c@{}cc@{\hspace{0pt}}cl}
        & \mLabel{\overbrace{\hspace{45pt}}^{\frac{n_\ssL +1}{2}}} &   \mLabel{\overbrace{\hspace{45pt}}^{\frac{n_\ssR -1}{2}}} & & \\
        \begin{block}{(c@{\hspace{0pt}}cc@{\hspace{0pt}}c)l}
        & \varepsilon^{-1} \hat X_{Y_\ssL\,[\sL]} & -X_{\sR|Y_\ssR\, [\sL]}^{\phantom{\sR|Y_\ssR}\,[\sR]} &&\mLabel{\Big\} \frac{n_\ssL -1}{2}}  \\
        & X_{\sR|Y_\ssR}& X_{Y_\ssR}^{[\sR]} &&\mLabel{\Big\} \frac{n_\ssR +1}{2}} \\
    \end{block}
  \end{blockarray}\,.
\end{equation}
where, with $i\in L$ and $p\in R$ (for readability we raise the matrix labels),
\begin{equation}
 X^{Y_\ssR}_{pq}=\frac1{\sigma_{pq}}\,,\qquad X^{\sR|Y_\ssR}_{ip} = \frac1{\sigma_{\sR p}}\,,\qquad \hat X^{Y_\ssR}_{ij} = x_{i\sL}x_{j\sL}\,X^{Y_\ssR}_{ij} = \frac{x_{i\sL}x_{j\sL}}{x_{ij}}\,.
\end{equation}
The leading order term in $\det X_Y$ is thus given by
\begin{equation}\label{eq:fact_X_prelim}
 \det X_Y = \varepsilon^{-\frac{n_\ssL - 1}{2}}\sum_{p\in R}(-1)^{1+p}\det \hat X_{Y_\ssL\cup\{p\}\, [\sL]}\,\det X_{Y_\ssR\,[p]}^{\phantom{Y_\ssR}[\sR]}\,,
\end{equation}
where the subscript $\det \hat X_{Y_\ssL\cup\{p\}}$ indicates the $(n_\sL+1)/2$ square matrix constructed from $\hat X_{Y_\ssL}$ and the additional row $p$ of $X_{\sR|Y_\ssR}$. As usual, we use square brackets to denote the removal of the respective rows and columns. We may now expand this determinant along the row $p$,
\begin{equation}
 \det \hat X_{Y_\ssL\cup\{p\}\, [\sL]} = \prod_{j\in L}x_{j\sL} \sum_{\bar i \in \overline{Y}_\sL}\frac{(-1)^{1+\bar i}}{\sigma_{\sR p}\,x_{\bar i\sL}}\,\det X_{Y_\sL\,[\sL]}^{\phantom{Y_\sL}[\bar i]} = -\frac{ \prod_{j\in L}x_{j\sL} }{\sigma_{\sR p}}\,\det X_{Y_\ssL}\,.
\end{equation}
Here, we used $X^{\sR|Y_\ssR}_{ip} = \frac1{\sigma_{\sR p}}=\sigma_{\sR p}^{-1}$, and the additional  factor of  $x_{\bar i\sL}^{-1}$ originates from factoring out the product $ \prod_{j\in L}x_{j\sL} $. In the last equality, we noted that the factors conspire to let us recover the full determinant $\det X_{Y_\ssL}$.  Inserting this identity back into \eqref{eq:fact_X_prelim}, we get the following factorization property for $\det X_{Y}$;
\begin{equation}\label{eq:fact_X_Y_app}
 \det X_Y =- \varepsilon^{-\frac{n_\ssL - 1}{2}}\,\prod_{j\in L}x_{j\sL}\,\det X_{Y_\ssL}\,\det X_{Y_\ssR}\,.
\end{equation}
One observation worth mentioning is that the factorization of $X_Y$ is solely responsible for the power-counting in the degeneration parameter $\varepsilon$. This is in line with what we expect, since $U^{(1,0)}$ (and also $U^{(0,1)}$) remaining of order one throughout the degeneration. \\

On the other hand, it is precisely this property that naively obscures the factorization properties of $\det U_Y$: since all components remain of order one, we do not expect to find a natural factorization corresponding to the two subspheres. However, the combination
\begin{equation}
 U_{ip}U_{jq}-U_{iq}U_{jp} = \frac{\la u_i u_j\ra\,\la u_pu_q\ra}{\sigma_{\sR p}\sigma_{\sR q}}\sim \varepsilon\,,
\end{equation}
is actually of subleading order in $\varepsilon$. Here, we have used that the denominators become independent of $i$ and $j$, as well as a Schouten identity in the $u$'s. This in turn ensures with $Y$ chosen as above,
\begin{equation}
 U_Y=   \begin{blockarray}{c@{}cc@{\hspace{0pt}}cl}
        & \mLabel{\overbrace{\hspace{45pt}}^{\frac{n_\ssL +1}{2}}} &   \mLabel{\overbrace{\hspace{45pt}}^{\frac{n_\ssR -1}{2}}} & & \\
        \begin{block}{(c@{\hspace{0pt}}cc@{\hspace{0pt}}c)l}
        & U_{Y_\ssL\,[\sL]} & -\hat U_{Y_\ssL|Y_\ssR\, [\sL]}^{\phantom{Y_\ssL|Y_\ssR}[\sR]} &&\mLabel{\Big\} \frac{n_\ssL -1}{2}}  \\
        & \hat U_{Y_\ssL|Y_\ssR}& U_{Y_\ssR}^{[\sR]} &&\mLabel{\Big\} \frac{n_\ssR +1}{2}} \\
    \end{block}
  \end{blockarray}\,.
\end{equation}
the leading order term in $\det U_Y$ can have at most one entry from the  off-diagonal blocks, i.e. the determinant factorizes similarly to $\det X_Y$, 
\begin{equation}
 \det U_Y = \sum_{p\in Y_\sR}(-1)^{1+p}\det U_{Y_\ssL\cup\{p\}\,[\sL]}\,\det U_{Y_\ssR\,[p]}^{\phantom{Y_\ssR}[\sR]}\,.
\end{equation}
Here the subscripts are defined in complete analogy to the $X$ above. We can thus follow the same strategy as before, and expand $\det U_{Y_\ssL\cup\{p\}\,[\sL]}$ in the additional row $p$, 
\begin{equation}
 \det U_{Y_\ssL\cup\{p\}\,[\sL]} = \sum_{\bar i \in \overline{Y}_\ssL}(-1)^{1+\bar i}\frac{\la u_{\bar i} u_p\ra}{\sigma_{\sR p}}\,\det U_{Y_\ssL\,[\sL]}^{\phantom{Y_\ssL}[\bar i]}\,.
\end{equation}
As before, this expression can actually be resummed to give the full $\det U_{Y_\ssL}$, which relies on the Schouten identity
\begin{equation}
 \la u_{\bar i} w_\sL\ra\,\la u_p u_\sR \ra=  \la u_{\bar i} u_p\ra \,\la w_\sL u_\sR\ra +O\big(\varepsilon^{3/2}\big)=\varepsilon^{1/2} \frac{\la u_{\bar i} u_p\ra}{\la \epsilon_\sL \epsilon_\sR\ra} +O\big(\varepsilon^{3/2}\big)\,.
\end{equation}
Using this, we recover the full determinant $\det U_{Y_\ssL}$,
\begin{equation}
 \det U_{Y_\ssL\cup\{p\}\,[\sL]} = i\,\la \epsilon_\sL \epsilon_\sR\ra\,\frac{\la u_p u_\sR\ra}{\sigma_{p\sR}}\sum_{\bar i \in \overline{Y}_\ssL}(-1)^{1+\bar i}\frac{\la w_{\bar i} w_\sL\ra}{x_{\bar i\sL}}\,\det U_{Y_\ssL\,[\sL]}^{\phantom{Y_\ssL}[\bar i]}=i\,\la \epsilon_\sL \epsilon_\sR\ra\,\frac{\la u_p u_\sR\ra}{\sigma_{p\sR}}\,\det U_{Y_\ssL}\,,
\end{equation}
which in turn gives the following factorization property for $\det U_Y$;
\begin{equation}\label{eq:fact_U_Y_app}
 \det U_Y = i\,\la \epsilon_\sL \epsilon_\sR\ra\,\det U_{Y_\ssL}\,\det U_{Y_\ssR}\,.
\end{equation}
Combining the factorization properties \eqref{eq:fact_X_Y_app} and \eqref{eq:fact_U_Y_app} for $ \det X_Y$ and $ \det U_Y$ with the independence of the choice of $Y_\sL$ and $Y_\sR$ ensured by \eqref{eq:lemma_kai_v2} (and proven in \cite{Roehrig:2017wvh}) then gives the factorization property of \cref{lemma:fact_pfU}.\hfill$\Box$

\subsection{Recursion 3 to 4 points}\label{sec:3-4points}
We show here how the BCFW shift defined in \eqref{eq:BCFW_shift} 
allows us to construct the four point amplitude from the three point in $\mathcal{N}=(1,1)$ super Yang-Mills. Having shown in \cref{sec:bdy} that the boundary terms vanish, the standard recursion procedure gives:\\
\begin{equation}
    A_4(1234)=A_{3}(\hat{1},2,P)_{a\dot{a}}\frac{1}{s_{12}}A_{3}(K,3,\hat{4})^{a\dot{a}}\,,
\end{equation}
with $k_P=-k_K=\hat{k}_1+k_2$. We have shifted here particles $1$ and $4$. The contraction between the little group indices of particles $P$ and $K$ comes from summing over the polarization states of the propagating particle, as prescribed by the BCFW procedure, to yield the numerator of the propagator. Taking the result we obtained for the three point amplitude we can write this expression as:
\begin{equation}\label{amp4=33}
    A_4(1234)=
    \frac{1}{s_{12}}
    (\la \epsilon_1m_1\ra\la \epsilon_2m_2\ra w_{Pa}+\mathrm{cyc.})
    (\la \epsilon_3m_3\ra\la \epsilon_4m_4\ra w_K^{a}+\mathrm{cyc.})\times (\mathrm{antifundamental})\,,
\end{equation}
where the contribution of antifundamental spinors is analogous to the two factors in parenthesis, only with tilded variables. All the variables $m$ and $w$ are defined with respect to shifted spinors, i.e. $m_1=m_{\hat{1}}$ but we omit the hats to make the expressions more readable. \eqref{amp4=33} can be expanded into:
\begin{align*}
     A_4(1234)=
    \frac{1}{s_{12}}\big(&\hat{1}_m2_m3_m\hat{4}_m\la w_Pw_K\ra
    +(\hat{1}_m2_w3_m\hat{4}_w+\hat{1}_m2_w3_w\hat{4}_m+\hat{1}_w2_m3_w\hat{4}_m+\hat{1}_w2_m3_m\hat{4}_w)\la m_Pm_K\ra\\&
    +(\hat{1}_m2_m3_m\hat{4}_w+\hat{1}_m2_m3_w\hat{4}_m)\la w_Pm_K\ra
    +(\hat{1}_w2_m3_m\hat{4}_m+\hat{1}_m2_w3_m\hat{4}_m)\la m_Pw_K\ra\big)\\&
    \times (\mathrm{antifundamental})\,.
\end{align*}
We have used a shorthand notation: $i_m=\la\epsilon_im_i\ra$ and $i_w=\la\epsilon_iw_i\ra$.\\
The computation of this amplitude is carried out in \cite{Cheung:2009dc}. One needs to specify the little group objects $m$ and $w$ for the internal particles $P,K$. Since $k_P=-k_K$, we can fix $\kappa_{pA}=i\kappa_{kA}$ and $\kappa_{p}^A=i\kappa_k^A$. Then $m_P,\tilde{m}_P$ are defined by \eqref{defkappa} 
and $w,\tilde{w}$ are their inverses. We can then write:
\begin{align*}
    (k_1\wedge k_2)_A^B&=m_{Pa}\tilde{m}_{P\dot{a}}\kappa_{PA}^a\kappa_P^{B\dot{a}}\\
    &=-m_{Pa}\tilde{m}_{P\dot{a}}\kappa_{KA}^a\kappa_K^{B\dot{a}}\,.
\end{align*}
Contracting with $\hat{\kappa}_{i\dot{c}}^A\hat{\kappa}_{Bjc}\tw_{i\dot{d}}w_{jd}\epsilon^{\dot{c}\dot{d}}\epsilon^{cd}$, where $i,j=3$ or $4$:
\begin{equation}
    m_{Pa}\tilde{m}_{P\dot{a}}m_{K}^{a}\tilde{m}_{K}^{\dot{a}}=-(k_1\wedge k_2)_A^B\hat{\kappa}_{i\dot{c}}^A\hat{\kappa}_{Bjc}\tw_{i\dot{d}}w_{jd}\epsilon^{\dot{c}\dot{d}}\epsilon^{cd}=-s_{14}\,.
\end{equation}
Exploiting this property one can impose
\begin{equation}
    \la m_P w_K\ra=0\,,
\end{equation}
and choose normalizations so that:
\begin{equation}
    w_K=\frac{m_P}{\sqrt{-s_{14}}}\qquad \tilde{w}_K=\frac{\tilde{m}_P}{\sqrt{-s_{14}}}
\end{equation}
The four point amplitude above then becomes:\\
\begin{multline}
     A_4(1234)=
    \frac{1}{s_{12}}\frac{1}{\la m_Pm_K\ra}\big(\hat{1}_m2_m3_m\hat{4}_m
    -s_{14}(\hat{1}_m2_w3_m\hat{4}_w+\hat{1}_m2_w3_w\hat{4}_m+\hat{1}_w2_m3_w\hat{4}_m\\+\hat{1}_w2_m3_m\hat{4}_w)
    \times (\mathrm{antifundamental})\,.
\end{multline}    
One can then check that:
\begin{equation}
    \hat{1}_m2_m3_m\hat{4}_m
    -s_{14}(\hat{1}_m2_w3_m\hat{4}_w+\hat{1}_m2_w3_w\hat{4}_m+\hat{1}_w2_m3_w\hat{4}_m\\+\hat{1}_w2_m3_m\hat{4}_w)=
    \la\hat{1}23\hat{4}\ra\,,
\end{equation}
by projecting it on the base $m_i,\,w_i$. This gives:
\begin{equation}
    A_4(1234)=\frac{1}{s_{12}s_{14}}\la\hat{1}23\hat{4}\ra[\hat{1}23\hat{4}]=\frac{\la1234\ra[1234]}{s_{12}s_{14}}\,,
\end{equation}
where the second equality follows from the invariance of the polarization spinors under the shift.

\bibliography{../twistor-bib}
\bibliographystyle{JHEP}

\end{document}